%%%%%%%%%%%%%%%%%%  tex macros for preprints, cm version %%%%%%%%%%%%%%
%         (P. Ginsparg <ginsparg@lanl.gov>, last updated 7/94)
%         hypertex extensions (still provisional), 7/26/94
%	  Some modifications by C.R.Mafra, 2012

%comment out this line to restore non-hyper functionality
%\input hyperbasics

\input amssym.tex % for blackboard bold

\def\unredoffs{}
\tolerance=1000\hfuzz=2pt
\catcode`\@=11 % This allows us to modify PLAIN macros.
\ifx\hyperdef\UNd@FiNeD\def\hyperdef#1#2#3#4{#4}\def\hyperref#1#2#3#4{#4}\def\href#1#2{#2}\fi
\magnification=1200\unredoffs\baselineskip=16pt plus 2pt minus 1pt
\def\Date#1{\vfill\leftline{#1}\tenpoint\supereject%
\footline={\hss\tenrm\hyperdef\hypernoname{page}\folio\folio\hss}}%
% (restores pagenumbers)

%%%%%% Hour:Minute %%%%%%%%%%%%%%%%%
{\count255=\time\divide\count255 by 60 \xdef\hourmin{\number\count255}
 \multiply\count255 by-60\advance\count255 by\time
 \xdef\hourmin{\hourmin:\ifnum\count255<10 0\fi\the\count255}
}
\def\date{\number\day.\number\month.\number\year\ at \hourmin}

%%%%%%%%%%%% Draft mode %%%%%%%%%%%%%
% puts date/time on each page in big mode, writes labels in margins

% use \nolabels to get rid of eqn, ref, and fig labels in draft mode
\def\nolabels{\def\wrlabeL##1{}\def\eqlabeL##1{}\def\reflabeL##1{}}
\def\writelabels{\def\wrlabeL##1{\leavevmode\vadjust{\rlap{\smash%
{\line{{\escapechar=` \hfill\rlap{\sevenrm\hskip.03in\string##1}}}}}}}%
\def\eqlabeL##1{{\escapechar-1\rlap{\sevenrm\hskip.05in\string##1}}}%
\def\reflabeL##1{\noexpand\llap{\noexpand\sevenrm\string\string\string##1}}}
\nolabels

% tagged sec numbers
\global\newcount\secno \global\secno=0
\global\newcount\meqno \global\meqno=1
\def\s@csym{}

%%%%%%%%% Section %%%%%%%%%%%%%
\def\newsec#1\par{\global\advance\secno by1%
{\toks0{#1}\message{(\the\secno. \the\toks0)}}%
\global\subsecno=0\eqnres@t\let\s@csym\secsym\xdef\secn@m{\the\secno}\noindent
{\bf\hyperdef\hypernoname{section}{\the\secno}{\the\secno.} #1}%
\writetoca{{\string\hyperref{}{section}{\the\secno}{\bf \the\secno\quad}} {\bf #1}}\par%
\nobreak\medskip\nobreak\noindent\ignorespaces}
\def\eqnres@t{\xdef\secsym{\the\secno.}\global\meqno=1\bigbreak\bigskip}
\def\sequentialequations{\def\eqnres@t{\bigbreak}}\xdef\secsym{}

%%%%%%%% Subsection %%%%%%%%%%%
\global\newcount\subsecno \global\subsecno=0
\def\subsec#1\par{\global\advance\subsecno by1%
{\toks0{#1}\message{(\s@csym\the\subsecno. \the\toks0)}}%
\global\subsubsecno=0%
\ifnum\lastpenalty>9000\else\bigbreak\fi
\noindent{\it\hyperdef\hypernoname{subsection}{\secn@m.\the\subsecno}%
{\secn@m.\the\subsecno.} #1}\writetoca{\string\hskip1.45cm
{\string\hyperref{}{subsection}{\secn@m.\the\subsecno}{\secn@m.\the\subsecno.}}
{#1}}\par\nobreak\medskip\nobreak\noindent\ignorespaces}

%%%%%%%%%%%%%%% Subsubsection %%%%%%%%%%%%%%%%%%%%%%%%%%%%%%%%%%%%
\global\newcount\subsubsecno \global\subsubsecno=0
\def\subsubsec#1\par{\global\advance\subsubsecno by1%
{\toks0{#1}\message{(\secn@m.\the\subsecno.\the\subsubsecno. \the\toks0)}}%
\global\subsubsubsecno=0%
\ifnum\lastpenalty>9000\else\bigbreak\fi
\noindent{\it\hyperdef\hypernoname{subsubsection}{\secn@m.\the\subsecno\the\subsubsecno}%
{\secn@m.\the\subsecno.\the\subsubsecno.} #1}
%%% Add Subsubsections to Index
%\writetoca{\string\quad{\string\hyperref{}{subsubsection}{\the\secno\the\subsecno\the
%\subsubsecno}{\baselineskip=9pt\it\the\secno.\the\subsecno.\the\subsubsecno.}}
% {\baselineskip=9pt\it\ #1}}
\par\nobreak\medskip\nobreak\noindent\ignorespaces}

%%%%%%%%%%%%%%% Subsubsubsection %%%%%%%%%%%%%%%%%%%%%%%%%%%%%%%%%%%%
\global\newcount\subsubsubsecno \global\subsubsubsecno=0
\def\subsubsubsec#1\par{\global\advance\subsubsubsecno by1%
{\toks0{#1}\message{(\secn@m.\the\subsecno.\the\subsubsecno.\the\subsubsubsecno \the\toks0)}}%
\ifnum\lastpenalty>9000\else\bigbreak\fi
\noindent{\it\hyperdef\hypernoname{subsubsection}{\secn@m.\the\subsecno\the\subsubsecno\the\subsubsubsecno}%
{\secn@m.\the\subsecno.\the\subsubsecno.\the\subsubsubsecno.} #1}%
\par\nobreak\medskip\nobreak\noindent\ignorespaces}

%%%%%%%%% sections with automatic labels %%%%%%%%%%%%%%%%%%%%%%%%%%

%%%%%%%%% section with label %%%%%%%%%%%%%%%%%%%%%%%%%%%%%%%%
\def\newnewsec#1#2\par{\global\advance\secno by1%
{\toks0{#2}\message{(\secn@m. \the\toks0)}}%
\global\subsecno=0\eqnres@t\let\s@csym\secsym\xdef\secn@m{\the\secno}\noindent
\ifnum\lastpenalty>9000\else\bigbreak\fi
\noindent{\bf\hyperdef\hypernoname{section}{\secn@m}{\secn@m.} #2}%
\writetoca{{\string\hyperref{}{section}{\the\secno}{\bf \the\secno\quad}} {\bf #2}}
%define section label
\DefWarn#1%
\xdef#1{\noexpand\hyperref{}{section}{\the\secno}%
{\the\secno}}\writedef{#1\leftbracket#1}\wrlabeL{#1=#1}%
\par\nobreak\medskip\nobreak\noindent\ignorespaces}

%%%%%%%%% subsection with label %%%%%%%%%%%%%%%%%%%%%%%%%%%%%%%%
\def\newsubsec#1#2\par{\global\advance\subsecno by1%
{\toks0{#2}\message{(\secn@m.\the\subsecno. \the\toks0)}}%
\global\subsubsecno=0%
\ifnum\lastpenalty>9000\else\bigbreak\fi
\noindent{\it\hyperdef\hypernoname{subsection}{\secn@m.\the\subsecno}%
{\secn@m.\the\subsecno.} #2}
%define section label
\DefWarn#1%
\xdef#1{\noexpand\hyperref{}{subsection}{\secn@m.\the\subsecno}%
{\secn@m.\the\subsecno}}\writedef{#1\leftbracket#1}\wrlabeL{#1=#1}%
\writetoca{\string\hskip1.45cm
{\string\hyperref{}{subsection}{\secn@m.\the\subsecno}{\secn@m.\the\subsecno.}}
{#2}}%
\par\nobreak\medskip\nobreak\noindent\ignorespaces}

%%%%%%%%% subsubsection with label %%%%%%%%%%%%%%%%%%%%%%%%%%%%%%%%
\def\newsubsubsec#1#2\par{\global\advance\subsubsecno by1%
{\toks0{#2}\message{(\secn@m.\the\subsecno.\the\subsubsecno. \the\toks0)}}%
\global\subsubsubsecno=0%
\ifnum\lastpenalty>9000\else\bigbreak\fi
\noindent{\it\hyperdef\hypernoname{subsubsection}{\secn@m.\the\subsecno\the\subsubsecno}%
{\secn@m.\the\subsecno.\the\subsubsecno.} #2}
%define section label
\DefWarn#1%
\xdef#1{\noexpand\hyperref{}{subsubsection}{\secn@m.\the\subsecno.\the\subsubsecno}%
{\secn@m.\the\subsecno.\the\subsubsecno}}\writedef{#1\leftbracket#1}\wrlabeL{#1=#1}%
\par\nobreak\medskip\nobreak\noindent\ignorespaces}

%%%%%%%%% subsubsubsection with label %%%%%%%%%%%%%%%%%%%%%%%%%%%%%%%%
\def\newsubsubsubsec#1#2\par{\global\advance\subsubsubsecno by1%
{\toks0{#2}\message{(\secn@m.\the\subsecno.\the\subsubsecno.\the\subsubsubsecno \the\toks0)}}%
\ifnum\lastpenalty>9000\else\bigbreak\fi
\noindent{\it\hyperdef\hypernoname{subsubsection}{\secn@m.\the\subsecno\the\subsubsecno\the\subsubsubsecno}%
{\secn@m.\the\subsecno.\the\subsubsecno.\the\subsubsubsecno.} #2}
%define section label
\DefWarn#1%
\xdef#1{\noexpand\hyperref{}{subsubsubsection}{\secn@m.\the\subsecno.\the\subsubsecno.\the\subsubsubsecno}%
{\secn@m.\the\subsecno.\the\subsubsecno.\the\subsubsubsecno}}\writedef{#1\leftbracket#1}\wrlabeL{#1=#1}%
\par\nobreak\medskip\nobreak\noindent\ignorespaces}

%%%%%%% Appendix %%%%%%%%%%%%%%
\def\appendix#1#2{\global\meqno=1\global\subsecno=0\global\subsubsecno=0\xdef\secsym{\hbox{#1.}}%
\bigbreak\bigskip\noindent{\bf Appendix \hyperdef\hypernoname{appendix}{#1}%
{#1.} #2}{\toks0{(#1. #2)}\message{\the\toks0}}%
\xdef\s@csym{#1.}\xdef\secn@m{#1}%
\writetoca{{\string\hyperref{}{appendix}{#1}{\bf {#1}\quad}} {\bf #2}}%
\par\nobreak\medskip\nobreak}

% \eqn\label{a+b=c}   gives displayed equation, numbered consecutively within sections.
% \eqnn, \eqna        define labels in advance, use \eqna\label before an eqalign and
%                     later \label a, \label b etc inside eqalign to get (2.3a), (2.3b) etc
%
\def\checkm@de#1#2{\ifmmode{\def\f@rst##1{##1}\hyperdef\hypernoname{equation}%
{#1}{#2}}\else\hyperref{}{equation}{#1}{#2}\fi}
\def\eqnn#1{\DefWarn#1\xdef #1{(\noexpand\relax\noexpand\checkm@de%
{\s@csym\the\meqno}{\secsym\the\meqno})}%
\wrlabeL#1\writedef{#1\leftbracket#1}\global\advance\meqno by1}
\def\f@rst#1{\c@t#1a\em@ark}\def\c@t#1#2\em@ark{#1}
\def\eqna#1{\DefWarn#1\wrlabeL{#1$\{\}$}%
\xdef #1##1{(\noexpand\relax\noexpand\checkm@de%
{\s@csym\the\meqno\noexpand\f@rst{##1}1}{\hbox{$\secsym\the\meqno##1$}})}
\writedef{#1\numbersign1\leftbracket#1{\numbersign1}}\global\advance\meqno by1}
\def\eqn#1#2{\DefWarn#1%
\xdef #1{(\noexpand\hyperref{}{equation}{\s@csym\the\meqno}%
{\secsym\the\meqno})}$$#2\eqno(\hyperdef\hypernoname{equation}%
{\s@csym\the\meqno}{\secsym\the\meqno})\eqlabeL#1$$%
\writedef{#1\leftbracket#1}\global\advance\meqno by1}
\def\xeqn{\expandafter\xe@n}\def\xe@n(#1){#1}
\def\xeqna#1{\expandafter\xe@n#1}
\def\eqns#1{(\e@ns #1{\hbox{}})}
\def\e@ns#1{\ifx\UNd@FiNeD#1\message{eqnlabel \string#1 is undefined.}%
\xdef#1{(?.?)}\fi{\let\hyperref=\relax\xdef\next{#1}}%
\ifx\next\em@rk\def\next{}\else%
\ifx\next#1\xeqn#1\else\def\n@xt{#1}\ifx\n@xt\next#1\else\xeqna#1\fi
\fi\let\next=\e@ns\fi\next}
\def\DefWarn#1{}
%
% footnotes
\newskip\footskip\footskip14pt plus 1pt minus 1pt %sets footnote baselineskip
\def\footnotefont{\ninepoint}\def\f@t#1{\footnotefont #1\@foot}
\def\f@@t{\baselineskip\footskip\bgroup\footnotefont\aftergroup\@foot\let\next}
\setbox\strutbox=\hbox{\vrule height9.5pt depth4.5pt width0pt}
\global\newcount\ftno \global\ftno=0
\def\foot{\global\advance\ftno by1\def\foot@rg{\hyperref{}{footnote}%
{\the\ftno}{\the\ftno}\xdef\foot@rg{\noexpand\hyperdef\noexpand\hypernoname%
{footnote}{\the\ftno}{\the\ftno}}}\footnote{$^{\foot@rg}$}}
%
%
%     \ref\label{text}
% generates a number, assigns it to \label, generates an entry.
% To list the refs on a separate page,  \listrefs
%
\global\newcount\refno \global\refno=1
\newwrite\rfile
\def\ref{[\hyperref{}{reference}{\the\refno}{\the\refno}]\nref}
\def\nref#1{\DefWarn#1%
\xdef#1{[\noexpand\hyperref{}{reference}{\the\refno}{\the\refno}]}%
\writedef{#1\leftbracket#1}%
\ifnum\refno=1\immediate\openout\rfile=\jobname.refs\fi
\chardef\wfile=\rfile\immediate\write\rfile{\noexpand\item{[\noexpand\hyperdef%
\noexpand\hypernoname{reference}{\the\refno}{\the\refno}]\ }%
\reflabeL{#1\hskip.31in}\pctsign}\global\advance\refno by1\findarg}
%	horrible hack to sidestep tex \write limitation
\def\findarg#1#{\begingroup\obeylines\newlinechar=`\^^M\pass@rg}
{\obeylines\gdef\pass@rg#1{\writ@line\relax #1^^M\hbox{}^^M}%
\gdef\writ@line#1^^M{\expandafter\toks0\expandafter{\striprel@x #1}%
\edef\next{\the\toks0}\ifx\next\em@rk\let\next=\endgroup\else\ifx\next\empty%
\else\immediate\write\wfile{\the\toks0}\fi\let\next=\writ@line\fi\next\relax}}
\def\striprel@x#1{} \def\em@rk{\hbox{}}
\def\lref{\begingroup\obeylines\lr@f}
\def\lr@f#1#2{\DefWarn#1\gdef#1{\let#1=\UNd@FiNeD\ref#1{#2}}\endgroup\unskip}
\def\semi{;\hfil\break}
\def\addref#1{\immediate\write\rfile{\noexpand\item{}#1}} %now unnecessary
\def\listrefs{\vfill\supereject\immediate\closeout\rfile\writestoppt
\baselineskip=\footskip\centerline{{\bf References}}\bigskip{\parindent=20pt%
\frenchspacing\escapechar=` \input \jobname.refs\vfill\eject}\nonfrenchspacing}
\def\startrefs#1{\immediate\openout\rfile=\jobname.refs\refno=#1}
\def\xref{\expandafter\xr@f}\def\xr@f[#1]{#1}
\def\refs#1{\count255=1[\r@fs #1{\hbox{}}]}
\def\r@fs#1{\ifx\UNd@FiNeD#1\message{reflabel \string#1 is undefined.}%
\nref#1{need to supply reference \string#1.}\fi%
\vphantom{\hphantom{#1}}{\let\hyperref=\relax\xdef\next{#1}}%
\ifx\next\em@rk\def\next{}%
\else\ifx\next#1\ifodd\count255\relax\xref#1\count255=0\fi%
\else#1\count255=1\fi\let\next=\r@fs\fi\next}
%

%
% this is ugly, but moore insists
\newwrite\ffile\global\newcount\figno \global\figno=1
\def\fig{fig.~\hyperref{}{figure}{\the\figno}{\the\figno}\nfig}
\def\nfig#1{\DefWarn#1%
\xdef#1{fig.~\noexpand\hyperref{}{figure}{\the\figno}{\the\figno}}%
\writedef{#1\leftbracket fig.\noexpand~\xfig#1}%
\ifnum\figno=1\immediate\openout\ffile=\jobname.figs\fi\chardef\wfile=\ffile%
{\let\hyperref=\relax
\immediate\write\ffile{\noexpand\medskip\noexpand\item{Fig.\ %
\noexpand\hyperdef\noexpand\hypernoname{figure}{\the\figno}{\the\figno}. }
\reflabeL{#1\hskip.55in}\pctsign}}\global\advance\figno by1\findarg}
\def\xfig{\expandafter\xf@g}\def\xf@g fig.\penalty\@M\ {}
\def\figs#1{figs.~\f@gs #1{\hbox{}}}
\def\f@gs#1{{\let\hyperref=\relax\xdef\next{#1}}\ifx\next\em@rk\def\next{}\else
\ifx\next#1\xfig #1\else#1\fi\let\next=\f@gs\fi\next}
%
%% because TeXlive 2011 is buggy wrt to tikz pictures with plain TeX..
\def\figin{\epsfcheck\figin}\def\figins{\epsfcheck\figins}
\def\epsfcheck{\ifx\epsfbox\UnDeFiNeD
\message{(NO epsf.tex, FIGURES WILL BE IGNORED)}
\gdef\figin##1{\vskip2in}\gdef\figins##1{\hskip.5in}% blank space instead
\else\message{(FIGURES WILL BE INCLUDED)}%
\gdef\figin##1{##1}\gdef\figins##1{##1}\fi}
\def\figinsert{\goodbreak\topinsert}
\def\ifig#1#2#3{\DefWarn#1\xdef#1{fig.~\the\figno}
\writedef{#1\leftbracket fig.\noexpand~\the\figno}%
\figinsert\figin{\centerline{#3}}
\smallskip
\leftskip=0pt \rightskip=0pt
\baselineskip12pt\noindent
{{\bf Fig.~\the\figno}\ \ninepoint #2}
\medskip
\global\advance\figno by1\par\endinsert}
%%%%%%%%%%%%%%%%%%%%%%%%%%%%%%%%%%%%%%%%%%%%%%%%%%%%%%%%%
\newwrite\lfile
{\escapechar-1\xdef\pctsign{\string\%}\xdef\leftbracket{\string\{}
\xdef\rightbracket{\string\}}\xdef\numbersign{\string\#}}
\def\writedefs{\immediate\openout\lfile=label.defs \def\writedef##1{%
{\let\hyperref=\relax\let\hyperdef=\relax\let\hypernoname=\relax
 \immediate\write\lfile{\string\checkdef\string##1\rightbracket}}}}%
\def\writestop{\def\writestoppt{\immediate\write\lfile{\string\pageno
 \the\pageno\string\startrefs\leftbracket\the\refno\rightbracket
 \string\def\string\secsym\leftbracket\secsym\rightbracket
 \string\secno\the\secno\string\meqno\the\meqno}\immediate\closeout\lfile}}
\def\writestoppt{}\def\writedef#1{}

% Section, subsection and appendix labels %
% Note that there must be a blanck line after \newsec,\subsec and before \seclab,\subseclab!
\def\seclab#1\par{\DefWarn#1%
\xdef #1{\noexpand\hyperref{}{section}{\the\secno}{\the\secno}}%
\writedef{#1\leftbracket#1}\wrlabeL{#1=#1}\par%
\nobreak\medskip\nobreak\noindent\ignorespaces}
\def\subseclab#1\par{\DefWarn#1%
\xdef #1{\noexpand\hyperref{}{subsection}{\the\secno.\the\subsecno}%
{\the\secno.\the\subsecno}}\writedef{#1\leftbracket#1}\wrlabeL{#1=#1}\par%
\nobreak\medskip\nobreak\noindent\ignorespaces}
\def\subsubseclab#1\par{\DefWarn#1%
\xdef#1{\noexpand\hyperref{}{subsubsection}{\the\secno.\the\subsecno.\the\subsubsecno}%
{\the\secno.\the\subsecno.\the\subsubsecno}}\writedef{#1\leftbracket#1}\wrlabeL{#1=#1}\par%
\nobreak\medskip\nobreak\noindent\ignorespaces}
\def\applab#1\par{\DefWarn#1%
\xdef#1{\noexpand\hyperref{}{appendix}{\secn@m}{\secn@m}}%
\writedef{#1\leftbracket#1}\wrlabeL{#1=#1}%
\par\nobreak\medskip\nobreak\noindent\ignorespaces}
\def\appsublab#1{\DefWarn#1%
\xdef #1{\noexpand\hyperref{}{appendix}{\secn@m.\the\subsecno}{\secn@m.\the\subsecno}}%
\writedef{#1\leftbracket#1}\wrlabeL{#1=#1}}
\newwrite\tfile \def\writetoca#1{}
\def\leaderfill{\leaders\hbox to 1em{\hss.\hss}\hfill}
% use this to write file with table of contents
\def\writetoc{\immediate\openout\tfile=\jobname.toc
   \def\writetoca##1{{\edef\next{\write\tfile{\noindent ##1
   \string\leaderfill{
% comment this line if you don't want hyperlinked page numbers on TOC
   \string\hyperref{}{page}{\noexpand\number\pageno}%
   {\noexpand\number\pageno}} \par}}\next}}
}
% and this lists table of contents on second pass
\newread\ch@ckfile
\def\listtoc{\immediate\closeout\tfile\immediate\openin\ch@ckfile=\jobname.toc
\ifeof\ch@ckfile\message{no file \jobname.toc, no table of contents this pass}%
\else\closein\ch@ckfile\centerline{\bf Contents}\nobreak\medskip%
{\baselineskip=16pt\footnotefont\parskip=0pt\catcode`\@=11\input\jobname.toc
\catcode`\@=12\bigbreak\bigskip}\fi}
\catcode`\@=12 % at signs are no longer letters
\def\tenpoint{\def\rm{\fam0\tenrm}% switch back to 10-point type
\textfont0=\tenrm \scriptfont0=\sevenrm \scriptscriptfont0=\fiverm
\textfont1=\teni  \scriptfont1=\seveni  \scriptscriptfont1=\fivei
\textfont2=\tensy \scriptfont2=\sevensy \scriptscriptfont2=\fivesy
\textfont\itfam=\tenit \def\it{\fam\itfam\tenit}\def\footnotefont{\ninepoint}%
\textfont\bffam=\tenbf \def\bf{\fam\bffam\tenbf}\def\sl{\fam\slfam\tensl}\rm}
\font\ninerm=cmr9 \font\sixrm=cmr6 \font\ninei=cmmi9 \font\sixi=cmmi6
\font\ninesy=cmsy9 \font\sixsy=cmsy6 \font\ninebf=cmbx9
\font\nineit=cmti9 \font\ninesl=cmsl9 \skewchar\ninei='177
\skewchar\sixi='177 \skewchar\ninesy='60 \skewchar\sixsy='60
\def\ninepoint{\def\rm{\fam0\ninerm}% switch to footnote font
\textfont0=\ninerm \scriptfont0=\sixrm \scriptscriptfont0=\fiverm
\textfont1=\ninei \scriptfont1=\sixi \scriptscriptfont1=\fivei
\textfont2=\ninesy \scriptfont2=\sixsy \scriptscriptfont2=\fivesy
\textfont\itfam=\ninei \def\it{\fam\itfam\nineit}\def\sl{\fam\slfam\ninesl}%
\textfont\bffam=\ninebf \def\bf{\fam\bffam\ninebf}\rm}
%
%---------------------------------------------------------------------
\hyphenation{anom-aly anom-alies coun-ter-term coun-ter-terms}

% Caption for inline tikzpictures
%\def\DefWarn#1{}
\def\tikzcaption#1#2{\DefWarn#1\xdef#1{Fig.~\the\figno}
\writedef{#1\leftbracket Fig.\noexpand~\the\figno}%
{
\smallskip
\leftskip=20pt \rightskip=20pt \baselineskip12pt\noindent
{{\bf Fig.~\the\figno}\ \ninepoint #2}
\bigskip
\global\advance\figno by1 \par}}

% convert numbers [1-9] to upper case letters [A-I]
\def\ntoalpha#1{%
\ifcase#1%
@%
\or A\or B\or C\or D\or E\or F\or G\or H\or I\or J\or K\or L\or M%
\fi
}

% Appendix label
\global\newcount\appno \global\appno=1
\def\applab#1{\xdef #1{\ntoalpha{\appno}}\writedef{#1\leftbracket#1}\wrlabeL{#1=#1}
\global\advance\appno by1}

% Clean up the title page definitions
\def\preprint#1 #2\par{\rightline{\vbox{\baselineskip12pt\hbox{#1}\hbox{#2}}}\vskip2cm}
% title with more than one line (note the blanck line in between)
%\title some line
%
%\tile another line
\def\title#1\par{\centerline{\bf #1}\nopagenumbers\pageno=0}
\def\author#1\par{\bigskip\bigskip\centerline{#1}}

\newcount\addressno

\def\email#1#2{%\unskip$^#1$
\footnote{\null}{\kern-\parindent \llap{$^#1$\hskip1pt}email: #2}}

% centermode for address lines
\def\startcenter{%
  \par
  \begingroup
  \leftskip=0pt plus 1fil
  \rightskip=\leftskip
  \parindent=0pt
  \parfillskip=0pt
}
\def\stopcenter{\endgroup}

\def\address{\bigskip%
  \ifnum\the\addressno=0\else\stopcenter\endgroup\fi
  \advance\addressno by 1%
  \begingroup
  \startcenter
  \it
  \obeylines
  \addressAux
}
\def\addressAux#1{#1}

% need to stop center mode and obeylines from address
\def\abstract{\stopcenter\endgroup\bigskip\bigskip\noindent}

% some sample definitions
\def\Dsl{\,\raise.15ex\hbox{/}\mkern-13.5mu D} %this one can be subscripted
\def\dsl{\raise.15ex\hbox{/}\kern-.57em\partial}
 
\def\boxeqn#1{\vcenter{\vbox{\hrule\hbox{\vrule\kern3pt\vbox{\kern3pt
	\hbox{${\displaystyle #1}$}\kern3pt}\kern3pt\vrule}\hrule}}}

 %pound sterling

\def\ap{{\alpha^{\prime}}}

\def\d{{\delta}}

\def\half{{1\over 2}}
\def\p{{\partial}}

\def\bar{\overline}
\def\({\left(}
\def\){\right)}

\def\cF{{\cal F}}
\def\cJ{{\cal J}}
\def\cK{{\cal K}}

\def\cY{{\cal Y}}
\def\cZ{{\cal Z}}

% blackboard bold

% primed summation symbol

% length of words, |P|

\def\perm#1{{\rm perm}#1}

\def\Im{\mathop{{\rm Im}}} %redefine plain TeX \Im..
% small inlined fractions, from the TeXbook
\def\sfrac#1/#2{\kern.1em\raise.5ex\hbox{\the\scriptfont0 #1}%
\kern-.1em/\kern-.15em\lower.25ex\hbox{\the\scriptfont0 #2}}

%shuffle product
\font\tenshuffle=shuffle10 \font\sevenshuffle=shuffle7 \font\fiveshuffle=shuffle7 at 5pt
\def\shuffle{{%
\def\Dshuffle{\mathbin{\hbox{\tenshuffle\char'001}}}%
\def\Sshuffle{\mathbin{\hbox{\sevenshuffle\char'001}}}%
\def\SSshuffle{\mathbin{\hbox{\fiveshuffle\char'001}}}%
\mathchoice{\Dshuffle}{\Dshuffle}{\Sshuffle}{\SSshuffle}}}

%\owedge

% From Knuth's \pfbox macro
\def\qed{\hbox{\hskip 3pt
%\lower2pt
\vbox{\hrule\hbox to 7pt{\vrule height 7pt\hfill\vrule}
\hrule}}\hskip3pt}

% do not display overfull marks
\overfullrule=0pt\relax

\frenchspacing

% DefWarn-like behavior for labels defined in advance
\def\checkdef#1#2{
\ifx\UndeFined#1%
	\def#1{#2}
%\immediate\write16{*** define label \string#1 by #2 ***}
\else
	\immediate\write16{*** BUG ***: the label \string#1 is already defined ***}
\fi
}
% define labels in advance
\newread\instream

\openin\instream= label.defs
\ifeof\instream\message{No labels in advance yet. Wait till next pass.}
\else\closein\instream \input label.defs
\fi
\writedefs

%%% References with hyperlinks to arxiv.org; both styles accepted
% Change arXiv to \arXiv ie
% [arXiv:hep-th/1234567].     --> [\arXiv:hep-th/1234567].
% [arXiv:1234.5678 [hep-th]]. --> [\arXiv:1234.5678 [hep-th]].
% Need to strip trailing [hep-th] (if present) to define valid URL
\def\arXiv:#1].{\hepthStrip#1 \nil}
\def\hepthStrip#1 #2\nil{\href{http://arxiv.org/abs/#1}{arXiv:#1 #2\unskip}].}

%%% Fraktur fonts for Berends-Giele components

\input amssym
\input epsf
\input localpaper.defs
%\input tikz

%\writelabels
%\draft

\title Towards the n-point one-loop superstring amplitude III:

\title One-loop correlators and their double-copy structure

\author
Carlos R. Mafra\email{\dagger}{c.r.mafra@soton.ac.uk}$^{\dagger}$ and
Oliver Schlotterer\email{\ddagger}{olivers@aei.mpg.de}$^{\ddagger,\ast}$

\address
$^\dagger$Mathematical Sciences and STAG Research Centre, University of Southampton,
Highfield, Southampton, SO17 1BJ, UK

\address
$^\ddagger$Max--Planck--Institut f\"ur Gravitationsphysik
Albert--Einstein--Institut, 14476 Potsdam, Germany

\address
$^\ast$Perimeter Institute for Theoretical Physics, Waterloo, ON N2L 2Y5, Canada

\abstract
In this final part of a series of three papers, we will assemble supersymmetric
expressions for one-loop correlators in pure-spinor superspace that are BRST
invariant, local, and single valued.
A key driving force in this construction is the generalization of a
so far unnoticed property at tree level;
the correlators have the symmetry structure akin
to {\it Lie polynomials}.
One-loop correlators up to seven points are presented in a variety of representations
manifesting different subsets of their defining properties.
These expressions are related via identities obeyed by the kinematic superfields and
worldsheet functions spelled out in the first two parts of this series and reflecting
a duality between the two kinds of ingredients.

Interestingly, the expression for the eight-point correlator following from our
method seems to capture correctly all the dependence on the worldsheet punctures
but leaves undetermined the coefficient of the holomorphic Eisenstein series
${\rm G}_4$. By virtue of chiral splitting, closed-string correlators follow
from the double copy of the open-string results.

\Date {December 2018}

%**************************************

\lref\wipI{
C.R.~Mafra and O.~Schlotterer,
``Towards the n-point one-loop superstring amplitude I:
Pure spinors and superfield kinematics'', [arXiv:1812.10969 [hep-th]]\semi
C.R.~Mafra and O.~Schlotterer,
``Towards the n-point one-loop superstring amplitude II:
Worldsheet functions and their duality to kinematics'', [arXiv:1812.10970 [hep-th]]\semi
C.R.~Mafra and O.~Schlotterer,
``Towards the n-point one-loop superstring amplitude III:
 One-loop correlators and their double-copy structure'', [arXiv:1812.10971 [hep-th]]
}

\lref\PSS{
  C.R.~Mafra,
  ``PSS: A FORM Program to Evaluate Pure Spinor Superspace Expressions,''
[arXiv:1007.4999 [hep-th]].
%%CITATION = arXiv:1007.4999%%
}
\lref\FORM{
	J.A.M.~Vermaseren,
	``New features of FORM,''
	arXiv:math-ph/0010025.
	%%CITATION = MATH-PH/0010025;%%
\semi
	M.~Tentyukov and J.A.M.~Vermaseren,
	``The multithreaded version of FORM,''
	arXiv:hep-ph/0702279.
	%%CITATION = HEP-PH/0702279;%%
}

\lref\GeyerELA{
  Y.~Geyer and R.~Monteiro,
  ``Gluons and gravitons at one loop from ambitwistor strings,''
JHEP {\bf 1803}, 068 (2018).
[arXiv:1711.09923 [hep-th]].
%%CITATION = QMUL-PH-17-29%%
}

\lref\GomezUHA{
  H.~Gomez, C.~R.~Mafra and O.~Schlotterer,
  ``Two-loop superstring five-point amplitude and $S$-duality,''
Phys.\ Rev.\ D {\bf 93}, no. 4, 045030 (2016).
[arXiv:1504.02759 [hep-th]].
%%CITATION = DAMTP-2015-20%%
}

\lref\MafraMJA{
  C.~R.~Mafra and O.~Schlotterer,
  ``Two-loop five-point amplitudes of super Yang-Mills and supergravity in pure spinor superspace,''
JHEP {\bf 1510}, 124 (2015).
[arXiv:1505.02746 [hep-th]].
%%CITATION = DAMTP-2015-25%%
}

\lref\MafraMCC{
  C.~R.~Mafra and O.~Schlotterer,
  ``Non-abelian $Z$-theory: Berends-Giele recursion for the $\alpha'$-expansion of disk integrals,''
JHEP {\bf 1701}, 031 (2017).
[arXiv:1609.07078 [hep-th]].
%%CITATION = arXiv:1609.07078%%
}

\lref\CarrascoYGV{
  J.~J.~M.~Carrasco, C.~R.~Mafra and O.~Schlotterer,
  ``Semi-abelian Z-theory: NLSM$+\phi^3$ from the open string,''
JHEP {\bf 1708}, 135 (2017).
[arXiv:1612.06446 [hep-th]].
%%CITATION = arXiv:1612.06446%%
}

\lref\Clavelli{
  L.~Clavelli, P.~H.~Cox and B.~Harms,
  ``Parity Violating One Loop Six Point Function in Type I Superstring Theory,''
  Phys.\ Rev.\ D {\bf 35} (1987) 1908.
  %%CITATION = PHRVA,D35,1908;%%
}

\lref\stanley{
	R.P. Stanley, ``Enumerative Combinatorics'', second edition, Cambridge University Press (2012).}

\lref\knuthconcrete{
	R. Graham, D.E. Knuth, and O. Patashnik,
 	``Concrete Mathematics: A Foundation for Computer Science'',
	Addison-Wesley Longman Publishing Co., Inc.,
	Boston, MA, USA, (1994).
}

\lref\stiefive{
  S.~Stieberger and T.R.~Taylor,
  ``NonAbelian Born-Infeld action and type 1. - heterotic duality 2: Nonrenormalization theorems,''
Nucl.\ Phys.\ B {\bf 648}, 3 (2003).
[hep-th/0209064].
%%CITATION = hep-th/0209064%%
}

\lref\lin{
	Z.~H.~Lin,
  	``One Loop Closed String Five Particle Fermion Amplitudes in the Covariant
  	Formulation,''
  	Int.\ J.\ Mod.\ Phys.\  A {\bf 5}, 299 (1990)
  	%%CITATION = IMPAE,A5,299;%%
  	\semi
  	Z.~H.~Lin, L.~Clavelli and S.~T.~Jones,
  	``Five Point Function in the Covariant Formulation of the Type I Superstring
  	Theory,''
  	Nucl.\ Phys.\  B {\bf 294}, 83 (1987).
  	%%CITATION = NUPHA,B294,83;%%
}

\lref\sen{
	J.~J.~Atick and A.~Sen,
  	``Covariant one loop fermion emission amplitudes in closed string theories,''
  	Nucl.\ Phys.\  B {\bf 293}, 317 (1987).
  	%%CITATION = NUPHA,B293,317;%%
}

\lref\tnn{
	D.E. Knuth, ``Two notes on notation'', Amer. Math. Monthly 99 (1992), no.\ 5, 403--422
	[math/9205211].
}

\lref\HeSPX{
  S.~He, O.~Schlotterer and Y.~Zhang,
  ``New BCJ representations for one-loop amplitudes in gauge theories and gravity,''
Nucl.\ Phys.\ B {\bf 930}, 328 (2018).
[arXiv:1706.00640 [hep-th]].
%%CITATION = arXiv:1706.00640%%
}

\lref\GreenSW{
  M.~B.~Green, J.~H.~Schwarz and L.~Brink,
  ``N=4 Yang-Mills and N=8 Supergravity as Limits of String Theories,''
Nucl.\ Phys.\ B {\bf 198}, 474 (1982).
%%CITATION = CALT-68-880%%
}

\lref\KK{
	R.~Kleiss, H.~Kuijf,
  	``Multi - Gluon Cross-sections And Five Jet Production At Hadron Colliders,''
	Nucl.\ Phys.\  {\bf B312}, 616 (1989).
	%%CITATION = Print-88-0425 (LEIDEN)%%
}
\lref\TsuchiyaVA{
	A.~Tsuchiya,
  	``More on One Loop Massless Amplitudes of Superstring Theories,''
	Phys.\ Rev.\ D {\bf 39}, 1626 (1989).
	%%CITATION = TIT/HEP-135%%
}

\lref\TsuchiyaJOO{
  A.~G.~Tsuchiya,
  ``On new theta identities of fermion correlation functions on genus g Riemann surfaces,''
[arXiv:1710.00206 [hep-th]].
%%CITATION = arXiv:1710.00206%%
}

\lref\BernQJ{
  Z.~Bern, J.~J.~M.~Carrasco and H.~Johansson,
  ``New Relations for Gauge-Theory Amplitudes,''
Phys.\ Rev.\ D {\bf 78}, 085011 (2008).
[arXiv:0805.3993 [hep-ph]].
%%CITATION = arXiv:0805.3993%%
}

\lref\verlindes{
	E.P.~Verlinde and H.L.~Verlinde,
	``Chiral Bosonization, Determinants and the String Partition Function,''
	Nucl.\ Phys.\ B {\bf 288}, 357 (1987).
	%%CITATION = Print-86-1416 (UTRECHT)%%
}

\lref\threeloop{
	H.~Gomez and C.~R.~Mafra,
  	``The closed-string 3-loop amplitude and S-duality,''
	JHEP {\bf 1310}, 217 (2013).
	[arXiv:1308.6567 [hep-th]].
	%%CITATION = DAMTP-2013-50%%
}

\lref\Ree{
	R. Ree, ``Lie elements and an algebra associated with shuffles'',
	Ann. Math. {\bf 62}, No. 2 (1958), 210--220.
}
\lref\BGschocker{
	M. Schocker,
	``Lie elements and Knuth relations,'' Canad. J. Math. {\bf 56} (2004), 871-882.
	[math/0209327].
}

\lref\lothaire{
	Lothaire, M., ``Combinatorics on Words'',
	(Cambridge Mathematical Library), Cambridge University Press (1997).
}

\lref\NLSM{
	J.~J.~M.~Carrasco, C.~R.~Mafra and O.~Schlotterer,
  	``Abelian Z-theory: NLSM amplitudes and $\alpha$'-corrections from the open string,''
	JHEP {\bf 1706}, 093 (2017).
	[arXiv:1608.02569 [hep-th]].
	%%CITATION = arXiv:1608.02569%%
}

\lref\DPellis{
	F.~Cachazo, S.~He and E.Y.~Yuan,
	``Scattering of Massless Particles: Scalars, Gluons and Gravitons,''
	JHEP {\bf 1407}, 033 (2014).
	[arXiv:1309.0885 [hep-th]].
	%%CITATION = arXiv:1309.0885%%
}

\lref\GreenTV{
  M.~B.~Green and M.~Gutperle,
  ``Effects of D instantons,''
Nucl.\ Phys.\ B {\bf 498}, 195 (1997).
[hep-th/9701093].
%%CITATION = hep-th/9701093%%
}

\lref\greenloop{
	M.~B.~Green and J.~H.~Schwarz,
  	``Supersymmetrical Dual String Theory. 3. Loops and Renormalization,''
	Nucl.\ Phys.\ B {\bf 198}, 441 (1982).
	%%CITATION = CALT-68-873%%
}

\lref\Richards{
	D.~M.~Richards,
	``The One-Loop Five-Graviton Amplitude and the Effective Action,''
	JHEP {\bf 0810}, 042 (2008).
	[arXiv:0807.2421 [hep-th]].
	%%CITATION = arXiv:0807.2421%%
}

\lref\fiveptNMPS{
	C.R.~Mafra and C.~Stahn,
  	``The One-loop Open Superstring Massless Five-point Amplitude with the Non-Minimal Pure Spinor Formalism,''
	JHEP {\bf 0903}, 126 (2009).
	[arXiv:0902.1539 [hep-th]].
	%%CITATION = arXiv:0902.1539%%
}

\lref\SiegelYI{
	W.~Siegel,
  	``Superfields in Higher Dimensional Space-time,''
	Phys.\ Lett.\ B {\bf 80}, 220 (1979).
	%%CITATION = HUTP-78/A030%%
}

\lref\reutenauer{
	C.~Reutenauer,
	``Free Lie Algebras'', London Mathematical Society Monographs, 1993.
}

\lref\fourptpaper{
	C.R.~Mafra,
  	``Four-point one-loop amplitude computation in the pure spinor formalism,''
	JHEP {\bf 0601}, 075 (2006).
	[hep-th/0512052].
	%%CITATION = hep-th/0512052%%
}
\lref\mafraids{
	C.R.~Mafra,
  	``Pure Spinor Superspace Identities for Massless Four-point Kinematic Factors,''
	JHEP {\bf 0804}, 093 (2008).
	[arXiv:0801.0580 [hep-th]].
	%%CITATION = IFT-P-001-2008%%
}

\lref\GreenUJ{
  M.~B.~Green, J.~G.~Russo and P.~Vanhove,
  ``Low energy expansion of the four-particle genus-one amplitude in type II superstring theory,''
JHEP {\bf 0802}, 020 (2008).
[arXiv:0801.0322 [hep-th]].
%%CITATION = arXiv:0801.0322%%
}

\lref\DHokerGMR{
  E.~D'Hoker, M.~B.~Green and P.~Vanhove,
  ``On the modular structure of the genus-one Type II superstring low energy expansion,''
JHEP {\bf 1508}, 041 (2015).
[arXiv:1502.06698 [hep-th]].
%%CITATION = DAMTP-08-02-2015%%
}

\lref\MafraNWR{
	C.R.~Mafra and O.~Schlotterer,
	``One-loop superstring six-point amplitudes and anomalies in pure spinor superspace,''
	JHEP {\bf 1604}, 148 (2016).
	[arXiv:1603.04790 [hep-th]].
	%%CITATION = arXiv:1603.04790%%
}

\lref\BergWUX{
  M.~Berg, I.~Buchberger and O.~Schlotterer,
  ``From maximal to minimal supersymmetry in string loop amplitudes,''
JHEP {\bf 1704}, 163 (2017).
[arXiv:1603.05262 [hep-th]].
%%CITATION = arXiv:1603.05262%%
}

\lref\GregoriHI{
  A.~Gregori, E.~Kiritsis, C.~Kounnas, N.~A.~Obers, P.~M.~Petropoulos and B.~Pioline,
  ``R**2 corrections and nonperturbative dualities of N=4 string ground states,''
Nucl.\ Phys.\ B {\bf 510}, 423 (1998).
[hep-th/9708062].
%%CITATION = hep-th/9708062%%
}

\lref\BroedelTTA{
  J.~Broedel, O.~Schlotterer and S.~Stieberger,
  ``Polylogarithms, Multiple Zeta Values and Superstring Amplitudes,''
Fortsch.\ Phys.\  {\bf 61}, 812 (2013).
[arXiv:1304.7267 [hep-th]].
%%CITATION = DAMTP-2013-22%%
}

\lref\EOMbbs{
  C.R.~Mafra and O.~Schlotterer,
  ``Multiparticle SYM equations of motion and pure spinor BRST blocks,''
JHEP {\bf 1407}, 153 (2014).
[arXiv:1404.4986 [hep-th]].
%%CITATION = AEI-2014-011%%
}

\lref\oneloopbb{
  C.R.~Mafra and O.~Schlotterer,
  ``The Structure of n-Point One-Loop Open Superstring Amplitudes,''
JHEP {\bf 1408}, 099 (2014).
[arXiv:1203.6215 [hep-th]].
%%CITATION = AEI-2012-032%%
}

\lref\GrisaruPX{
  M.~T.~Grisaru, A.~E.~M.~van de Ven and D.~Zanon,
  ``Four Loop beta Function for the N=1 and N=2 Supersymmetric Nonlinear Sigma Model in Two-Dimensions,''
Phys.\ Lett.\ B {\bf 173}, 423 (1986).
%%CITATION = HUTP-86/A020%%
}

\lref\GrisaruVI{
  M.~T.~Grisaru and D.~Zanon,
  ``$\sigma$ Model Superstring Corrections to the Einstein-hilbert Action,''
Phys.\ Lett.\ B {\bf 177}, 347 (1986).
%%CITATION = HUTP-86/A046%%
}

\lref\partI{
  C.R.~Mafra and O.~Schlotterer,
  ``Cohomology foundations of one-loop amplitudes in pure spinor superspace,''
[arXiv:1408.3605 [hep-th]].
%%CITATION = arXiv:1408.3605%%
}
\lref\PScohomology{
  N.~Berkovits,
  ``Cohomology in the pure spinor formalism for the superstring,''
JHEP {\bf 0009}, 046 (2000).
[hep-th/0006003].
%%CITATION = hep-th/0006003%%
\semi
N.~Berkovits and O.~Chandia,
  ``Lorentz invariance of the pure spinor BRST cohomology for the superstring,''
Phys.\ Lett.\ B {\bf 514}, 394 (2001).
[hep-th/0105149].
%%CITATION = IFT-P-041-2001%%
}

\lref\nptString{
	C.~R.~Mafra, O.~Schlotterer and S.~Stieberger,
  ``Complete N-Point Superstring Disk Amplitude I. Pure Spinor Computation,''
Nucl.\ Phys.\ B {\bf 873}, 419 (2013).
[arXiv:1106.2645 [hep-th]].
%%CITATION = arXiv:1106.2645%%
}

\lref\MafraGIA{
  C.~R.~Mafra and O.~Schlotterer,
  ``Solution to the nonlinear field equations of ten dimensional supersymmetric Yang-Mills theory,''
Phys.\ Rev.\ D {\bf 92}, no. 6, 066001 (2015).
[arXiv:1501.05562 [hep-th]].
%%CITATION = AEI-2015-005%%
}

\lref\wittentwistor{
	E.~Witten,
	``Twistor-Like Transform In Ten-Dimensions,''
	Nucl.\ Phys.\  B {\bf 266}, 245 (1986).
	%%CITATION = NUPHA,B266,245;%%
}
\lref\psf{
 	N.~Berkovits,
	``Super-Poincare covariant quantization of the superstring,''
	JHEP {\bf 0004}, 018 (2000)
	[arXiv:hep-th/0001035].
	%%CITATION = JHEPA,0004,018;%%
}
\lref\MPS{
  N.~Berkovits,
  ``Multiloop amplitudes and vanishing theorems using the pure spinor formalism for the superstring,''
JHEP {\bf 0409}, 047 (2004).
[hep-th/0406055].
%%CITATION = hep-th/0406055%%
}

\lref\MafraIOJ{
C.~R.~Mafra and O.~Schlotterer,
  ``Double-Copy Structure of One-Loop Open-String Amplitudes,''
Phys.\ Rev.\ Lett.\  {\bf 121}, no. 1, 011601 (2018).
[arXiv:1711.09104 [hep-th]].
%%CITATION = arXiv:1711.09104%%
}
\lref\oneloopMichael{
 M.~B.~Green, C.~R.~Mafra and O.~Schlotterer,
  ``Multiparticle one-loop amplitudes and S-duality in closed superstring theory,''
JHEP {\bf 1310}, 188 (2013).
[arXiv:1307.3534 [hep-th]].
%%CITATION = AEI-2013-219%%	
}

\lref\DHokerPDL{
	E.~D'Hoker and D.~H.~Phong,
  	``The Geometry of String Perturbation Theory,''
	Rev.\ Mod.\ Phys.\  {\bf 60}, 917 (1988).
	%%CITATION = PUPT-1039%%
}
\lref\xerox{
	E.~D'Hoker and D.~H.~Phong,
  	``Conformal Scalar Fields and Chiral Splitting on Superriemann Surfaces,''
	Commun.\ Math.\ Phys.\  {\bf 125}, 469 (1989).
	%%CITATION = UCLA-88-TEP-38%%
}

\lref\GreenMN{
	M.~B.~Green, J.~H.~Schwarz and E.~Witten,
	``Superstring Theory. Vol. 2: Loop Amplitudes, Anomalies And Phenomenology,''
	Cambridge  University Press (1987).
}

\lref\AnomalyGreen{
	M.B.~Green and J.H.~Schwarz,
  	``The Hexagon Gauge Anomaly in Type I Superstring Theory,''
  	Nucl.\ Phys.\ B {\bf 255} (1985) 93.
  	%%CITATION = NUPHA,B255,93;%%
\semi
	M.B.~Green and J.H.~Schwarz,
  	``Anomaly Cancellation in Supersymmetric D=10 Gauge Theory and Superstring Theory,''
  	Phys.\ Lett.\ B {\bf 149} (1984) 117.
  	%%CITATION = PHLTA,B149,117;%%
}

\lref\BernUE{
  Z.~Bern, J.~J.~M.~Carrasco and H.~Johansson,
  ``Perturbative Quantum Gravity as a Double Copy of Gauge Theory,''
Phys.\ Rev.\ Lett.\  {\bf 105}, 061602 (2010).
[arXiv:1004.0476 [hep-th]].
%%CITATION = UCLA-10-TEP-102%%
}

\lref\BernYXU{
  Z.~Bern, J.~J.~Carrasco, W.~M.~Chen, H.~Johansson and R.~Roiban,
  ``Gravity Amplitudes as Generalized Double Copies of Gauge-Theory Amplitudes,''
Phys.\ Rev.\ Lett.\  {\bf 118}, no. 18, 181602 (2017).
[arXiv:1701.02519 [hep-th]].
%%CITATION = UCLA-17-TEP-101%%
}

\lref\FMS{
	D.~Friedan, E.J.~Martinec and S.H.~Shenker,
  	``Conformal Invariance, Supersymmetry and String Theory,''
  	Nucl.\ Phys.\ B {\bf 271} (1986) 93.
  	%%CITATION = NUPHA,B271,93;%%
}
\lref\expPSS{
	N.~Berkovits,
  	``Explaining Pure Spinor Superspace,''
  	[hep-th/0612021].
  	%%CITATION = hep-th/0612021%%
}

\lref\KawaiXQ{
  H.~Kawai, D.~C.~Lewellen and S.~H.~H.~Tye,
  ``A Relation Between Tree Amplitudes of Closed and Open Strings,''
Nucl.\ Phys.\ B {\bf 269}, 1 (1986).
%%CITATION = CLNS-85/667%%
}

\lref\NMPS{
  	N.~Berkovits,
  	``Pure spinor formalism as an N=2 topological string,''
	JHEP {\bf 0510}, 089 (2005).
	[hep-th/0509120].
	%%CITATION = hep-th/0509120%%
}
\lref\SiegelXJ{
	W.~Siegel,
  	``Classical Superstring Mechanics,''
	Nucl.\ Phys.\ B {\bf 263}, 93 (1986).
	%%CITATION = UCB-PTH-85-23%%
}

\lref\anomalypaper{
	N.~Berkovits and C.R.~Mafra,
	``Some Superstring Amplitude Computations with the Non-Minimal Pure Spinor Formalism,''
	JHEP {\bf 0611}, 079 (2006).
	[hep-th/0607187].
	%%CITATION = hep-th/0607187%%
}

\lref\GeyerBJA{
  Y.~Geyer, L.~Mason, R.~Monteiro and P.~Tourkine,
  ``Loop Integrands for Scattering Amplitudes from the Riemann Sphere,''
Phys.\ Rev.\ Lett.\  {\bf 115}, no. 12, 121603 (2015).
[arXiv:1507.00321 [hep-th]].
%%CITATION = arXiv:1507.00321%%
}

\lref\GeyerJCH{
  Y.~Geyer, L.~Mason, R.~Monteiro and P.~Tourkine,
  ``One-loop amplitudes on the Riemann sphere,''
JHEP {\bf 1603}, 114 (2016).
[arXiv:1511.06315 [hep-th]].
%%CITATION = CERN-PH-TH-2015-267%%
}

\lref\GeyerWJX{
  Y.~Geyer, L.~Mason, R.~Monteiro and P.~Tourkine,
  ``Two-Loop Scattering Amplitudes from the Riemann Sphere,''
Phys.\ Rev.\ D {\bf 94}, no. 12, 125029 (2016).
[arXiv:1607.08887 [hep-th]].
%%CITATION = CERN-TH-2016-172%%
}

\lref\GeyerXWU{
  Y.~Geyer and R.~Monteiro,
  ``Two-Loop Scattering Amplitudes from Ambitwistor Strings: from Genus Two to the Nodal Riemann Sphere,''
JHEP {\bf 1811}, 008 (2018).
[arXiv:1805.05344 [hep-th]].
%%CITATION = arXiv:1805.05344%%
}

\lref\Zfunctions{
{\tt http://repo.or.cz/Zfunctions.git}
}

\input labelI.defs
\input labelII.defs

\listtoc
\writetoc
\filbreak

%********************
\newsec{Introduction}

This is the third part of a series of papers \wipI\ towards
the derivation of one-loop correlators of massless open- and closed-superstring
states using techniques from the pure-spinor formalism \refs{\psf,\MPS}.
We often refer to section and equation numbers from part I \& part II and then
prefix these numbers by the corresponding roman numerals I and II.
The main result of this paper is the assembly of local one-loop correlators
in pure-spinor superspace \expPSS\ up to eight points. This will be done by combining
two main ingredients:
\medskip
\item{1.} local kinematic
building blocks introduced in part I that capture the essentials of
the pure-spinor zero-mode saturation rules and transform covariantly under the
BRST charge
\item{2.} worldsheet functions introduced in part II capturing the singularities generated
by OPE contractions among vertex operators. In particular, their monodromies as
the vertex positions are moved around the genus-one cycles also follow a notion
of ``covariance''. More precisely, the monodromies are described by a system of
equations that share the same properties of the so-called BRST invariants and
naturally lead to a {\it duality} between kinematics and worldsheet functions.
\medskip
\noindent
The fundamental guiding principle that will act as the recipe to combine these
two ingredients will correspond to the one-loop generalization of a symmetry
property obeyed by the analogous tree-level correlators derived in \nptString\
and reviewed in section~\secexperA\ below. More precisely, the tree-level
correlators are composed of products of Lie-symmetric kinematic building
blocks $V_{123 \ldots p}$ and shuffle-symmetric worldsheet functions $\cZ^{\rm
tree}_{123 \ldots p}=(z_{12}z_{23} \ldots z_{p-1,p})^{-1}$. Given the similar structure between
these symmetries and the composing elements in a
theorem of Ree concerning Lie polynomials \Ree, we dubbed the correlators
obtained in this way as having a {\it Lie-polynomial form\/}. We will see that
this line of reasoning leads to a key assumption of this paper, that the local
$n$-point one-loop correlators of the open superstring can be written as
\eqn\KintroIII{
\cK_n(\ell) = \sum_{r=0}^{n-4}{1\over r!}\Big(
V_{A_1}T^{m_1 \ldots m_r}_{A_2, \ldots, A_{r+4}}\cZ^{m_1 \ldots m_r}_{A_1, \ldots, A_{r+4}}
+ \big[12 \ldots n|A_1, \ldots,A_{r+4}\big]\Big) + \hbox{ corrections}\,.
}
Definitions of the kinematic building blocks $T^{m_1 \ldots m_r}_{A_2, \ldots,
A_{r+4}}$ and the worldsheet functions $\cZ^{m_1 \ldots m_r}_{A_1, \ldots,
A_{r+4}}$ can be found in part I and part II, respectively\foot{The
worldsheet functions can also be downloaded from \Zfunctions\ as text files
in {\tt FORM} \FORM\ format.}. The notation for the
permutations in terms of partitions of words addresses the kind of permutations
resulting from the interplay between shuffles and Lie symmetries, and is
explained in the appendix~\stirlingapp. As discussed in part II, a beneficial
side effect of requiring shuffle symmetry for the worldsheet functions is that
the resulting functions automatically contain non-singular pieces that are
invisible to an OPE analysis. Lie symmetries in turn refer to the generalized
Jacobi identities satisfied by the kinematic building blocks, in lines with the
Bern--Carrasco--Johansson duality between color and kinematics \BernQJ.

The notions of locality, BRST invariance and single-valuedness will then lead to a
discussion for why the ``+ corrections'' are needed starting at $n\geq 7$
points. In section~\explsec, a multitude of representations for the correlators with 
$n=4,5,6,7$ (including the ``+ corrections'' at $n=7$) will be given that expose different 
subsets of their properties. While the $n=8$ correlator following from the proposal \KintroIII\
satisfies many non-trivial constraints, it fails to be BRST invariant by terms
proportional to the holomorphic Eisenstein series ${\rm G}_4$. In the future, we
expect to address this challenging leftover problem in order to extend our results 
to arbitrary numbers of points.

Section \loopintsec\ is dedicated to manifesting the modular properties of the open-
and closed-string correlators by integrating out the loop momentum. We will relate
a double-copy structure of the open-string correlators \MafraIOJ\ to the low-energy 
limit of the closed-string amplitudes. This incarnation of the duality between kinematics
and worldsheet functions is checked in detail up to multiplicity seven, and we describe
the problems and perspectives in the quest for an $n$-point generalization at the
end of section \loopintsec.

%**********************************************************************************
\newnewsec\secexper One-loop correlators of the open superstring: general structure

In this section, we set the stage for assembling one-loop correlators $ {\cal K}_n(\ell)$ from the
system of kinematic building blocks and worldsheet functions introduced in part I and II. By 
their definition in section \chiralsplitsec, correlators $ {\cal K}_n(\ell)$
carry the kinematic dependence of one-loop open-string amplitudes among $n$ massless states
\eqn\againopen{
{\cal A}_n  =
\sum_{\rm top} C_{\rm top} \int_{D_{\rm top}}\!\!\!\!
d\tau \, dz_2 \, d z_3 \,\ldots\, d z_{n} \, \int d^{D} \ell \ |{\cal I}_n(\ell)|\,
\langle {\cal K}_n(\ell)\rangle\,,
}
with $\langle\ldots \rangle$ denoting the zero-mode integration prescription of
the pure-spinor formalism \psf. The integration domains $D_{\rm top}$ for the
modular parameter $\tau$ and vertex positions $z_j$ are tailored to the
topologies of a cylinder or a M\"obius strip with associated color factors
$C_{\rm top}$, see \GreenMN\ for details. The integration over loop momenta
$\ell$ is an integral part of the chiral-splitting method
\refs{\verlindes,\DHokerPDL,\xerox}, which allows to derive massless
closed-string one-loop amplitudes from an integrand of double-copy form
\eqn\againclosed{
{\cal M}_n  =
 \int_{{\cal F}}
d^2\tau \, d^2z_2 \, d^2 z_3 \, \ldots \, d^2 z_{n} \,
\int d^{D} \ell \ |{\cal I}_n(\ell)|^2 \,
\langle {\cal K}_n(\ell)\rangle \, \langle\tilde{\cal K}_n(-\ell)\rangle\,,
}
with ${\cal F}$ denoting the fundamental domain for inequivalent tori w.r.t.\ the modular group.
Both of \againopen\ and \againclosed\ involve the universal one-loop Koba--Nielsen factor
\eqn\IIIKNfactor{
{\cal I}_n(\ell) \equiv  \exp\Big( \sum^n_{i<j} s_{ij} \log \theta_1(z_{ij},\tau)
+  \sum_{j=1}^n  z_j(\ell\cdot k_j) + {\tau \over 4\pi i} \ell^2 \Big)\,,
}
with lightlike external momenta $k_j$, where we use the shorthands
\eqn\shorthd{
s_{ij}\equiv k_i \cdot k_j \, , \ \ \ \ \ \ z_{ij} \equiv z_i - z_j
}
and conventions where $2\ap =1$ for open strings and ${\ap \over 2}=1$ for
closed ones.

In trying to calculate multiparticle one-loop amplitudes using the pure-spinor
prescription \onepresc, one soon realizes that most efforts tend to be hampered
by the complicated nature of the $b$-ghost \bghost. This difficulty, however,
motivates a less direct approach which illuminates the structure of the answer
in a somewhat unexpected way; the organizing principle will be drawn from the
tree-level correlators of \nptString.

%*******************************************************
\newsubsec\secexperA Lessons from tree-level correlators

Recall that $n$-point open-string tree amplitudes in the pure-spinor formulation
require the evaluation of the following $n$-point correlation function \psf,
\eqn\nptfcttree{
\langle \! \langle V_1(z_1) \prod_{j=2}^{n-2}U_j(z_j)V_{n-1}(z_{n-1}) V_n(\infty) \rangle
\! \rangle_{{\rm tree}} \equiv \langle {\cal K}_n^{\rm tree}  \rangle
\prod_{i<j}^n |z_{ij}|^{s_{ij}}\,,
}
see \Vvertex\ and \integratedU\ for the vertex operators $V_i$ and $U_j$.
The definition of the tree-level {\it correlators} $\cK_n^{\rm tree}$ on the
right-hand side is analogous to that of one-loop correlators $\cK_n(\ell)$, cf.\ \theampA.
The idea is to strip off the universal factors of $ |z_{ij}|^{s_{ij}}$ from the 
path integral, i.e.\ the tree-level analogue of the one-loop Koba--Nielsen factor \IIIKNfactor. 
The computation of the correlators ${\cal K}_n^{\rm tree}$ boils down 
to using the CFT rules of the pure-spinor formalism to perform OPE contractions among 
the vertex operators in \nptfcttree.

One of the crucial steps in the calculation of
\nptString\ was showing that the multiparticle
vertex operators $V_P$ \EOMbbs\ could be used
as the fundamental building blocks of the correlator; for
example, in terms of the function ${\cal Z}^{\rm tree}_{P}$ defined by
\eqn\shufAagain{
\cZ^{\rm tree}_{123\ldots p} \equiv {1\over z_{12} z_{23}\ldots z_{p-1,p}}\,,
}
we have
\eqnn\experB
$$\eqalignno{
V_1(z_1)U_2(z_2)&\cong  V_{12} {\cal Z}^{\rm tree}_{12}
&\experB\cr
V_1(z_1)U_2(z_2)U_3(z_3)&\cong   V_{123}\cZ^{\rm tree}_{123}
+  V_{132}\cZ^{\rm tree}_{132}\,,
}$$
where the symbol $\cong$ is a reminder that the above relations are valid up to
total derivatives and BRST-exact quantities. As reviewed in section~\shufflesec,
the accompanying functions exhibit shuffle symmetry such as $\cZ^{\rm tree}_{12}
+ \cZ^{\rm tree}_{21} = 0$, $\cZ^{\rm tree}_{123} - \cZ^{\rm tree}_{321} = 0$
and $\cZ^{\rm tree}_{123} + \cZ^{\rm tree}_{213} + \cZ^{\rm tree}_{231} = 0$
dual to the Lie symmetries $V_{12}{=}-V_{21}, \ V_{123}=-V_{213}$ and
$V_{123}+V_{231}+V_{312}=0$, cf.\ \genjac.  When combined with \experB, these
symmetries lead to the following generalization:
\eqn\experC{
V_1(z_1)\prod_{i=1}^{n}U_{a_i}(z_{a_i}) \cong \sum_{|A|=n}  V_{1A} {\cal Z}^{\rm
tree}_{1A}\,,\qquad \cZ^{\rm tree}_{A\shuffle B} = 0\,,\quad \forall
A,B\neq\emptyset\,,
}
which eventually gives rise to the solution found in \nptString. As detailed in section \wordssec,
the summation range $|A|=n$ in \experC\ refers to the $n!$ words $A$ formed by permutations of
$a_1a_2\ldots a_{|A|}$ with $|A|=n$.

At this point one may realize that the right-hand side of \experC\ has
the structure of a {\it Lie polynomial} \refs{\reutenauer,\Ree} and that
the expressions for the $n$-point correlators at tree level obtained in
\nptString\ can be written in terms of their products. More precisely,
$\cK_n^{\rm tree}$ is given by two copies of \experC\ with $n{-}2$ deconcatenations
$AB=23\ldots n{-}2$ and an overall permutation over $(n{-}3)!$ letters for a total of $(n{-}2)!$ terms:
\eqn\experF{
{\cal K}_{n}^{\rm tree} =\!\!\!\!\!\! \sum_{AB=23 \ldots n-2}\!\!\!\!\!\!\!\!
\big(V_{1A} \cZ^{\rm tree}_{1A}\big)\big(V_{n-1,B}\cZ^{\rm tree}_{n-1,B}\big) V_n +
\perm(23 \ldots n{-}2)\,.
}
For example ($\cZ^{\rm tree}_i\equiv 1$),
\eqnn\experE
$$\eqalignno{
{\cal K}_3^{\rm tree}&= V_1 V_2 V_3\,, \ \ \ \ \ \
{\cal K}_4^{\rm tree} = V_{12} {\cal Z}^{\rm tree}_{12} V_3 V_4
+ V_{1} V_{32} {\cal Z}^{\rm tree}_{32} V_4\,,   &\experE\cr
{\cal K}_5^{\rm tree} &=
\big(V_{123}\cZ^{\rm tree}_{123} + V_{132}\cZ^{\rm tree}_{132}\big) V_4 V_5
+ V_{1}\big(V_{423}\cZ^{\rm tree}_{423} + V_{432}\cZ^{\rm tree}_{432}\big)V_5\cr
&\quad{}+ \big(V_{12} {\cal Z}^{\rm tree}_{12}\big)\big( V_{43}{\cal Z}^{\rm tree}_{43}\big) V_5
+ \big(V_{13} \cZ^{\rm tree}_{13}\big)\big( V_{42}\cZ^{\rm tree}_{42}\big)
V_5\,.
}$$
Note that
$V_{123} {\cal Z}^{\rm tree}_{123} +  V_{132} {\cal Z}^{\rm tree}_{132}$
in \experE\ is symmetric in $1,2,3$ even though only two
out of $3!$ permutations are spelled out.
This is a consequence of the Lie-polynomial structure of the correlator\foot{
This follows from the identity $\sum_{A}{1\over |A|}\cZ_{A}V_A =
\sum_{B}\cZ_{iB}V_{iB}$ \Ree.};
the right-hand side of \experC\ is permutation
symmetric in $1,a_1,a_2,\ldots,a_{|A|}$
even though only the weaker symmetry in $a_1,a_2,\ldots,a_{|A|}$ is manifest.

The Lie-polynomial structure of the building blocks in the tree-level correlator
\experF\ motivates us to search for a similar organization of the one-loop
correlators.

%***************************************************
\newsubsec\secexperB Assembling one-loop correlators

Let us summarize what we have seen in part I and II in order to
better understand the motivation behind the
general form of the one-loop correlators $\cK_n(\ell)$ to be proposed shortly.
\medskip
\item{$\bullet$}
In section~\SYMsec, we reviewed the definition of local superfields that satisfy
generalized Jacobi identities and, in section~\LocalBBsec, we showed how they
can be assembled in several classes of local building blocks.

\item{$\bullet$}
In section~\dualitysec, we constructed functions composed of
the expansion coefficients of the Kronecker--Eisenstein series
that obey shuffle symmetries when the vertex insertion
points are permuted.
\medskip
\noindent Let us thread the above points together in view of the tree-level structure
discussed above. Firstly, since the short-distance singularities within the
correlator are independent on the global properties of the Riemann surface,
the shuffle symmetries of the worldsheet functions should also be a property
of the worldsheet functions at one loop. And secondly, the shuffle symmetry
obeyed by the functions are the driving force in the Lie-polynomial
organization of the tree-level correlators with local kinematic building
blocks. When taken together these points suggest that the {\it superfields and
worldsheet functions of one-loop correlators have the same symmetry structure
of Lie polynomials.} This realization will lead to a beautiful organization of
superstring one-loop correlators.

%************************************************************************
\newsubsec\secexperC The Lie-polynomial structure of one-loop correlators

The additional zero modes at genus one, in particular the availability of loop
momenta, allow for a significantly richer system of kinematic building blocks
as compared to the tree-level kinematics $V_{1A}V_{n-1,B}V_n$ in \experF. Also
their accompanying worldsheet functions must accommodate the different OPE
singularities and powers of loop momentum characteristic to each zero-mode
saturation pattern, see e.g.\ \justMULT\ and \WfourMULT. The corresponding Lie
polynomials will therefore differ with respect to these features but will preserve
their mathematical characterization as sums over products of shuffle- and Lie-symmetric
objects.

To exemplify and to give a preview of what is
ahead, the four-, five-, and six-point correlators at one loop
will be written as products of local kinematic building blocks $T^{m_1\ldots}_{A,B,\ldots}$
(cf.\ sections \secTAA\ to \secTCC) and worldsheet functions 
(cf.\ section \bootsec) as follows,
\eqnn\experFexamp
$$\eqalignno{
{\cal K}_4(\ell)&= V_1 T_{2,3,4} \cZ_{1,2,3,4}\,, &\experFexamp\cr
{\cal K}_5(\ell)&= V_{1} T^m_{2,3,4,5}  \cZ^m_{1,2,3,4,5} 
+\big[ V_{12}T_{3,4,5}  \cZ_{12,3,4,5}
+ (2\leftrightarrow 3,4,5) \big]\cr
&+\big[ V_{1}T_{23,4,5}  \cZ_{1,23,4,5} + (2,3|2,3,4,5) \big]\,, \cr
{\cal K}_6(\ell)&= {1\over 2}V_{1} T^{mn}_{2,3,4,5,6}  \cZ^{mn}_{1,2,3,4,5,6}
%- \big[V_1 J_{2|3,4,5,6} \cZ_{2|1,3,4,5,6}  + (2\leftrightarrow 3,4,5,6) \big]
\cr
&+\big[ V_{12}T^m_{3,4,5,6}  \cZ^m_{12,3,4,5,6} + (2\leftrightarrow 3,4,5,6) \big] \cr
&+\big[ V_{1}T^m_{23,4,5,6}  \cZ^m_{1,23,4,5,6} + (2,3|2,3,4,5,6) \big] \cr
&+ \big[ V_{123} T_{4,5,6} \cZ_{123,4,5,6}
+V_{132} T_{4,5,6} \cZ_{132,4,5,6} + (2,3|2,3,4,5,6) \big] \cr
&+ \big[ (V_{12} T_{34,5,6} \cZ_{12,34,5,6} + {\rm cyc}(2,3,4))
+ (2,3,4|2,3,4,5,6) \big] \cr
&+  \big[ (V_{1} T_{2,34,56} \cZ_{1,2,34,56} +{\rm cyc}(3,4,5))
+ (2\leftrightarrow3,4,5,6) \big]\cr
&+  \big[ V_{1} T_{234,5,6} \cZ_{1,234,5,6}  +  V_{1} T_{243,5,6} \cZ_{1,243,5,6} 
+ (2,3,4|2,3,4,5,6) \big]\,,
}$$
where $m,n,p,\ldots=0,1,\ldots,9$ denote Lorentz-vector indices. As in part I and II,
the separation of words $A,B,\ldots$ through a comma in a subscript indicates that
the parental object is symmetric under $A\leftrightarrow B$, e.g.\ $T_{A,B,C}=T_{B,A,C}=T_{A,C,B}$.
The generalized Jacobi symmetries of $V_P$ then apply to all of $A,B,\ldots$, e.g.\ 
$T_{234,5,6}+{\rm cyc}(2,3,4)=0$. Moreover, $+(a_1,\ldots,a_p|a_1,\ldots ,a_{p+q})$ 
refers to summing over all the ${p+q \choose p}$ subsets of $a_1,\ldots ,a_{p+q}$ involving $p$ 
elements $a_i$ in the place of $a_1,\ldots,a_p$.

The Lie-polynomial form of the correlator \experFexamp\ is also
convenient for obtaining different representations. For example,
after rewriting $\sum_A V_{iA}\cZ_{iA} = \sum_{A,B} V_{iA}\d_{A,B}\cZ_{iB}$
one can use \InvSPhi\ to obtain the {\it trading identity\/},
\eqn\tradingId{
\sum_A V_{iA}\cZ_{iA}=\sum_{A} M_{iA} Z^{(s)}_{iA}\ .
}
Shuffle symmetric Berends--Giele currents $M_B$ and Lie-symmetric worldsheet
functions $Z^{(s)}_{B}$ are defined in \BGmap\ and \cZToZ, respectively,
and \tradingId\ can be easily generalized to any number of words.
This kind of manipulation played a key role in expressing the tree-level
correlator in terms of an $(n{-}3)!$ basis of worldsheet functions in \nptString.

We will be concerned with the particulars of the expressions \experFexamp\ in
section~\explsec; for the moment we note that their growing number of terms
calls for a more convenient notation. In the subsequent discussion we will
distill the combinatorial properties of these permutation sums and propose an
intuitive notation for them.

%********************************************************
\newsubsubsec\stirlingsec Stirling cycle permutation sums

In order to grasp the combinatorics of \experFexamp, note that the symmetries of
the Lie polynomial $\sum_A V_A\cZ_A$ imply that only $(p{-}1)!$ permutations are
independent for words of length $p$. This is true for each word $A_i$ in terms
such as $V_{A_1}T_{A_2,A_3,A_4}\cZ_{A_1,A_2,A_3,A_4}$. For an $n$-point
correlator these words $A_i$ must encompass all particle labels, that is
$|A_1|+|A_2|+|A_3|+|A_4|=n$. Therefore the sums of
$V_{A_1}T_{A_2,A_3,A_4}\cZ_{A_1, A_2,A_3,A_4}$ in the correlators of
\experFexamp\ can be interpreted as being all the permutations of $n$ labels
that are composed of $4$ cycles, or $p$ cycles in the general case of tensorial
$V_{A_1}T^{m_1m_2\ldots}_{A_2,A_3,\ldots,A_p}$. This is the characterization of
the {\it Stirling cycle numbers\/}\foot{We are following the terminology and
notation proposed in \tnn; they are commonly known as the {\it Stirling numbers
of the first kind}.} ${n\stirling p}$.

Using the above interpretation, the
scalar building blocks in \experFexamp\ are generated by the following
combinatorial notation
\eqn\experH{
{\cal K}_n(\ell) \, \big|_{V_A T_{B,C,D}} = V_A T_{B,C,D} \cZ_{A,B,C,D} +
\big[12 \ldots n|A,B,C,D\big]\,,
}
where $+\big[12 \ldots n|A,B,C,D\big]$ indicates a sum over the
{\it Stirling cycle permutations\/} of the set $\{1,2, \ldots,n\}$, defined
in the appendix~\stirlingapp. As a consequence of this definition,
each term of \experH\ has leg one as the first letter of $A$, cf.\ \cycord.

Similarly, the vector contribution to ${\cal K}_5(\ell)$ and ${\cal
K}_6(\ell)$ in \experFexamp\ follows the same combinatorial pattern as the
scalars and is captured by extending the Stirling cycle
permutations to five slots in a similar manner,
\eqn\experI{
{\cal K}_n(\ell) \, \big|_{V_A T^m_{B,C,D,E}} = V_A T^m_{B,C,D,E}
\cZ^m_{A,B,C,D,E} +  \big[12 \ldots n|A,B,C,D,E\big]\,.
}
The generalization of the above sums to more slots is straightforward.

%************************************************
\newsubsubsec\LiePolsec Unrefined Lie polynomials

The Stirling cycle permutations allow for a straightforward
generalization of the correlators in \experFexamp\ to multiplicities
$n\ge 4$,
\eqn\dzeroGeneral{
\cK^{(0)}_n(\ell) = \sum_{r=0}^{n-4}{1\over r!}\Big(
V_{A_1}T^{m_1 \ldots m_r}_{A_2, \ldots, A_{r+4}}\cZ^{m_1 \ldots m_r}_{A_1, \ldots, A_{r+4}}
+ \big[12 \ldots n|A_1, \ldots,A_{r+4}\big]\Big)\,,
}
where the summand with $r=0$ and $r=1$ reproduces \experH\ and \experI, respectively.
The reason for the superscript in $\cK^{(0)}_n(\ell)$ will become clear below, and this is
related to the corrections in \KintroIII. Expanding the sum yields,
\eqnn\experJ
$$\eqalignno{
\cK^{(0)}_n(\ell)&=
V_{A_1} T_{A_2,A_3,A_4} \cZ_{A_1, A_2,A_3,A_4}
+ \big[12 \ldots n|A_1, \ldots,A_4\big] &\experJ\cr
&+{1\over 1!} V_{A_1} T^{m_1}_{A_2, \ldots,A_5} \cZ^{m_1}_{A_1, A_2,\ldots,A_5}
+ \big[12 \ldots n|A_1, \ldots,A_5\big]\cr
&+{1\over 2!} V_{A_1} T^{m_1m_2}_{A_2, \ldots,A_6} \cZ^{m_1m_2}_{A_1,A_2, \ldots,A_6}
+\big[12 \ldots n|A_1, \ldots,A_6\big]\cr
&\hskip70pt \vdots \cr
&+{1\over (n{-}4)!} V_{A_1} T^{m_1 \ldots m_{n-4}}_{A_2, \ldots,A_n}
\cZ^{m_1 \ldots m_{n-4}}_{A_1, \ldots,A_n}
+\big[12 \ldots n|A_1,A_2, \ldots,A_n\big] \, .
}$$
We will see in section \explsec\ that \dzeroGeneral\
gives the correct form of the one-loop correlators up to and including six
points, i.e., $\cK_n(\ell) = \cK_n^{(0)}(\ell)$ for $n\le 6$.
By ``correct'' we mean that the resulting correlators satisfy a number
of requirements detailed in section \FinalAssemblysec, the most stringent ones 
being BRST invariance and single-valuedness. 

So the question to consider is whether the expression \experJ\ provides
the complete answer for correlators with seven or more external states.
Unfortunately this is not the case; the explicit construction of the
seven-point correlator indicates that the proposal \experJ\ needs
to be amended by terms involving superfields with higher degrees of
refinement defined in section \RefBBsec.
This will be done below and leads to an expression for ${\cal K}_7$
that passes all consistency checks. At eight points and beyond, however, the
appearance of Eisenstein series in the correlators cannot be determined by
the methods in this work. Hence, we will only propose an expression for ${\cal K}_8$ 
up to an unknown kinematic factor multiplying ${\rm G}_4$ while completely fixing 
its dependence on the $z_j$.

%*******************************************
\subsubsec Including refined building blocks

The reason why $\cK_n^{(0)}(\ell)$ in \experJ\ cannot be the full expression for the one-loop
correlator for $n\ge7$ is related to BRST invariance; it is not difficult to
show that the seven-point instance is not BRST invariant using the
worldsheet functions discussed in part II. However, the desired
invariance can still be achieved by adding corrections containing {\it refined}
superfields $J_{A|B,C,D,E}$ and their tensorial generalizations, cf.\ section \RefBBsec. 
The patterns encountered at multiplicities seven and eight suggest the following
organization; the $n$-point correlator contains contributions with varying
degree $d$ of refinement according to,
\eqn\experK{
{\cal K}^{\rm Lie}_n(\ell) \equiv
\sum_{d=0}^{\lfloor {n-4\over 2} \rfloor} (-1)^d \cK^{(d)}_n(\ell) \, .
}
The alternating minus sign in \experK\ is chosen for later convenience. The $d=0$ contribution
is given by \dzeroGeneral\ for $n\ge4$, while the first instance
of refined corrections with $d=1$ could already appear in the six-point correlator, 
\eqn\thezero{
\cK^{(1)}_6(\ell) = V_1 J_{2|3,4,5,6} \cZ_{2|1,3,4,5,6}  + (2\leftrightarrow3,4,5,6)\, .
}
However, as detailed in section \sixwssec, the accompanying functions
$\cZ_{2|1,3,4,5,6}$ can be chosen
to vanish, i.e.\ $\cK^{(1)}_6(\ell) =0$. Therefore, the first non-vanishing
contribution to \experK\ with $d=1$ occurs at seven points,
\eqnn\experM
$$\eqalignno{
\cK^{(1)}_7(\ell) &=
V_1 J^m_{2|3,4,5,6,7} \cZ^m_{2|1,3,4,5,6,7}  + (2\leftrightarrow 3,4,5,6,7) &\experM \cr
&+ \big[ V_{12}J_{3|4,5,6,7}\cZ_{3|12,4,5,6,7}
+ V_{13}J_{2|4,5,6,7}\cZ_{2|13,4,5,6,7} +(2,3|2,3,\ldots,7) \big] \cr
&+ \big[ V_{1}J_{23|4,5,6,7}\cZ_{23|1,4,5,6,7}  +(2,3|2,3,\ldots,7) \big]
\cr
&+ \big[( V_{1}J_{2|34,5,6,7}\cZ_{2|1,34,5,6,7}+{\rm cyc}(2,3,4))  +(2,3,4|2,3,\ldots,7) \big]\,,
}$$
see \sevNotorious\ for the refined worldsheet functions $\cZ_{A|B,\ldots}$.
Similarly, eight points give rise to the first non-vanishing instance of $d=2$,
\eqn\experN{
\cK^{(2)}_8(\ell)=
V_1 J_{2,3|4,5,6,7,8}\, \cZ_{2,3|1,4,5,6,7,8}  + (2,3|2,3,4,5,6,7,8)\,.
}
These expressions generalize to $n\ge7$ points at generic degree $d$ of
refinement as
\eqnn\ddGeneral
$$\eqalignno{
\cK^{(d)}_n(\ell) &= \sum_{r=0}^{n-4-2d}{1\over r!}\Big(
\big(V_{A_1}J^{m_1 \ldots m_r}_{A_2, \ldots,A_{d+1}|A_{d+2}, \ldots, A_{r+4+2d}}
\cZ^{m_1 \ldots m_r}_{A_2, \ldots,A_{d+1}|A_1,A_{d+2}, \ldots, A_{r+4+2d}}&\ddGeneral\cr
&\hskip50pt + (A_2, \ldots,A_{d+1}|A_2, \ldots,A_{r+4+2d})\big)
 + \big[12 \ldots n|A_1, \ldots,A_{r+4+2d}\big]\Big)\,.
}$$
More explicitly, expanding the sum in \ddGeneral\ for the case $d=1$ yields,
\eqnn\experO
$$\eqalignno{
\cK^{(1)}_n(\ell)&=
\big(V_{A_1} J_{A_2|A_3, \ldots,A_6} \cZ_{A_2|A_1,A_3, \ldots,A_6}
+(A_2{\leftrightarrow }A_3,\ldots,A_6)\big)
+ \big[1 \ldots n|A_1, \ldots,A_6\big] &\experO \cr
&{} +{1\over 1!}\big(V_{A_1} J^m_{A_2|A_3, \ldots,A_7} \cZ^m_{A_2|A_1,A_3, \ldots,A_7}
+(A_2{\leftrightarrow} A_3,\ldots,A_7)\big)
+ \big[1 \ldots n|A_1, \ldots,A_7\big]\cr
&{} +{1\over 2!}\big(V_{A_1} J^{mn}_{A_2|A_3, \ldots,A_8}
\cZ^{mn}_{A_2|A_1,A_3, \ldots,A_8}
+(A_2{\leftrightarrow} A_3,\ldots,A_8)\big)
+ \big[1 \ldots n|A_1, \ldots,A_8\big]\cr
&\hskip70pt \vdots  \cr
&{} +{1\over (n{-}6)!} \big(V_{A_1} J^{m_1 \ldots m_{n-6}}_{A_2|A_3, \ldots,A_n}
\cZ^{m_1 \ldots m_{n-6}}_{A_2|A_1,A_3, \ldots,A_n}
+(A_2{\leftrightarrow} A_3,\ldots,A_n)\big)
+\big[1 \ldots n|A_1, \ldots,A_n\big]\,.
}$$
The collection of $\cK^{(d)}_n(\ell)$ with $d=0,1,\ldots,\lfloor {n-4\over 2} \rfloor$ 
summarized by \experK\ makes up the bulk of the open-string one-loop correlators
and will be referred to as its Lie-series part. The expressions \ddGeneral\ for $\cK^{(d)}_n(\ell)$
with $d\geq 1$ furnish a large class of the corrections in \KintroIII. We will see that 
up to and including eight points, the BRST variation of \experK\ is purely anomalous (it is written in 
terms of the anomalous superfields $Y$, see section \anoBBsecI) and it is natural to conjecture that this 
behavior is valid for arbitrary $n$.

%************************************************************************
\newsubsubsec\QGeneralsec BRST variation of the Lie-polynomial correlator

By BRST covariance of their kinematic building blocks in section~\LocalBBsec, the $Q$ variations of the
above $\cK_n^{\rm Lie}(\ell)$ boil down to ghost-number four superfields
$V_{A_1} V_{A_2} T^{m_1 \ldots}_{A_3,A_4,\ldots}$ and
$V_{A_1} Y^{m_1 \ldots}_{A_2,A_3,\ldots}$. As will be detailed in the next section, the
coefficients of these ghost-number four combinations read as follows in the simplest
non-vanishing variations,
\eqnn\QcKsample
$$\eqalignno{
-Q{\cal K}^{\rm Lie}_5(\ell) \big|_{V_1V_2 T_{3,4,5}}
&= k_2^m \cZ^m_{1,2,3,4,5} {+}  s_{21}\cZ_{21,3,4,5}
{+}s_{23}\cZ_{1,23,4,5} {+}s_{24}\cZ_{1,24,3,5} {+}s_{25}\cZ_{1,25,3,4} \cr
%%%
-Q{\cal K}^{\rm Lie}_6(\ell) \big|_{  V_1 V_2 T^m_{3,4,5,6} } &=
k_2^n \cZ^{mn}_{1,2,3,4,5,6} - k_2^m \cZ_{2|1,3,4,5,6} + \bigl[s_{21}\cZ^m_{21,3,4,5,6} +
(1\leftrightarrow 3,4,5,6)\bigr]
\cr
-Q{\cal K}^{\rm Lie}_6(\ell) \big|_{  V_{12} V_3 T_{4,5,6} } &=   k_3^m {\cal Z}^m_{12,3,4,5,6} {+} s_{31} {\cal Z}_{312,4,5,6} {-} s_{32} {\cal Z}_{321,4,5,6} 
{+} \big[ s_{34} {\cal Z}_{12,34,5,6} {+} (4{\leftrightarrow} 5,6)\big]  \cr
-Q{\cal K}^{\rm Lie}_6(\ell) \big|_{  V_1 V_2 T_{34,5,6} } &=
k_2^m \cZ^m_{1,2,34,5,6} {+} s_{23}\cZ_{1,234,5,6} {-} s_{24}\cZ_{1,243,5,6} {+}
\bigl[s_{21}\cZ_{21,34,5,6} {+} (1{\leftrightarrow} 5,6)\bigr]
\cr
-Q{\cal K}^{\rm Lie}_6(\ell) \big|_{  V_1 V_{23} T_{4,5,6} } &=  k_{23}^m \cZ^m_{1,23,4,5,6} - \cZ_{2|1,3,4,5,6} + \cZ_{3|1,2,4,5,6}  &\QcKsample\cr
& \ \ \  +
\bigl[s_{31}\cZ_{231,4,5,6} - s_{21}\cZ_{321,4,5,6}
+(1\leftrightarrow 4,5,6)\bigr]   \, ,
}$$
as well as
\eqnn\QcKsampleA
$$\eqalignno{
-Q{\cal K}^{\rm Lie}_6(\ell) \big|_{  V_1 Y_{2,3,4,5,6} } &= {1\over 2} \cZ^{mm}_{1,2,3,4,5,6} - \big[  \cZ_{2|1,3,4,5,6}+ (2\leftrightarrow 3,4,5,6) \big] \, ,  &\QcKsampleA
}$$
where ${\cal K}^{\rm Lie}_{n=5,6}(\ell)$ already furnish the complete correlators ${\cal K}_n(\ell)$.
Note that we have disregarded the vanishing of $\cZ_{2|1,3,4,5,6}$ for later convenience,
and one can compactly absorb the Mandelstam invariants in \QcKsample\ into the $S[A,B]$ map
defined in \Smap, e.g.\ $s_{31} {\cal Z}_{312,4,5,6} -  s_{32} {\cal Z}_{321,4,5,6}=\cZ_{S[3,12],4,5,6}$.
Based on these examples and analogous observations on $Q\cK_n^{\rm Lie}(\ell)$ for higher values of $n$,
it is possible infer a general pattern and propose closed formulae. We organize the general conjecture
on the BRST variation of the correlator \experK\ into the following Stirling permutation sums,
\eqnn\QcKgeneral
$$\eqalignno{
Q\cK_n^{\rm Lie}(\ell) &=
-\sum_{r=0}^{n-5} {1\over r!}
T^{(0,r)}_{A_1|A_2,\ldots,A_{r+5}}
+ \big[12 \ldots n|A_1,\ldots,A_{r+5}\big] &\QcKgeneral\cr
%%%
&\quad{}+\sum_{r=0}^{n-6} {1\over r!}
Y^{(0,r)}_{A_1|A_2,\ldots,A_{r+6}}
+ \big[12 \ldots n|A_1,\ldots,A_{r+6}\big]\cr
&\quad{}-\sum_{r=0}^{n-7} {1\over r!}
T^{(1,r)}_{A_1|A_2,\ldots,A_{r+7}}
+ \big[12 \ldots n|A_1,\ldots,A_{r+7}\big]\cr
&\quad{}+\sum_{r=0}^{n-8} {1\over r!}
Y^{(1,r)}_{A_1|A_2,\ldots,A_{r+8}}
+ \big[12 \ldots n|A_1,\ldots,A_{r+8}\big]\cr
&\quad{}+\;\cdots\, ,
}$$
where the suppressed terms $T^{(d,r)}$ and $Y^{(d,r)}$ in $\ldots$ refer to higher 
degree of refinement $d\geq 2$ and start to contribute at $n=9$. The case $r=0$ 
is understood as containing no vector indices in the superfields, and a
upper negative integer in the sum must be discarded; $\sum^{-i}_{r=0}( \ldots) \to0$.
The shorthands $T^{(d,r)}$ contain $T$-like\foot{Recall that the $J_{A| \ldots}$ building
block is naturally identified as a $d=1$ refined version of $T$.}
building blocks, and their definitions at refinement $d=0,1$
\eqnn\QLiedZero
$$\eqalignno{
T^{(0,r)}_{A_1|A_2,\ldots,A_{r+5}} &\equiv
V_{A_1}V_{A_2}T^{m_1 \ldots m_r}_{A_3,\ldots,A_{r+5}}
\Theta^{(0)\,m_1 \ldots m_r}_{A_2|A_1,A_3,\ldots,A_{r+5}}
+ (A_2\leftrightarrow A_3, \ldots,A_{r+5})\,, &\QLiedZero\cr
T^{(1,r)}_{A_1|A_2,\ldots,A_{r+7}}&\equiv
\Big( \big[ V_{A_1}V_{A_2}J^{m_1 \ldots m_r}_{A_3|A_4,\ldots,A_{r+7}}
\Theta^{(1)\,m_1 \ldots m_r}_{A_2|A_3|A_1,A_4,\ldots,A_{r+7}}
+ (A_3\leftrightarrow A_4, \ldots,A_{r+7}) \big] \cr
&\hskip30pt + (A_2\leftrightarrow A_3, \ldots,A_{r+7})\Big)
}$$
admit an obvious generalization to higher values of $d$.
Similarly, the shorthands $Y^{(d,r)}$ contain anomalous superfields $Y$ with
degree of refinement $d$, see equations \Wanondef\ and \Yrefdef,
and their definitions at $d=0,1$,
\eqnn\Yddefs
$$\eqalignno{
Y^{(0,r)}_{A_1|A_2,\ldots,A_{r+6}} &\equiv
V_{A_1}Y^{m_1 \ldots m_r}_{A_2,A_3,\ldots,A_{r+6}}
\Xi^{(0)\,m_1 \ldots m_r}_{A_1|A_2,\ldots,A_{r+6}}\,,&\Yddefs\cr
Y^{(1,r)}_{A_1|A_2,\ldots,A_{r+8}}&\equiv
V_{A_1}\big[Y^{m_1 \ldots m_r}_{A_2|A_3,\ldots,A_{r+8}}
\Xi^{(1)\,m_1 \ldots m_r}_{A_1|A_2|A_3,\ldots,A_{r+8}}
+ (A_2\leftrightarrow A_3, \ldots,A_{r+8})\big]\, ,
}$$
suggest their analogues at $d\geq 2$. Finally, the shorthands $\Theta^{(d)}$ and $\Xi^{(d)}$ 
stand for the following linear combinations of worldsheet functions with degree $d$ of refinement
that capture the right-hand sides of \QcKsample,
\eqnn\ThetaSix
$$\eqalignno{
\Theta^{(0)\,m_1 m_2\ldots m_r}_{A|B_1,B_2,\ldots,B_{r+4}}&\equiv
k_A^p \cZ^{pm_1 m_2\ldots m_r}_{A,B_1,B_2,\ldots,B_{r+4}} + \big[
\cZ^{m_1 m_2\ldots m_r}_{S[A,B_1],B_2,\ldots,B_{r+4}}
+(B_1\leftrightarrow B_2,\ldots,B_{r+4})\big]\cr
&- k_A^{(m_1}\cZ^{m_2 \ldots m_r)}_{A|B_1, \ldots,B_{r+4}}
- \sum_{A=XY}\big(\cZ^{m_1 m_2 \ldots m_r}_{X|Y,B_1, \ldots,B_{r+4}}
- (X\leftrightarrow Y)\big)\,, &\ThetaSix\cr
\Theta^{(1)\,m_1 m_2\ldots m_r}_{A|B|B_1,B_2,\ldots,B_{r+5}}&\equiv
- k_A^p \cZ^{pm_1 \ldots m_r}_{B|A,B_1, \ldots,B_{r+5}}
- \cZ^{m_1 \ldots m_r}_{S[A,B]|B_1, \ldots,B_{r+5}}\cr
&- \big[\cZ^{m_1 \ldots m_r}_{B|S[A,B_1], \ldots,B_{r+5}}
+ (B_1\leftrightarrow B_2,\ldots,B_{r+5})\big]\cr
& + k_A^{(m_1}\cZ^{m_2 \ldots m_r)}_{A,B|B_1, \ldots,B_{r+5}}
+ \sum_{A=XY}\big(\cZ^{m_1 m_2 \ldots m_r}_{X,B|Y,B_1, \ldots,B_{r+5}} - (X\leftrightarrow Y)\big)\,,
}$$
(recall that $S[A,B]$ denotes the S-map defined in \Smap), and
\eqnn\XiSix
$$\eqalignno{
\Xi^{(0)\,m_1 m_2\ldots m_r}_{A_1|B_1,\ldots,B_{r+5}}&\equiv
-\half \cZ^{ppm_1 \ldots m_r}_{A_1,B_1, \ldots,B_{r+5}}
+ \big[\cZ^{m_1\ldots m_r}_{B_1|A_1,B_2, \ldots,B_{r+5}} + (B_1\leftrightarrow
B_2, \ldots,B_{r+5})\big]\,,\quad{}&\XiSix\cr
\Xi^{(1)\,m_1 m_2\ldots m_r}_{A_1|A_2|B_1,\ldots,B_{r+6}}&\equiv
\half \cZ^{ppm_1 \ldots m_r}_{A_2|A_1,B_1, \ldots,B_{r+6}}%\cr
%&\qquad{}
- \big[\cZ^{m_1\ldots m_r}_{A_2,B_1|A_1,B_2, \ldots,B_{r+6}}
+ (B_1\leftrightarrow B_2, \ldots, B_{r+6})\big]\,.
}$$
Hence, after modding out by Lie symmetries of the superfields, combining \QcKgeneral\ and
\QLiedZero\ identifies $\Theta^{(d)\,m_1 m_2\ldots m_r}_{A_2|B_1,\ldots,B_d|A_1, A_3,\ldots}$
to be the coefficient of $V_{A_1} V_{A_2} J^{m_1\ldots m_r}_{B_1,\ldots,B_d|A_3,\ldots}$
in $Q {\cal K}_n^{\rm Lie}$. Similarly, $\Xi^{(d)\,m_1 m_2\ldots m_r}_{A_1|B_1,\ldots,B_d|A_2,\ldots}$ 
turns out to be the coefficient of $V_{A_1} Y^{m_1\ldots m_r}_{B_1,\ldots,B_d|A_2,\ldots}$
by \QcKgeneral\ and \Yddefs.

Two comments are in order here. First, notice that the presentation of the
BRST variation as a Stirling permutation sum (with the conventions of the
appendix~\stirlingapp) is essential to fix the ambiguity of $V_AV_B=-V_BV_A$ in
matching the $V_AV_B$ products in \QLiedZero\ to $\Theta^{(d)}_{B|A, \ldots}$.
For example, the conventions of the appendix~\stirlingapp\ fix the relative
ordering between the cycles $(1)(234)$ in the permutation sum such that we get
$V_{1}V_{234}T_{5,6,7}\,\Theta^{(0)}_{234|1,5,6,7}$ rather than
$V_{234}V_{1}T_{5,6,7}\,\Theta^{(0)}_{1|234,5,6,7}$. And second, although a
bit surprising, the BRST variation leads to crossing-symmetric definitions
such as $\Theta^{(0)}_{A|B,C,D,E}$ in $B,C,D$ and $E$; in other words, the
worldsheet functions multiplying $V_{1234}V_5T_{6,7,8}$ are related by a
relabeling of those that multiply $V_1V_2T_{3456,7,8}$.

The monodromy variations of section~\DQdualsec\ and the elliptic identities of
section \newdualitysec\ imply that the definitions \ThetaSix\ are generalized 
elliptic integrands (GEIs); $D\Theta^{(d)}=0$. Moreover, by inserting the solutions of 
the bootstrap procedure in section \bootsec\ up to $n=8$ points, the GEIs $\Theta^{(d)}$ 
are in fact found to vanish up to total derivatives.
The coefficients $\Xi^{(d)}$ of the anomalous terms, however, turn out to be non-zero.
Instead, the trace relations among worldsheet functions discussed in
section~\tracesZsec\ simplify the explicit expressions of $\Xi^{(d)}$ from \XiSix\ 
at $n\leq 7$ points to a single term. In summary, we obtain
\eqn\genTheta{
\Theta^{(0)\,m_1 m_2\ldots m_r}_{A|B_1,B_2,\ldots,B_{r+4}}\cong0\,, \ \ \ \ 
\Theta^{(1)\,m_1 m_2\ldots m_r}_{A|B|B_1,B_2,\ldots,B_{r+4}}\cong0\,,  \ \ \ \ n\leq 8 \, ,
}
and
\eqn\genThetaXi{
\Xi^{(0)\,m_1 m_2\ldots m_r}_{A_1|B_1,\ldots,B_{r+6}}\cong
- \cZ_{A_1|B_1,\ldots,B_{r+6}}\,, \ \ \ \ 
\Xi^{(1)\,m_1 m_2\ldots m_r}_{A_1|A_2|B_1,\ldots,B_{r+6}}\cong
\cZ_{A_1,A_2|B_1, \ldots,B_{r+6}}\,, \ \ \ \ n\leq 7 \, ,
}
see section \brstsec\ for the eight-point examples of $\Xi^{(d)}$.

The simplest examples of \ThetaSix\ are given by (see \QcKsample\ for the former two),
\eqnn\exThetaSix
$$\eqalignno{
\Theta^{(0)}_{1|2,3,4,5} & = k_1^p \cZ^p_{1,2,3,4,5} + \big[s_{12}\cZ_{12,3,4,5}
+ (2\leftrightarrow3,4,5)\big]\,,&\exThetaSix\cr
\Theta^{(0)}_{12|3,4,5,6} & = k_{12}^p \cZ^p_{12,3,4,5,6}
+ \big[s_{23}\cZ_{123,4,5,6} - s_{13}\cZ_{213,4,5,6} +
(3\leftrightarrow4,5,6)\big]\cr
&\quad{} - \cZ_{1|2,3,4,5,6} + \cZ_{2|1,3,4,5,6}\,,\cr
\Theta^{(1)}_{1|2|3,4,5,6,7} & = - k_1^p\cZ^p_{2|1,3,4,5,6,7} - s_{12}\cZ_{12|3,4,5,6,7}
- \big[s_{13}\cZ_{2|13,4,5,6,7} + (3\leftrightarrow4,5,6,7)\big]\,,
}$$
and one can verify from the expressions for $\cZ$
in section \bootsec\ that these linear combinations indeed yield total derivatives.
For more examples, see the appendix~\allThetasapp.

%***********************************
\subsubsec Anomalous Lie polynomials

Given the non-vanishing expressions for $\Xi^{(d)}$ in \genThetaXi,
the Lie-series part $\cK_n^{\rm Lie}(\ell)$ of the correlators from
\experK\ is not BRST invariant for $n\ge 7$. More precisely, we have $T^{(d,r)}=0$
but $Y^{(d,r)}\neq 0$ in \QcKgeneral. Fortunately, these non-vanishing terms are purely
anomalous and suggest to add corrections containing exclusively anomalous
superfields of the form $Y^{m_1\ldots}_{A_1,\ldots,A_d|B_1,\ldots}$, see
\Wanondef, \Ytensordef\ and \refWc. Therefore our proposal for the one-loop
correlator becomes,
\eqn\experQ{
{\cal K}_n(\ell)=
\cK^{\rm Lie}_n(\ell) + {\cal K}_n^Y(\ell)\,,
}
for some $\cK_n^Y(\ell)$ to be determined.
Such an anomaly sector ${\cal K}_n^Y(\ell)$ is
plausible by the kinematic identities of section~\CJACOBIsec, as they mix
anomalous and non-anomalous terms.
Up to and including six points, we have
\eqn\experR{
{\cal K}_{n}^Y(\ell) = 0\,,\quad\hbox{for $n\leq6$} \, .
}
From multiplicity seven on, we need to find an expression
for the
anomaly sector ${\cal K}_{n}^Y(\ell)$ such that the full
correlator satisfies the criteria summarized below.
Even though we will find the proper $\cK_n^Y(\ell)$ in the seven-point example 
of section~\explsec, this is done case-by-case, so it would be desirable
to understand the general pattern behind them.

%*****************************************************************
\newsubsec\FinalAssemblysec Final assembly of one-loop correlators

The general form of the one-loop correlators \experQ\ was suggested by analogy
with the Lie-polynomial structure observed at tree level \nptString. The one-loop
correlators ${\cal K}_n(\ell)$ are expressions in the cohomology \PScohomology\ of pure-spinor
superspace that depend on the loop momentum $\ell^m$ and the zero modes of the
pure spinor $\lambda^\alpha$ and of the superspace coordinate $\theta^\alpha$.
Moreover, they are also expanded in terms of worldsheet functions that have to
be integrated over the vertex operator insertions points as well as over the
moduli space that parametrize the different genus-one surfaces. Given this setting, the final
assembly of one-loop correlators $\cK_n(\ell)$ as defined in \againopen\ must
satisfy the following fundamental requirements:
\medskip
\item{1.} The correlator must be in the cohomology of the BRST operator;
\item{2.} The correlator must be a single-valued function with respect to both
$z_i$ and $\ell^m$;
\item{3.} The correlator must admit a local representation;
\item{4.} The correlator must be manifestly\foot{The symmetry with respect to
leg $1$ is not manifest in the prescription \onepresc\ and therefore can be
verified only up to total $\tau$ derivatives originating from BRST integration
by parts \MPS.} symmetric in the labels
$(2,3, \ldots,n)$.
\medskip
\noindent These conditions arise from general CFT considerations applied to
the pure-spinor amplitude prescription \onepresc, and they are compatible with
the tree-level arguments that led to the Lie-polynomial proposal \experQ. The 
notion of single-valuedness in 2.\ is defined in \MonFinv, and 3.\ refers to the 
absence of kinematic poles $s_{P}^{-1}$ in a local representation of ${\cal K}_n$.
The combination of 1.\ and 3.\ turns out to be particularly constraining: Any 
BRST-invariant linear combination of the building blocks of section \LocalBBsec\
has been checked to vanish in the cohomology at $5\leq n \leq 8$ points (see
appendix \localapp\ for further details).
Therefore there is no freedom of adding BRST-invariant local terms multiplying single-valued
functions at these multiplicities.

In the next section we write down explicit examples of one-loop correlators
fulfilling the above criteria up to seven points. Moreover, we propose an
expression at eight points with mild violations of 1. and 3.: Its BRST
variation vanishes only up to local terms proportional to the Eisenstein series
of modular weight four, ${\rm G}_4$, and certain terms in the anomaly sector
${\cal K}_8^Y$ violate locality. We expect that the eight-point proposal
to be given in section \EightPointsec\ differs from the correct correlator ${\cal K}_8$
by ${\rm G}_4$ multiplying an unknown kinematic factor, i.e. it correctly captures
all dependence on the $z_i$.

%************************************************************************
\newnewsec\explsec One-loop correlators of the open superstring: examples

We will now apply all the techniques developed in the previous sections to
obtain explicit expressions for the one-loop correlators of the open superstring
in a manifestly supersymmetric fashion. The correlators at four, five, six and
seven points meet all the requirements described in section~\FinalAssemblysec,
and we will elaborate on the aforementioned issues with the eight-point correlator below.

%********************************
\newsubsec\secnineone Four points

The four-point correlator is uniquely determined by the zero-mode
integration over the pure-spinor variables and it was firstly
computed by Berkovits in \MPS. Using the definition \TABCdef\ its
correlator can be written as the manifestly local
pure-spinor superspace expression
\eqn\fourptcorr{
{\cal K}_4(\ell) = V_1 T_{2,3,4}\,.
}
Note that there are no worldsheet singularities among the vertex
positions nor an explicit dependence on the loop momentum $\ell^m$.
This is in accordance with the general discussion in section~\twotwosec\ that
a $n$-point correlator ${\cal K}_n(\ell)$ is a polynomial in loop
momenta of degree $n{-}4$ and that the maximum number of
OPE contractions is also $n{-}4$. It has been shown in
\mafraids\ using BRST cohomology identities in pure-spinor superspace
that the one-loop correlator \fourptcorr\ is proportional to its tree-level
counterpart \experE,
\eqn\looptree{
\langle V_1 T_{2,3,4}\rangle = s_{12}s_{23} A^{\rm SYM}(1,2,3,4)\,.
}
Therefore it reproduces the well-known \refs{\GreenMN,\greenloop} supersymmetric completion of $t_8 F^4$
and the one-loop amplitudes of Brink, Green and Schwarz with bosonic external 
states \GreenSW.

%********************************
\newsubsec\secninetwo Five points

The reasoning behind the derivation of the five-point correlator will be
presented in detail as it constitutes the prototype for similar derivations at
higher points. Not surprisingly, the outcome of the following analysis is in
accordance with the general features of one-loop correlators summarized in
section~\secexperC.

As discussed in section \twotwosec, the pure-spinor prescription \MPS\
implies that the five-point correlator ${\cal K}_5(\ell)$ is a polynomial of
degree one in $\ell$ with at most one OPE singularity. Therefore the
correlator is composed of two classes of terms containing: $(i)$ one OPE
contraction, $(ii)$ one loop momentum. Let us consider them in turn.

%******************
\subsubsec The OPEs

The two inequivalent OPEs $V_1(z_1)U_2(z_2)$ and
$U_2(z_2)U_3(z_3)$ can be derived from \allOPEs\ and give rise to two-particle vertex operators
\ranktwoU\ and \Vonetwo,
\eqnn\twopartvertA
$$\eqalignno{
V_1(z_1)U_2(z_2)&\rightarrow  g^{(1)}_{12} V_{12}(z_2)\,, \qquad
U_2(z_2)U_3(z_3)\rightarrow  g^{(1)}_{23} U_{23}(z_3)\,,&\twopartvertA
}$$
where $g^{(w)}_{ij}\equiv g^{(w)}(z_i{-}z_j,\tau)$ refer to expansion coefficients of
the Kronecker--Eisenstein series, see \EisKron, with $g^{(1)}(z,\tau)= \partial_z \log \theta_1(z,\tau)$.
In both cases the zero-mode integration for $d_\alpha$ and $N_{mn}$ only admits the
$b$-ghost sector $b^{(4)}$ defined in section~\twotwosec\ and yields $T_{A,B,C}$
according to
the multiplicity-agnostic rule \justMULT.
In assembling all the ten OPE channels we obtain
\eqn\allOPEfive{
 {\cal K}_5(\ell)\big|_{\rm OPE} = \big[g^{(1)}_{12} V_{12} T_{3,4,5} + (2\leftrightarrow 3,4,5)\big]
 +\big[g^{(1)}_{23} V_{1} T_{23,4,5} + (2,3|2,3,4,5)\big]\,.
}

%*************************************
\subsubsec Adjoining the loop momentum

Five points is the first instance where a loop momentum can be extracted from 
the external vertices or the $b$-ghost. According to the discussion of
section~\vecsec, the relevant $b$-ghost sectors are $b^{(4)}$ and $b^{(2)}$,
and they give rise to the schematic contributions $\ell_m V_1 A_2^m T_{3,4,5}$
and $\ell_m V_1 W^m_{2,3,4,5}$, respectively. BRST covariance fixes their
relative coefficients to
\eqn\allellfive{
\cK_5(\ell)\big|_{\ell} = \ell_m V_1 T^m_{2,3,4,5}\,,
}
see \TmABCDdef. By adjoining the contribution \allOPEfive\ from OPEs, one arrives at
the following final expression for the five-point correlator anticipated in section \antiKfive:
\eqnn\completefive
$$\eqalignno{
{\cal K}_5(\ell)  &= {\cal K}_5(\ell)\big|_{\ell} + {\cal K}_5(\ell) \big|_{\rm OPE}
&\completefive \cr
&= \ell_m V_1 T^m_{2,3,4,5}+ \big[g^{(1)}_{12} V_{12} T_{3,4,5} {+} (2\leftrightarrow 3,4,5)\big]
 +\big[g^{(1)}_{23} V_{1} T_{23,4,5} {+} (2,3|2,3,4,5)\big]\,.
}$$
It will be rewarding to rewrite the correlator \completefive\ in a slightly more
abstract manner, since the higher-point generalization will become more
natural in this way. The correlator lines up with the Lie-polynomial structure of \dzeroGeneral,
\eqnn\fiveLie
$$\eqalignno{
\cK_5(\ell) &=
 V_{A_1} T^m_{A_2, \ldots,A_5} \cZ^m_{A_1, \ldots,A_5}
+ \big[12345|A_1, \ldots,A_5\big] &\fiveLie\cr
&\hskip1.35pt{}+ V_{A_1} T_{A_2, \ldots,A_4} \cZ_{A_1, \ldots,A_4}
+ \big[12345|A_1, \ldots,A_4\big]\,,
}$$
where the notation for the permutations is explained after \experH\ and in the
appendix~\stirlingapp. Expanding the above permutations leads
to the following 
${5\stirling 5}+{5\stirling 4}=1+10=11$ terms,
\eqnn\fivecorr
$$\eqalignno{
{\cal K}_5(\ell)& = V_1 T^m_{2,3,4,5} \cZ^m_{1,2,3,4,5} &\fivecorr\cr
&{}+V_{12} T_{3,4,5} \cZ_{12,3,4,5}  + (2{\leftrightarrow} 3,4,5)\cr
&{}+V_1 T_{23,4,5}\cZ_{1,23,4,5} + (2,3|2,3,4,5)\,.
}$$
In comparing \fivecorr\ with \completefive\ we can read off the following
$\cZ$-functions,
\eqn\compansatz{
{\cal Z}_{12,3,4,5} = g^{(1)}_{12}\,, \qquad
{\cal Z}^m_{1,2,3,4,5}= \ell^m\,,
}
which correspond to the functions \fiveSol\ studied in section~\fivewssec.
As this example demonstrates, the presentation of the correlator as the Lie-polynomial
\fiveLie\ organizes the worldsheet functions in a way that manifests
the parallels with the kinematic building blocks as highlighted in section~\dualitysec.

In summary, the five-point one-loop correlator \fivecorr\ is a manifestly local
expression of superfields that was obtained using general arguments based on
the amplitude prescription of the pure-spinor formalism. If we want to argue that
it is also the {\it correct} correlator, it must be BRST invariant and single-valued
as well.

%*************************
\subsubsec BRST invariance

It is straightforward to use the BRST variations of the local
building blocks -- \exampOne, \QTs\ and \QTms\ -- to obtain the $n=5$ instance
of the general BRST variation \QcKgeneral. Indeed, a short calculation yields
\eqnn\brstfive
$$\eqalignno{
Q{\cal K}_5(\ell)
&=- V_1V_2 T_{3,4,5}\Big[ k_2^m \cZ^m_{1,2,3,4,5} + \big[
s_{21}\cZ_{21,3,4,5} + (1\leftrightarrow3,4,5)
\big]\Big]
 + (2\leftrightarrow 3,4,5)\cr
&=- V_1V_2T_{3,4,5}\Theta^{(0)}_{2|1,3,4,5} +(2\leftrightarrow3,4,5)\,, &\brstfive\cr
}$$
where in the second line we used the shorthand defined in \ThetaSix.
At first sight \brstfive\ appears to be different than zero, but
luckily this particular arrangement of integrands turns out to be
a total worldsheet derivative,
\eqn\totder{
k_2^m \cZ^m_{1,2,3,4,5} + \bigl[s_{21}\cZ_{21,3,4,5} +
(1\leftrightarrow3,4,5)\bigr]=
%\Theta_{2|1,3,4,5} =
(\ell\cdot k_2) + \bigl[
s_{21}\gg1(2,1) + (1\leftrightarrow3,4,5)\bigr]
 \cong 0\,, 
}
where we used the expansions \compansatz\ and the identity \zderivzero. Therefore the
five-point correlator \fivecorr\ is BRST invariant.

%***************************
\subsubsec Single-valuedness

From the discussion in section~\doublysec, it follows that the monodromies
around the $A$-cycle vanish for any combination of $\ell^m$ and $g^{(n)}_{ij}$, 
so the correlator \fivecorr\ will be single valued if its monodromies around the
$B$-cycle also vanish. In this case, the variations \covD\ yield
\eqnn\singlefive
$$\eqalignno{
D{\cal K}_5(\ell) &= \Omega_1\Big( k_1^m V_1T^m_{2,3,4,5}
+ \big[ V_{12}T_{3,4,5} + 2\leftrightarrow 3,4,5\big]\Big)&\singlefive\cr
&{}+\Omega_2\Big( k_2^m V_1T^m_{2,3,4,5} +
V_{21}T_{3,4,5}
+ \big[ V_{1}T_{23,4,5} + 3\leftrightarrow 4,5\big]\Big) +
(2\leftrightarrow 3,4,5)\,,\cr
}$$
see section \monodromysec\ for the linearized-monodromy operator $D$. Note that 
the superspace expressions that multiply the formal variables $\Omega_i$ for $i=1, \ldots,5$ 
in the definition \covC\ of $D$ are BRST-closed and local. However, as discussed in the 
appendix \localapp, the BRST cohomology is empty 
for local superspace expressions and therefore the above combinations must 
be BRST-exact. In fact, one can show via \QJex\ and \deltasecE\ that
\eqn\singleFiveAgain{
D\cK_5(\ell)=\Omega_1 QJ_{1|2,3,4,5} + \big[\Omega_2
(QD_{1|2|3,4,5}-\Delta_{1|2,3,4,5})
+ (2\leftrightarrow3,4,5)\big]\cong0\,.
}
Since the anomalous superfield $\Delta_{1|2,3,4,5}$ was shown to
be BRST-exact in \partI, the monodromy variation \singleFiveAgain\ vanishes
in the cohomology of the pure-spinor superspace (indicated by $\cong0$), and
the correlator \fivecorr\ is therefore single-valued.

%******************************************************************
\subsubsec Duality between worldsheet functions and BRST invariants

The vanishing of \totder\ is a clear indication of the duality between
worldsheet functions and BRST invariants discussed in section~\dualitysec\ and
pointed out in \MafraIOJ; it corresponds to the BRST-exact linear combination of
superfields in \singlefive\ under the replacement \MTocZ,
\eqnn\McZdualfive
$$\eqalignno{
0&\cong k_2^m V_1T^m_{2,3,4,5} +
V_{21}T_{3,4,5}
+ V_{1}T_{23,4,5}
+ V_{1}T_{24,3,5}
+ V_{1}T_{25,3,4}
&\McZdualfive \cr
\Longleftrightarrow\ \  0 &\cong
k_2^m Z^{(s)m}_{1,2,3,4,5} + Z^{(s)}_{21,3,4,5} 
+ Z^{(s)}_{23,1,4,5}
+ Z^{(s)}_{24,1,3,5}
+ Z^{(s)}_{25,1,3,4} \,,
}$$
where the Lie-symmetric worldsheet functions $Z^{(s)}$ have been introduced in section \Liesymsec. 
The superspace expression in the first line of \McZdualfive\
vanishes because it is BRST exact, see \deltasecE, while the integrand in the
second line vanishes because it is a total worldsheet derivative, see
\zderivzero. This correspondence between BRST invariance and monodromy invariance  
is a central example of the {\it duality} between pure-spinor-superspace
expressions and one-loop worldsheet functions. In fact, further
investigation of such relations led to the discussions presented in
section~\dualitysec.

%*****************************************************
\subsubsec Different representations of the five-point correlator

Since the correlator \fivecorr\ is local, single-valued and BRST invariant, it
meets the criteria of section~\FinalAssemblysec\ to be the open-superstring 
five-point correlator. We will now exploit the
properties of both the superspace expressions and the worldsheet functions
to rewrite it in various ways that manifest different subsets of these fundamental properties.

%*************************************************************************************
\newsubsubsubsec\fiveCZsec The $C\cdot\cZ$ representation: manifesting BRST invariance

Integration-by-parts identities can be used to yield a manifestly BRST closed representation of
the correlator: first rewrite \fivecorr\ in terms of Berends--Giele currents $M_A$ and $M_{B,C,D}$ 
associated with $V_A$ and $T_{B,C,D}$ using the trading identity \tradingId\ of the Lie polynomial as
\eqnn\fivecorrM
$$\eqalignno{
{\cal K}_5(\ell)& = M_1 M^m_{2,3,4,5} \cZ^{m}_{1,2,3,4,5}
+ \big[ M_{12} M_{3,4,5} s_{12}\cZ_{12,3,4,5}  + (2\leftrightarrow 3,4,5)
\big] &\fivecorrM\cr
&+ M_1 M_{23,4,5}s_{23}\cZ_{1,23,4,5} + (2,3|2,3,4,5)\,.
}$$
Next, the integration-by-parts identity \totder\ can be used
to eliminate all functions of the form $\cZ_{1i,A,B,C}$ with $i\neq\emptyset$
(i.e.\ all of $g^{(1)}_{12},g^{(1)}_{13},g^{(1)}_{14},g^{(1)}_{15}$).
Doing this leads to
\eqnn\fiveIBPed
$$\eqalignno{
\cK_5(\ell) &= \cZ^{m}_{1,2,3,4,5}\Big(M_1M^m_{2,3,4,5} + \big[k_2^m
M_{12}M_{3,4,5} + (2\leftrightarrow3,4,5) \big]\Big) &\fiveIBPed\cr
&\quad{}+\big[ s_{23}\cZ_{1,23,4,5}(M_1M_{23,4,5} + M_{12}M_{3,4,5} -
M_{13}M_{2,4,5}) + (2,3|2,3,4,5) \big]\,.
}$$
In this way, the terms inside the round brackets build up the
Berends--Giele expansions of the BRST invariants from \scalarCs\ and \vectorCs\
such that \fiveIBPed\ becomes
\eqn\brstfivecorr{
\cK_5(\ell) =
C^m_{1|2,3,4,5} \cZ^{m}_{1,2,3,4,5}
+ \big[C_{1|23,4,5}\, s_{23}\cZ_{1,23,4,5}  + (2,3|2,3,4,5)\big]\,.
}
Since $C^m_{1|A,B,C,D}$ and $C_{1|A,B,C}$ are both BRST closed,
\brstfivecorr\ constitutes a manifestly BRST invariant
representation of the local correlator \fivecorr.

%**************************************************************************************
\newsubsubsubsec\fiveTEsec The $T\cdot E$ representation: manifesting single-valuedness

Since the five-point correlator \fivecorr\ is single valued, it is worthwhile to spell out a
representation that manifests this property. To do this, we rewrite the terms containing 
a factor of $V_{1A}$ with non-empty $A$ using the BRST cohomology identity
\eqn\Vtwoid{
V_{12}T_{3,4,5} \cong k_2^m V_1 T^m_{2,3,4,5} + \big[V_1T_{23,4,5}+
(3\leftrightarrow4,5)\big]\,,
}
which follows from \deltasecE\ and the BRST-exactness of
$\Delta_{1|2,3,4,5}$. Doing this replacement in the correlator
\fivecorr\ and collecting terms leads to
\eqnn\fiveEsingle
$$\eqalignno{
\cK_5(\ell) &= V_1 T^m_{2,3,4,5}\Big(
\cZ^m_{1,2,3,4,5} + \big[k_2^m \cZ_{12,3,4,5} +
(2\leftrightarrow 3,4,5)\big] \Big) &\fiveEsingle\cr
&+ \big[V_1T_{23,4,5}\big(\cZ_{1,23,4,5} + \cZ_{12,3,4,5} - \cZ_{13,2,4,5}\big)
+ (2,3|2,3,4,5)\big]\,.
}$$
The combinations of $\cZ$-functions in the round brackets can be identified with the
GEIs $E^m_{1|2,3,4,5}$ and $E_{1|23,4,5}$ from \fiveEls\ and \fiveElls, respectively.
Using these functions, the correlator \fiveEsingle\ takes the manifestly
single-valued form:
\eqn\fiveTE{
\cK_5(\ell) = V_1 T^m_{2,3,4,5} E^m_{1|2,3,4,5} + \big[V_1T_{23,4,5}E_{1|23,4,5}
+ (2,3|2,3,4,5)\big]\,.
}
This representation reverses the roles of worldsheet functions and kinematic factors
in comparison to \brstfivecorr\foot{This becomes particularly
transparent by introducing $Z^{(s) m}_{1,2,3,4,5}=\cZ^{m}_{1,2,3,4,5}$ and
$Z^{(s)}_{1,23,4,5}=s_{23}\cZ_{1,23,4,5}$ in \brstfivecorr.}: Manifest BRST invariance is traded for 
manifest monodromy invariance. 

%***********************************************************************
\newsubsubsubsec\fiveCEsec The $C\cdot E$ representation: manifesting BRST invariance \& single-valuedness

The five-point correlator can also be rewritten such as to
manifest both BRST invariance and single-valuedness. To this effect we
eliminate $\cZ^m_{1,2,3,4,5} = E^m_{1|2,3,4,5}- [ k_2^m \cZ_{12,3,4,5} +
(2\leftrightarrow 3,4,5) ]$ as well as $\cZ_{1,23,4,5} =E_{1|23,4,5} -\cZ_{12,3,4,5} +\cZ_{13,2,4,5} $ 
from \brstfivecorr\ and use the BRST cohomology identity
\deltasecE\ to obtain
\eqn\covK{
\cK_5(\ell) = C^m_{1|2,3,4,5}E^m_{1|2,3,4,5}
+ [ C_{1|23,4,5} s_{23}E_{1|23,4,5}  + (2,3|2,3,4,5) ]\,,
}
which reproduces the double-copy expression for the
five-point correlator proposed in \MafraIOJ\ and manifests both 
BRST invariance and single-valuedness.

%************************************
\subsubsec Summary of representations

As shown above, there are multiple Lie-polynomial representations of
the five-point correlator according to which features are chosen
to be manifested:
\eqnn\allfives
$$\eqalignno{
{\cal K}_5(\ell)& = V_1 T^m_{2,3,4,5} \cZ^m_{1,2,3,4,5}
+ \big[ V_{12} T_{3,4,5} \cZ_{12,3,4,5}  + (2 \leftrightarrow 3,4,5)
\big] &\allfives\cr
&\quad{}+ \big[ V_1 T_{23,4,5}\cZ_{1,23,4,5} + (2,3|2,3,4,5) \big]\,,\cr
\cK_5(\ell) &= V_1 T^m_{2,3,4,5} E^m_{1|2,3,4,5} + \big[V_1T_{23,4,5}E_{1|23,4,5}
+ (2,3|2,3,4,5)\big]\,,\cr
\cK_5(\ell) &=
 C^m_{1|2,3,4,5} \cZ^m_{1,2,3,4,5}
+ \big[ C_{1|23,4,5} s_{23}\cZ_{1,23,4,5} + (2,3|2,3,4,5) \big]\,,\cr
{\cal K}_5(\ell) &=C^m_{1|2,3,4,5} E^m_{1|2,3,4,5}
+ [ C_{1|23,4,5} s_{23} E_{1|23,4,5} + (2,3|2,3,4,5) ]\,.
}$$
In addition to the above, the single-valued representation of the five-point
correlator obtained by explicit integration over the loop momentum will be
presented in section~\loopintsec.

We remark that the one-loop five-point amplitude in the open superstring has
been computed with the RNS and GS formalisms for states in the Neveu-Schwarz
sector \refs{\TsuchiyaVA,\stiefive,\Richards} and in the Ramond sector
\refs{\lin,\sen}. Manifestly supersymmetric expressions were obtained in
\fiveptNMPS\ using the non-minimal pure-spinor formalism \NMPS\ and
later in \oneloopbb\ using the minimal pure-spinor formalism.

%*********************************
\newsubsec\secninethree Six points

We will now show that the general formulas summarized in section~\secexperC\ give
rise to the correct six-point one-loop correlator. The Lie-polynomial form of the six-point 
correlator is given by
\eqnn\sixLie
$$\eqalignno{
\cK_6(\ell) &=
\half V_{A_1} T^{mn}_{A_2, \ldots,A_6} \cZ^{mn}_{A_1, \ldots,A_6}
+\big[123456|A_1, \ldots,A_6\big] &\sixLie\cr
&\hskip8.5pt{}+ V_{A_1} T^m_{A_2, \ldots,A_5} \cZ^m_{A_1, \ldots,A_5}
+ \big[123456|A_1, \ldots,A_5\big]\cr
&\hskip8.5pt{}+ V_{A_1} T_{A_2, \ldots,A_4} \cZ_{A_1, \ldots,A_4}
+ \big[123456|A_1, \ldots,A_4\big]\,,
}$$
with the following worldsheet functions as derived in section~\sixwssec,
\eqnn\newgsAgain
$$\eqalignno{
\cZ_{123,4,5,6}&= g^{(1)}_{12}g^{(1)}_{23} + g^{(2)}_{12} +
g^{(2)}_{23} - g^{(2)}_{13}\,, &\newgsAgain\cr
\cZ_{12,34,5,6}&= g^{(1)}_{12}g^{(1)}_{34}
+ g^{(2)}_{13} + g^{(2)}_{24}
- g^{(2)}_{14} - g^{(2)}_{23}\,,\cr
\cZ^m_{12,3,4,5,6}&= \ell^m g^{(1)}_{12} 
+ (k_2^m - k_1^m)g^{(2)}_{12}
+ \big[ k_3^m (g^{(2)}_{13} - g^{(2)}_{23}) + (3\leftrightarrow
4,5,6)\big]\,,\cr
\cZ^{mn}_{1,2,3,4,5,6}&= \ell^m\ell^n +
\bigl[( k_1^{m}k_2^{n}+k_1^{n}k_2^{m}) g^{(2)}_{12} + (1,2|1,2,3,4,5,6)
\bigr]\,.
}$$
Note that a possible contribution of a $d=1$ refined sector according to \ddGeneral\
is suppressed since the monodromy variations \sixMon\ are compatible with $\cZ_{1|2,3,4,5,6} = 0$.
The explicit expansion of the Stirling cycle permutations in \sixLie\ generates a total of
${6\stirling 6}+{6\stirling 5}+{6\stirling 4}=1+15+85=101$ terms,
\eqnn\sixloc
$$\eqalignno{
\cK_6(\ell) &= {1\over2}
V_1 T^{mn}_{2,3,4,5,6}\cZ_{1,2,3,4,5,6}^{mn} &\sixloc\cr
&+  V_{12} T^m_{3,4,5,6} \cZ^m_{12,3,4,5,6}
+ (2\leftrightarrow 3,4,5,6)\cr
&+
 V_1 T^m_{23,4,5,6}\cZ^m_{1,23,4,5,6}  + (2,3|2,3,4,5,6) \cr
&+V_{123}T_{4,5,6} \cZ_{123,4,5,6} +V_{132} T_{4,5,6} \cZ_{132,4,5,6}
+ (2,3|2,3,4,5,6) \cr
& + V_1 T_{234,5,6} \cZ_{1,234,5,6}
+ V_1 T_{243,5,6} \cZ_{1,243,5,6}
+ (2,3,4|2,3,4,5,6)\cr
& + \big[\big(V_{12} T_{34,5,6}\cZ_{12,34,5,6} + {\rm cyc}(2,3,4)\big)
 + (2,3,4|2,3,4,5,6)\big]\cr
 &+\big[\bigl(V_{1}T_{2,34,56} \cZ_{1,2,34,56}   + {\rm
 cyc}(3,4,5)\bigr)+ (2\leftrightarrow 3,4,5,6) \big]\,, \cr
}$$
where the indicated permutations defined in \sumbin\ act on a line-by-line basis.

We will now prove that the six-point correlator \sixLie\ is both single-valued
and BRST invariant, in accordance with the expectations outlined in
section~\FinalAssemblysec.

%*************************
\subsubsec BRST invariance

As anticipated in section \QGeneralsec, the BRST algebra of the building blocks leads to the following 
$Q$-variation of the correlator \sixLie,
\eqnn\BRSTsixTwo
$$\eqalignno{
Q\cK_6(\ell) &= -\half V_1 Y_{2,3,4,5,6} \cZ^{mm}_{1,2,3,4,5,6}&\BRSTsixTwo\cr
&\quad{}
- V_1V_2 T^m_{3,4,5,6}\Theta^{(0)\,m}_{2|1,3,4,5,6}
- V_{12}V_{3}T_{4,5,6}\Theta^{(0)}_{3|12,4,5,6} + (2\leftrightarrow3,4,5,6)\cr
&\quad{} - \big[V_1V_2T_{34,5,6}\Theta^{(0)}_{2|1,34,5,6}
+ (3,4|3,4,5,6)\big]+(2\leftrightarrow3,4,5,6)\cr
&\quad{} - V_1V_{23}T_{4,5,6}\Theta^{(0)}_{23|1,4,5,6} + (2,3|2,3,4,5,6)\,,
}$$
where the shorthands $\Theta^{(0)\,m \ldots}_{1|A,\ldots}$ were
defined in \ThetaSix. After discarding the vanishing refined $\cZ$-function,
they are given by
\eqnn\DeltaZs
$$\eqalignno{
\Theta^{(0)\,m}_{2|1,3,4,5,6} &=
k_2^n \cZ^{mn}_{1,2,3,4,5,6} + \bigl[s_{21}\cZ^m_{21,3,4,5,6} +
(1\leftrightarrow 3,4,5,6)\bigr]\cong0\,, &\DeltaZs\cr
\Theta^{(0)}_{3|12,4,5,6} &= k_3^m {\cal Z}^m_{12,3,4,5,6} + s_{31} {\cal Z}_{312,4,5,6} -  s_{32} {\cal Z}_{321,4,5,6} 
+ \big[ s_{34} {\cal Z}_{12,34,5,6} + (4\leftrightarrow 5,6)\big] \cong 0
\, ,\cr
\Theta^{(0)}_{2|1,34,5,6} &=
k_2^m \cZ^m_{1,2,34,5,6} + s_{23}\cZ_{1,234,5,6} - s_{24}\cZ_{1,243,5,6} +
\bigl[s_{21}\cZ_{21,34,5,6} + (1\leftrightarrow 5,6)\bigr]\cong0\,, \cr
\Theta^{(0)}_{23|1,4,5,6} &=
k_{23}^m \cZ^m_{1,23,4,5,6} +
\bigl[s_{31}\cZ_{231,4,5,6} - s_{21}\cZ_{321,4,5,6}
+(1\leftrightarrow 4,5,6)\bigr]\cong0\,,
}$$
and conspire to total derivatives in $z_j$. Therefore 
the BRST variation is proportional
to the trace $\cZ^{mm}_{1,2,3,4,5,6}$,
\eqn\QsixLie{
Q\cK_6(\ell) =  -\half V_1 Y_{2,3,4,5,6} \cZ^{mm}_{1,2,3,4,5,6}
=-  2\pi i \, V_1 Y_{2,3,4,5,6}  {\p\over\p\tau}\log{\cal I}_6(\ell) \cong0\,,
}
where the total $\tau$ derivative of the Koba--Nielsen
factor has been identified in \traceB. Thus, the BRST variation is a
boundary term in moduli space \FMS, and the usual mechanism of anomaly
cancellation \AnomalyGreen\ implies that the amplitudes computed from
the correlator \sixLie\ are BRST invariant.

%************************************************************************************
\newsubsubsubsec\sixCZsec The $C\cdot\cZ$ representation: manifesting BRST invariance

Now that BRST invariance of the six-point correlator is proven, let us rewrite it
using the BRST invariants from section~\BRSTpseudosec, in a similar spirit as done
with the five-point correlator in the previous section. There are different ways
to achieve this, one uses the trading identity \tradingId\ to rewrite
the Lie polynomial \sixLie\ as,
\eqnn\KXM
$$\eqalignno{
\cK_6(\ell) &=
\half M_{A_1} M^{mn}_{A_2, \ldots,A_6} Z^{(s),mn}_{A_1, \ldots,A_6}
+\big[123456|A_1, \ldots,A_6\big] &\KXM\cr
&\hskip8.35pt{}+ M_{A_1} M^m_{A_2, \ldots,A_5} Z^{(s),m}_{A_1, \ldots,A_5}
+ \big[123456|A_1, \ldots,A_5\big]\cr
&\hskip8.35pt{}+ M_{A_1} M_{A_2, \ldots,A_4} Z^{(s)}_{A_1, \ldots,A_4}
+ \big[123456|A_1, \ldots,A_4\big]\,,
}$$
where $Z^{(s)}$ is defined in \cZToZ.
The idea now is to exploit the fact that terms of the form
$M_1M^{m \ldots}_{A,B, \ldots}$, which feature the single-particle Berends--Giele current $M_1$,
are the leading terms in the expansion of the BRST (pseudo-)invariants from
section~\BRSTinvsec. Therefore they can be rewritten as
\eqn\rMone{
M_1 M^{m \ldots}_{A,B, \ldots} = C^{m \ldots}_{1|A,B, \ldots} + \cdots,
}
where the terms in the ellipsis on the right-hand side are
linear combinations of $M_{1A}M^{m \ldots}_{B, \ldots}$ with $A\neq\emptyset$
that uniquely follow from the definition of the BRST pseudo-invariants
in \scalarCs\ to \tenssixpt. Plugging in the above expressions into the correlator \KXM\ yields
\eqnn\pseudoLieSix
$$\eqalignno{
\cK_6(\ell) &=
\half C^{mn}_{1|A_1, \ldots,A_5} Z^{(s)mn}_{1,A_1, \ldots,A_5}
+\big[23456|A_1, \ldots,A_5\big] &\pseudoLieSix\cr
&\hskip8.35pt{}+ C^m_{1|A_1, \ldots,A_4} Z^{(s)m}_{1,A_1, \ldots,A_4}
+ \big[23456|A_1, \ldots,A_4\big]\cr
&\hskip8.35pt{}+ C_{1|A_1, \ldots,A_3} Z^{(s)}_{1,A_1, \ldots,A_3}
+ \big[23456|A_1, \ldots,A_3\big]\,.
}$$
To arrive at \pseudoLieSix\
the following three topologies of terms
(and their permutations) were discarded as they are
total derivatives:
\eqnn\Thetasix
$$\displaylines{
s_{34}M_{12}M_{34,5,6}\Theta^{(0)}_{2|1,34,5,6}\cong0\,,\qquad
M_{12}M^m_{3,4,5,6}\Theta^{(0)\,m}_{2|1,3,4,5,6}\cong0\,,\hfil\Thetasix\hfilneg\cr
M_{123}M_{4,5,6}\Big(k_3^m\Theta^{(0)\,m}_{2|1,3,4,5,6}
+ s_{12}\Theta^{(0)}_{3|12,4,5,6}
+ \big[s_{34}\Theta^{(0)}_{2|1,34,5,6}+(4\leftrightarrow5,6)\big]\Big)\cong0\,.
}$$
Expanding the Stirling cycle permutations in \pseudoLieSix\ yields
the following
${5\stirling 5}+{5\stirling 4}+{5\stirling 3} = 1+10+35 = 46$ terms,
\eqnn\pseudoSix
$$\eqalignno{
\cK_6(\ell) & =
\half C^{mn}_{1|2,3,4,5,6} Z^{(s)mn}_{1,2,3,4,5,6}
+\bigl[ C^m_{1|23,4,5,6} Z^{(s)m}_{1,23,4,5,6} +
(2,3|2,3,4,5,6)\bigr] &\pseudoSix\cr
&+\bigl[ C_{1|234,5,6} Z^{(s)}_{1,234,5,6}
+ C_{1|243,5,6} Z^{(s)}_{1,243,5,6} + (2,3,4 | 2,3,4,5,6) \big]\cr
&+\bigl[
C_{1|23,45,6} Z^{(s)}_{1,23,45,6}
+ C_{1|24,35,6} Z^{(s)}_{1,24,35,6}
+ C_{1|25,34,6} Z^{(s)}_{1,25,34,6} + (6\leftrightarrow 2,3,4,5)
\bigr]\,.
}$$
Note one important difference between the expansion above and an earlier representation; 
unlike the local Lie polynomial \sixLie\ in which six labels are distributed among the available 
slots, in the non-local representation \pseudoLieSix\ only five labels participate
in the Stirling permutations. Like this, the initially $101$ terms
in \sixloc\ conspire to the considerably smaller number of $46$ terms in \pseudoSix.

As a consistency check, we note that
the scalar $C_{1|A,B,C}$ and vectorial $C^m_{1|A,B,C,D}$ are manifestly
BRST closed while the BRST variation of the two-tensor $C^{mn}_{1|A,B,C,D,E}$
is proportional to $\d^{mn}$ \partI. Hence, we arrive at the same conclusion as in \QsixLie\
\eqn\BRSTMZsix{
Q\cK_6(\ell) = - \half V_1 Y_{2,3,4,5,6} \cZ^{mm}_{1,2,3,4,5,6}
= -  2\pi i \, V_1 Y_{2,3,4,5,6}  {\p\over\p\tau}\log{\cal I}_6(\ell)  \cong0\,,
}
since $Z^{(s)\,mn}_{1,2,3,4,5,6}=\cZ^{mn}_{1,2,3,4,5,6}$ follows from \cZToZ.

%*******************************************
\newsubsubsec\Dsixproofsec Single-valuedness

To prove that the correlator \sixLie\ is single-valued it is sufficient to show 
that its integration-by-parts-equivalent representation \pseudoSix\ is single-valued.
After a tedious calculation using the monodromy variations \sixMon\ one gets
\eqn\sixDK{
D\cK_6(\ell) = \Omega_1 \d\cK_6^{(1)} +\Omega_2\d\cK_6^{(2)} + \cdots
+\Omega_6\d\cK_6^{(6)}\,,
}
where
\eqnn\expdK
$$\eqalignno{
\d\cK_6^{(1)} & =
k_1^n C^{mn}_{1|2,3,4,5,6}E^m_{1|2,3,4,5,6}
+ \big[k^m_1 s_{23}C^m_{1|23,4,5,6} E_{1|23,4,5,6} +
(2,3|2,3,4,5,6)\big] &\expdK\cr
\d\cK_6^{(2)} & = k_2^m C^{mn}_{1|2,3,4,5,6} E^n_{2|1,3,4,5,6} +
\big[s_{23}C^m_{1|23,4,5,6}E^m_{2|1,3,4,5,6}+(3\leftrightarrow4,5,6)\big]\cr
& \! \! \! \! \! \! \! \! +\big[k_2^m s_{34}C^m_{1|2,34,5,6}E_{2|1,34,5,6} + (3,4|3,4,5,6)\big]\cr
&\! \! \! \! \! \! \! \!  +\big[\big(s_{23}s_{45}C_{1|23,45,6}E_{2|1,3,45,6} +
s_{23}s_{34}C_{1|234,5,6}E_{2|1,34,5,6}+{\rm cyc}(3,4,5)\big) +
(3\leftrightarrow4,5,6)\big] \, ,
}$$
and the other $\d\cK_6^{(i)}$ for $i=3,4,5,6$ are obtained from relabeling of
$\d\cK_6^{(2)}$ under $(2\leftrightarrow3)$, $(2\leftrightarrow4)$ and so forth.
The structural difference between $\d\cK_6^{(1)}$ and $\d\cK_6^{(j)}$ for
$j\neq1$ arises from the choice of basis for the BRST
pseudo-invariants which singles out leg number $1$ in $C^{m \ldots}_{1|A,
\ldots}$. To expose this, one uses the kinematic change-of-basis identities dual to \cobA\
and \cobB\ \partI\ (also see section \dualbasissec\ and the appendix~\changebasisapp) to rewrite $\d\cK_6^{(2)}$
in a basis of $C^{m \ldots}_{2|A, \ldots}$ to obtain\foot{The anomalous term in the change-of-basis identity 
$C^{mn}_{1|2,3,4,5,6}= 
\delta^{mn} {\cal Y}_{21,3,4,5,6}+C^{mn}_{2|1,3,4,5,6}+\ldots$ \partI\ has already 
been discarded from \deltatwo\ since the accompanying
GEI $k_2^mE^n_{2|1,3,4,5,6}$ vanishes upon contraction with $\delta^{mn}$.}
\eqn\deltatwo{
\d\cK_6^{(2)}=k_2^m C^{mn}_{2|1,3,4,5,6}E^n_{2|1,3,4,5,6}
+ \big[k^m_2 s_{13}C^m_{2|13,4,5,6} E_{2|13,4,5,6} +
(1,3|1,3,4,5,6)\big]\,,
}
which is clearly the relabeling of $\d\cK_6^{(1)}$ under $1\leftrightarrow2$.
Therefore, it suffices to demonstrate the vanishing of $\d\cK_6^{(1)}$ to prove
that the correlator \sixLie\ is single-valued. 

To show that $\d\cK_6^{(1)}$ vanishes, we use the kinematic
BRST cohomology identities \partI
\eqnn\tenone
$$\eqalignno{
k_1^m C^m_{1|23,4,5,6}&\cong -  P_{1|2|3,4,5,6} + P_{1|3|2,4,5,6} - \Delta_{1|23,4,5,6} &\tenone
\cr
k^n_1 C^{mn}_{1|2,3,4,5,6} &\cong
- \big[k^m_2 P_{1|2|3,4,5,6} + (2\leftrightarrow 3,4,5,6)\big]
- \Delta^m_{1|2,3,4,5,6}\,, 
}$$
that follow from \sixptQDs\ and can be used to bring \expdK\ into the following form
\eqnn\sixdelone
$$\eqalignno{
\d\cK_6^{(1)}&\cong - \Delta^m_{1|2,3,4,5,6}E^m_{1|2,3,4,5,6}
- \big[\Delta_{1|23,4,5,6}s_{23}E_{1|23,4,5,6} + (2,3|2,3,4,5,6)\big]  &\sixdelone\cr
&-\Big\{ P_{1|2|3,4,5,6} \big(k_2^m E^m_{1|2,3,4,5,6} + \big[s_{23}E_{1|23,4,5,6} +
(3\leftrightarrow4,5,6)\big] \big)
+ (2\leftrightarrow 3,4,5,6) \Big\} \,.
}$$
The coefficients of $P_{1|2|3,4,5,6}$ in the second line in turn
conspire to total derivatives,
\eqn\extvanish{
k_2^m E^m_{1|2,3,4,5,6} + \big[s_{23}E_{1|23,4,5,6} +
(3\leftrightarrow4,5,6)\big]\cong0\,,
}
see section \thissubsec.
Finally, combining the relabelings of the first line of \sixdelone, we arrive at
\eqnn\DsixFinal
$$\eqalignno{
-D\cK_6(\ell) &\cong \Omega_1\Big(\Delta^m_{1|2,3,4,5,6}E^m_{1|2,3,4,5,6} + \big[
\Delta_{1|23,4,5,6} s_{23} E_{1|23,4,5,6} + (2,3|2,3,4,5,6)\big]\Big)\cr
&\quad{}+(1\leftrightarrow2,3,4,5,6)\,.&\DsixFinal
}$$
As reviewed in section~\Deltasec, the unrefined anomalous building blocks $\Delta^{m_1
\ldots}_{1|A_1, \ldots}$ are BRST exact, so the monodromy variation
\DsixFinal\ vanishes in the cohomology of the pure-spinor BRST charge, finishing
the proof that the six-point correlator \sixLie\ is single valued. By the interplay of the
cohomology identity \tenone\ and the GEI relation \extvanish, our proof of $D\cK_6(\ell) \cong0$
constitutes an illuminating showcase of the duality between kinematics and worldsheet functions. 

We note that there are other ways to prove the single-valuedness of the
six-point correlator \sixLie. One such proof, given in section \loopintsec,
follows by explicitly integrating the loop momentum from the correlator while
verifying that only the single-valued functions $f^{(n)}_{ij}$ in \NHKron\ build up in the
outcome. Another proof, presented in the appendix~\LocalMonapp, uses
manipulations involving the manifestly-local representation $T\cdot \cZ$.
However, in exploiting the BRST invariance of the correlator in its $C\cdot
\cZ$ representation the proof above is considerably simpler than the others.

%********************************************************************************************************
\newsubsubsubsec\sixCEsec The $C\cdot E$ representation: manifesting BRST invariance \& single-valuedness

As another application of the duality between kinematics and worldsheet functions,
we shall now derive a manifestly BRST-invariant and single-valued representation
of the six-point correlator. The idea is to start from the $C\cdot\cZ$ representation \pseudoSix\
and to exploit the dual
\eqn\dualofthis{
\cZ^{m \ldots}_{1,A,B, \ldots} = E^{m \ldots}_{1|A,B, \ldots} + \cdots
}
of \rMone: Each $\cZ$-function with leg one in a single-particle slot is taken as a leading
term of a GEI, see \SixElliptic, and the additional terms in the ellipsis of \dualofthis\ are of the form
$\cZ^{m\ldots}_{1C,D,\ldots}$ with $C\neq \emptyset$. In this way, a long sequence of BRST cohomology 
identities given in section~\CJACOBIsec\ leads to the following 
manifestly BRST-invariant and single-valued Lie-polynomial form of \pseudoSix,
\eqnn\pseudoEllipticSix
$$\eqalignno{
\cK_6(\ell) &=
\half C^{mn}_{1|A_1, \ldots,A_5} E^{(s),mn}_{1|A_1, \ldots,A_5}
+\big[23456|A_1, \ldots,A_5\big]
&\pseudoEllipticSix\cr
&\quad{}+ C^m_{1|A_1, \ldots,A_4} E^{(s)m}_{1|A_1, \ldots,A_4}
+ \big[23456|A_1, \ldots,A_4\big]\cr
&\quad{}+ C_{1|A_1, \ldots,A_3} E^{(s)}_{1|A_1, \ldots,A_3}
+ \big[23456|A_1, \ldots,A_3\big]\,\cr
&\quad{} - \big[P_{1|A_1|A_2, \ldots,A_5} E^{(s)}_{1|A_1|A_2, \ldots,A_5}
+ (A_1\leftrightarrow A_2, \ldots,A_5)\big]
+\big[23456|A_1, \ldots,A_5\big]\, .
}$$
The GEIs have been expressed in terms of the Lie symmetric $E^{(s)m \ldots}_{1|A, \ldots}$ defined in \KLTE,
and similar to \pseudoLieSix, only five legs participate in the Stirling permutations.
More explicitly, expanding the above sums over Stirling cycle permutations yields
\eqnn\lowenE
$$\eqalignno{
\cK_{6}(\ell) &={1\over 2}  C^{mn}_{1|2,3,4,5,6}E^{mn}_{1|2,3,4,5,6}
- \big[  P_{1|2|3,4,5,6}E_{1|2|3,4,5,6} + (2\leftrightarrow 3,4,5,6) \big]\cr
&{} +\big[ C^m_{1|23,4,5,6}\,s_{23} E^m_{1|23,4,5,6} +(2,3|2,3,4,5,6) \big]  \cr
&{} + [\big(C_{1|23,45,6}\,s_{23} s_{45} E_{1|23,45,6} + {\rm
cyc}(3,4,5)\big)+(6\leftrightarrow5,4,3,2)] &\lowenE\cr
&{} + \big[\big(C_{1|234,5,6}\,s_{23} s_{34} E_{1|234,5,6} + {\rm cyc}(2,3,4)\big)
 + (2,3,4|2,3,4,5,6)\big]\,,
}$$
and reproduces the double-copy expression for the six-point correlator proposed in \MafraIOJ.
The refined GEI $E_{1|2|3,4,5,6}$ arises from its expansion \covPSb\ in terms
of $\cZ$-functions and boils down to the $g^{(n)}_{ij}$ in \SixErefined. By the vanishing
of $\cZ_{2|1,3,4,5,6}$, the $C\cdot {\cal Z}$-representation \pseudoSix\ of the six-point correlator 
does not feature any analogue of the terms $P_{1|2|3,4,5,6}E_{1|2|3,4,5,6}$ in the first line of \lowenE.

Furthermore, from the trace relation \traceA\ among GEIs,
\eqnn\traceAagain
$$\eqalignno{
{1\over 2} \delta_{mn} E^{mn}_{1|2,3,4,5,6}
&=  \big[ E_{1|2|3,4,5,6} + (2\leftrightarrow 3,4,5,6) \big]
+ 2\pi i{\partial \over \partial \tau} \log{\cal I}_6(\ell)\,, &\traceAagain
}$$
one concludes that the BRST variation of \lowenE\ is a boundary term \MafraIOJ
\eqnn\lowenF
$$\eqalignno{
Q\cK_{6}(\ell) &= - V_1 Y_{2,3,4,5,6}  \Big( {1\over 2} E^{mm}_{1|2,3,4,5,6} - \big[ E_{1|2|3,4,5,6} + (2\leftrightarrow 3,4,5,6)\big] 
\Big) &\lowenF \cr
&= - 2\pi i \, V_1 Y_{2,3,4,5,6}  {\p\over\p\tau}\log{\cal I}_6(\ell) \cong 0\,,
}$$
as required by the anomaly cancellation condition.

%************************************************************************************
\newsubsubsubsec\sixTEsec The $T\cdot E$ representation: manifesting locality \& single-valuedness

The $C\cdot E$ representation \pseudoEllipticSix\ is not manifestly local, but
it is written in terms of GEIs manifesting
monodromy invariance. However, by construction, we know that
\pseudoEllipticSix\ is equivalent to the local representation \sixLie, so all
the non-localities within the pseudo-invariants $C$ and $P$ must be spurious.
In the following discussions we exploit this reasoning to find a new
representation that is both manifestly local and monodromy invariant.

We can do this starting from \lowenE, plugging in the
Berends--Giele expansion of the pseudo-invariants and separating terms
according to their kinematic poles. The non-local terms turn out to vanish (as will
be exemplified below) while the local terms conspire to produce the full
correlator $\cK_6(\ell)$. After going through the algebra we obtain the
following manifestly local and monodromy-invariant
form of the six-point correlator $\cK_6(\ell)$,
\eqnn\augF
$$\eqalignno{
\cK_6(\ell) &= {1\over 2} V_1 T^{mn}_{2,3,4,5,6}E^{mn}_{1|2,3,4,5,6}
- \big[V_1 J_{2|3,4,5,6} E_{1|2|3,4,5,6}
+(2\leftrightarrow 3,4,5,6) \big]\,,&\augF\cr
&+ \big[ V_1 T^m_{23,4,5,6} E^m_{1|23,4,5,6}  +(2,3|2,3,4,5,6) \big] \cr
&+\big[ V_1T_{234,5,6} E_{1|234,5,6} + V_1T_{243,5,6} E_{1|243,5,6} +(2,3,4|2,3,4,5,6) \big]\cr
&+ \big[ \big( V_1T_{2,34,56} E_{1|2,34,56} +{\rm cyc}(3,4,5) \big) +(2\leftrightarrow 3,4,5,6) \big]\,.
}$$
The non-local terms from the kinematic side turn out to vanish due to identities
obeyed by their accompanying worldsheet functions. For instance, one such class
of terms (featuring an uncancelled $s_{12}$ pole) is given by,
\eqn\zeroJ{
	  M_{12} T_{34,5,6}\big(
          k_2^m E^m_{1|2,34,5,6}
          +s_{23} E_{1|234,5,6}
          - s_{24}E_{1|243,5,6}
          +s_{25} E_{1|25,34,6}
          +s_{26} E_{1|26,34,5}
          \big)\cong0\,,
	  }
whose vanishing follows from one of the GEI relations \covW. Similarly,
one can check that all the other classes of non-local terms vanish as well.
In summary, the expressions \sixloc, \pseudoSix, \lowenE\ and \augF\ for
the six-point correlator generalize the four representations of the five-point
correlator in \allfives.

Note that the $T\cdot E$ representation \augF\ is related to the $C\cdot {\cal Z}$ representation \pseudoSix\ 
through the duality between kinematics and worldsheet functions: In order to see this, one needs to adjoin the
vanishing terms $-\big[{\cal Z}_{2|1,3,4,5,6} P_{1|2|3,4,5,6}+(2\leftrightarrow 3,4,5,6)\big]$ to the latter.

%*****************************************************
\newsubsubsec\SixCompsec Comparison with older results

To conclude the discussion of the six-point correlator, we make contact between the
above representations and a manifestly BRST invariant expression for the six-point amplitude 
that has been presented in \MafraNWR. Starting from the $C\cdot {\cal Z}$-representation \pseudoSix,
expanding the worldsheet functions and collecting terms yields,
\eqnn\tediousSix
$$\eqalignno{
{\cal K}_{6}(\ell) &= {1\over2} \ell_m \ell_n C^{mn}_{1|2,3,4,5,6}
+ \ell_m  \big[ s_{23}g^{(1)}_{23} C^m_{1|23,4,5,6} + (2,3|2,\ldots,6)
\big] &\tediousSix\cr
&+ \big[\big(s_{23}s_{34}g^{(1)}_{23}  g^{(1)}_{34} C_{1|234,5,6}
%+ s_{24}s_{34}g^{(1)}_{24} g^{(1)}_{43} C_{1|243,5,6}
%+ s_{23}s_{24}g^{(1)}_{32} g^{(1)}_{24} C_{1|324,5,6}
+ {\rm cyc}(2,3,4)\big)
+ (2,3,4 | 2,3,4,5,6) \big]
\cr
& + \big[ \big(s_{23}s_{45}g^{(1)}_{23} g^{(1)}_{45} C_{1|23,45,6}
%+ s_{24}s_{35}g^{(1)}_{24} g^{(1)}_{35} C_{1|24,35,6}
%+ s_{25}s_{34}g^{(1)}_{25} g^{(1)}_{34} C_{1|25,34,6}
+ {\rm cyc}(3,4,5)\big)
+ (6 \leftrightarrow 5,4,3,2) \big]  \cr
&+ \big[ g^{(2)}_{12} C_{1|2|3,4,5,6} + (2\leftrightarrow 3,4,5,6)
\big]+ \big[ g^{(2)}_{23} C_{1|(23)|4,5,6} + (2,3|2,3,4,5,6) \big]\,,
}$$
where we defined the following shorthands for the coefficients
of $g_{12}^{(2)}$ and $g_{23}^{(2)}$,
\eqnn\Ponetwo
$$\eqalignno{
C_{1|2|3,4,5,6} &\equiv k_1^m k_2^n C^{mn}_{1|2,3,4,5,6} +  \big[
s_{23}k^m_1 C^m_{1|23,4,5,6} + (3\leftrightarrow4,5,6)\big]\,, &\Ponetwo\cr
C_{1|(23)|4,5,6} &\equiv k_2^m k_3^n C^{mn}_{1|2,3,4,5,6}
+ s_{23} (k^m_3 - k^m_2)C^m_{1|23,4,5,6}\cr
&+ \big[s_{24} k^m_3 C^m_{1|24,3,5,6} + s_{34} k^m_2 C^m_{1|34,2,5,6}
+ (4\leftrightarrow 5,6)\big]\cr
%&+\Big[ s_{34}(s_{23}C_{1|234,5,6}
%{-}s_{24}C_{1|243,5,6}) %&\Ptwothree\cr
%%&\qquad{}
%+ s_{24}(s_{35}C_{1|24,35,6} {+} s_{36}C_{1|24,36,5}) +
%(4\leftrightarrow 5,6)\Big]\,.
%}$$
&+\big[ s_{34}s_{23}C_{1|234,5,6}+s_{23}s_{24} C_{1|324,5,6}
-s_{24}s_{34}C_{1|243,5,6} %&\Ptwothree\cr
%&\qquad{} +
+(4\leftrightarrow 5,6)\big]\cr
%%%
&+\big[  s_{24}s_{35}C_{1|24,35,6} + s_{25}s_{34}C_{1|25,34,6} +
(4,5|4,5,6)\Big]\,.
}$$
These combinations are easily seen to satisfy
\eqn\brstP{
QC_{1|2|3,4,5,6} = - s_{12}V_1Y_{2,3,4,5,6}\,,\qquad
QC_{1|(23)|4,5,6} = - s_{23}V_1Y_{2,3,4,5,6}\,,
}
so the BRST variation of \tediousSix\ reproduces the desired
Koba--Nielsen derivative in $\tau$. Using BRST cohomology identities one can show that
\eqnn\equivOld
$$\eqalignno{
C_{1|2|3,4,5,6}&\cong s_{12}P_{1|2|3,4,5,6}\,,&\equivOld\cr
C_{1|(23)|4,5,6}&\cong
 \half s_{23}\Bigl(P_{1|2|3,4,5,6} + P_{1|3|2,4,5,6} +
 (k_3^m - k_2^m)C^{m}_{1|23,4,5,6} \cr
& +
 \big[ s_{34} C_{1|234,5,6} + s_{24} C_{1|324,5,6}+(4\leftrightarrow
 5,6)\big]\Bigr)\,,
}$$
which will imply, after integration over the loop momentum in
section~\loopintsec, that \tediousSix\
gives rise to an equivalent version of the six-point pure-spinor
correlator expression of \MafraNWR\foot{To see the equivalence we note, in particular, equation (3.15) of
\MafraNWR.}. 

The bosonic six-point one-loop amplitude of the open superstring was computed in the RNS formalism,
see \refs{\TsuchiyaVA, \stiefive} for the parity even part and \Clavelli\ for the parity odd part.

%************************************
\newsubsec\SevenPointsec Seven points

Following the general structure of the one-loop correlator presented
in \experQ, the local seven-point correlator is proposed to be
\eqn\sevenLiecorr{
\cK_7(\ell) = \cK_7^{\rm Lie}(\ell) + \cK_7^{Y}(\ell)\,,
}
where $\cK_n^{\rm Lie}(\ell)$ is defined in \experK\ and the anomaly
sector $\cK_7^Y(\ell)$
will be determined below. The unrefined contribution to
$\cK_n^{\rm Lie}(\ell)=\cK_7^{(0)}(\ell) -\cK_7^{(1)}(\ell) $
follows the pattern of \experJ,
\eqnn\sevenLie
$$\eqalignno{
\cK_7^{(0)}(\ell) &=
{1\over 3!}V_{A_1} T^{mnp}_{A_2, \ldots,A_7}
\cZ^{mnp}_{A_1, \ldots,A_7} + \big[1234567|A_1, \ldots,A_7\big] &\sevenLie\cr
&{}+{1\over 2!}V_{A_1} T^{mn}_{A_2, \ldots,A_6}
\cZ^{mn}_{A_1, \ldots,A_6}+ \big[1234567|A_1, \ldots,A_6\big]\cr
&{} +
V_{A_1} T^m_{A_2, \ldots,A_5}
\cZ^m_{A_1, \ldots,A_5}+ \big[1234567|A_1, \ldots,A_5\big]\cr
&{}+
V_{A_1} T_{A_2, \ldots,A_4} \cZ_{A_1, \ldots,A_4} + \big[1234567|A_1,
\ldots,A_4\big]\,,
}$$
with a total number of terms given by
${7\stirling7}+{7\stirling6}+{7\stirling5}+{7\stirling4}=1+21+175+735=932$ (see
its explicit expansion in \sevenloc). The worldsheet functions entering \sevenLie\
and the subsequent equations are determined from
their monodromy variations. The solutions for the three topologies
of scalar ${\cal Z}$-functions, the two topologies of vectorial ones
and the tensorial ones can be found in \newsevengs, \SevVecOne,
\SevTopTwo\ and \seVTens, respectively.

The above $\cK_7^{(0)}(\ell)$ alone is not BRST invariant,
and this fact motivates the introduction of refined contributions $\cK_7^{(1)}(\ell)$ to \experK.
In fact, the general discussion of refined correlators $\cK_n^{(d)}(\ell)$ in section~\LiePolsec\
originated from the explicit findings of this example. The seven-point expression
\eqnn\sevRef
$$\eqalignno{
\cK_7^{(1)}(\ell) &=
 \big[V_{A_1} J^m_{A_2|A_3, \ldots,A_7}
\cZ^m_{A_2|A_1,A_3, \ldots,A_7}
+ (A_2\leftrightarrow A_3,\ldots,A_7)\big]
+ \big[1234567|A_1, \ldots,A_7\big]\cr
&\ \ \  + \big[V_{A_1} J_{A_2|A_3, \ldots,A_6} \cZ_{A_2|A_1,A_3, \ldots,A_6}
+ (A_2\leftrightarrow A_3,\ldots,A_6)\big]
+ \big[1234567|A_1, \ldots,A_6\big]\cr
%%%%
&=
V_1 J^m_{2|3,4,5,6,7} \cZ^m_{2|1,3,4,5,6,7}  + (2\leftrightarrow 3,4,5,6,7) &\sevRef \cr
&\ \ \ + \big[ V_{12}J_{3|4,5,6,7}\cZ_{3|12,4,5,6,7}  + V_{13}J_{2|4,5,6,7}\cZ_{2|13,4,5,6,7} +(2,3|2,3,\ldots,7) \big]
\cr
&\ \ \ + \big[ V_{1}J_{23|4,5,6,7}\cZ_{23|1,4,5,6,7}  +(2,3|2,3,\ldots,7) \big]
\cr
&\ \ \  + \big[( V_{1}J_{2|34,5,6,7}\cZ_{2|1,34,5,6,7}+{\rm cyc}(2,3,4))  +(2,3,4|2,3,\ldots,7) \big]
}$$
with $5{7\stirling6}+6{7\stirling7}=105+6=111$ terms in total lines up with
the general proposal \experO\ at refinement $d=1$. We have seen in \sevNotorious\ that
the three topologies of refined functions appearing in \sevRef\ are simple combinations of
\eqnn\sevNotoriousAgain
$$\eqalignno{
\cZ_{12|3,4,5,6,7}&= \p \gg2(1,2) + s_{12}\gg1(1,2)\gg2(1,2) - 3s_{12}\gg3(1,2) &\sevNotoriousAgain \cr
&\cong - 3s_{12}\gg3(1,2) + g^{(2)}_{12}( \ell \cdot k_2 + s_{23} g^{(1)}_{23}+ s_{24} g^{(1)}_{24}+\ldots
+ s_{27} g^{(1)}_{27}) \,.
}$$
However, the sum of the $d=0$ and $d=1$ correlators in \sevenLie\ and \sevRef\ is
still not enough to yield a BRST-invariant seven-point correlator, see the discussion
of $Q {\cal K}_n^{\rm Lie}(\ell)$ in section \QGeneralsec. This necessitates the
additional purely anomalous contribution to \sevenLiecorr\ given by
\eqn\cKDelta{
\cK_7^{Y}(\ell) = -\Delta_{1|2|3,4,5,6,7}\cZ_{12|3,4,5,6,7} +
(2\leftrightarrow3,4,5,6,7)\,,
}
where the anomalous superfield $\Delta_{1|2|3,\ldots,7}$ is defined in \deltasecC.
By the arguments of section \localdel, the components $\langle \Delta_{1|2|3,\ldots,7} \rangle$
cannot have any kinematic pole, so addition of \cKDelta\ does not spoil the locality
of the seven-point correlator \sevenLiecorr.

Note that the representation of $\cZ_{12|3,4,5,6,7}$ in the second line of \sevNotoriousAgain\
manifests that $ \cK_7(\ell)$ can be written without any derivatives $\partial g^{(m)}_{ij}$
or products $g^{(1)}_{ij}g^{(2)}_{ij}$ with coinciding arguments.  This observation should play
an important role for the transcendentality properties upon integration over $z_j$.

%*************************
\subsubsec BRST invariance

In order to show that the full correlator \sevenLie\ is BRST invariant, let us
first consider its non-anomalous part, $Q\cK_7^{\rm Lie}(\ell)$. This
computation can be organized according to the ghost-number four products of
superfields it generates; this general structure was anticipated in
section~\QGeneralsec\ but it is instructive to see it again in this particular
case:
\eqnn\QsevLie
$$\eqalignno{
Q\cK_7^{\rm Lie}&= -\half T^{(0,2)}_{1|2,3,4,5,6,7} - T^{(1,0)}_{1|2,3,4,5,6,7}&\QsevLie\cr
&- \big[T^{(0,1)}_{12|3,4,5,6,7} +(2\leftrightarrow3,4,5,6,7)\big]
- \big[T^{(0,1)}_{1|23,4,5,6,7} +(2,3|2,3,4,5,6,7)\big]\cr
&- \big[T^{(0,0)}_{123|4,5,6,7} + T^{(0,0)}_{132|4,5,6,7} +(2,3|2,3,4,5,6,7)\big]\cr
&- \big[T^{(0,0)}_{1|234,5,6,7} + T^{(0,0)}_{1|243,5,6,7} +(2,3,4|2,3,4,5,6,7)\big]\cr
&- \big[T^{(0,0)}_{12|34,5,6,7} + T^{(0,0)}_{13|24,5,6,7} + T^{(0,0)}_{14|23,5,6,7} +(2,3,4|2,3,4,5,6,7)\big]\cr
&- \big[ T^{(0,0)}_{1|23,45,6,7}  + T^{(0,0)}_{1|24,53,6,7}  + T^{(0,0)}_{1|25,34,6,7}  + (6,7|2,3,4,5,6,7) \big] \cr
&+Y^{(0,1)}_{1|2,3,4,5,6,7}+ \big[Y^{(0,0)}_{12|3,4,5,6,7} +(2\leftrightarrow3,4,5,6,7)\big] \cr
& + \big[Y^{(0,0)}_{1|23,4,5,6,7} +(2,3|2,3,4,5,6,7)\big]\, ,
}$$
where, following \QLiedZero, the non-anomalous building blocks are contained in
\eqnn\allsevTs
$$\eqalignno{
T^{(0,2)}_{1|2,3,4,5,6,7}&=
V_{1}V_{2}T^{m n}_{3,\ldots,7}\mkern1mu
\Theta^{(0)\,m n}_{2|1,3,\ldots,7}
+ (2\leftrightarrow3,4,5,6,7)\,, &\allsevTs\cr
T^{(0,1)}_{12|3,4,5,6,7}&=
V_{12}V_{3}T^{m}_{4,\ldots,7}\mkern1mu
\Theta^{(0)\,m}_{3|12,4,\ldots,7}
+ (3\leftrightarrow4,5,6,7)\,,\cr
T^{(0,1)}_{1|23,4,5,6,7}&=
V_{1}V_{23}T^{m}_{4,\ldots,7}\mkern1mu
\Theta^{(0)\,m}_{23|1,4,\ldots,7}
+ (23\leftrightarrow4,5,6,7)\,,\cr
T^{(0,0)}_{123|4,5,6,7}&=
V_{123}V_{4}T_{5,\ldots,7}\mkern1mu
\Theta^{(0)}_{4|123,5,\ldots,7}
+ (4\leftrightarrow5,6,7)\,,\cr
T^{(0,0)}_{12|34,5,6,7}&=
V_{12}V_{34}T_{5,\ldots,7}\mkern1mu
\Theta^{(0)}_{34|12,5,\ldots,7}
+ (34\leftrightarrow5,6,7)\,,\cr
T^{(0,0)}_{1|234,5,6,7}&=
V_{1}V_{234}T_{5,\ldots,7}\mkern1mu
\Theta^{(0)}_{234|1,5,\ldots,7}
+ (234\leftrightarrow5,6,7)\,,\cr
T^{(0,0)}_{1|23,45,6,7}  &= V_1 V_{23} T_{45,6,7}\mkern1mu
\Theta^{(0)}_{23|1,45,6,7}+(23\leftrightarrow 45,6,7)\,,\cr
T^{(1,0)}_{1|2,3,4,5,6,7}&=
V_{1}V_{2}\big[J_{3|4,\ldots,7}\mkern1mu
\Theta^{(1)}_{2|3|1,4,\ldots,7}
+ (3\leftrightarrow4,5,6,7)\big]
+ (2\leftrightarrow3,4,5,6,7)\,,
}$$
while the anomalous building blocks are contained in $Y^{(d,r)}$ given by \Yddefs,
\eqnn\allsevYs
$$\eqalignno{
Y^{(0,1)}_{1|2,3,4,5,6,7}&=
V_{1}Y^m_{2,3,4,5,6,7}\,\Xi^{(0) \, m}_{1|2,3,\ldots,7}
 \, &\allsevYs \cr
Y^{(0,0)}_{12|3,4,5,6,7}&=
V_{12}Y_{3,4,5,6,7}\,\Xi^{(0)}_{12|3,\ldots,7}\,,\cr
Y^{(0,0)}_{1|23,4,5,6,7}&=
V_{1}Y_{23,4,5,6,7}\,\Xi^{(0)}_{1|23,4,\ldots,7}\,.
}$$
It is evident from the above permutations that the general compact
expression \QcKgeneral\ leads to involved combinatorics resulting in many terms
present in the seven-point BRST variation \QsevLie, even when
written using the shorthands $\Theta^{(d)}$ and $\Xi^{(d)}$ defined
in \ThetaSix\ and \XiSix.
Fortunately,
the analysis of the outcome is also greatly simplified by this very same
organization, as it suffices to check only a handful of different {\it topologies}
of $\Theta^{(d)}$ and $\Xi^{(d)}$ rather than all their permutations. In fact,
it is straightforward to check that all $T^{(d,r)}$ terms above vanish due to
\eqnn\BRSTKsevA
$$\eqalignno{
V_1V_2T^{mn}_{3,4,5,6,7}\Theta^{(0)\,mn}_{2|1,3,4,5,6,7}&\cong 0,\qquad
V_1V_2T^{m}_{34,5,6,7}\Theta^{(0)\,m}_{2|1,34,5,6,7}\cong 0,&\BRSTKsevA\cr
V_1V_{23}T^{m}_{4,5,6,7}\Theta^{(0)\,m}_{23|1,4,5,6,7} &\cong 0,\qquad
V_1V_2T_{34,56,7}\Theta^{(0)}_{2|1,34,56,7} \cong0,\cr
V_1V_2T_{345,6,7}\Theta^{(0)}_{2|1,345,6,7} &\cong0,\qquad
V_1 V_{23}T_{45,6,7}\Theta^{(0)}_{23|1,45,6,7} \cong0,\cr
V_1V_{234}T_{5,6,7}\Theta^{(0)}_{234|1,5,6,7} &\cong0,\qquad
V_1V_2J_{3|4,5,6,7}\Theta^{(1)}_{2|3|1,4,5,6,7}\cong0\,,
%V_{12}V_{3}T_{45,6,7}\Theta^{(0)}_{3|12,45,6,7} &\cong0\,,\qquad
%V_{12}V_{34}T_{5,6,7}\Theta^{(0)}_{34|12,5,6,7} \cong0\cr
%V_{123}V_{4}T_{5,6,7}\Theta^{(0)}_{4|123,5,6,7} &\cong0\,,\qquad
%V_{12}V_3T^{m}_{4,5,6,7}\Theta^{(0)\,m}_{3|12,4,5,6,7}&\cong 0,\qquad
}$$
whose explicit expansions in terms of shuffle-symmetric functions $\cZ$ can be
found in the appendix~\allThetasapp. The coefficients of $V_{1A}$ with $A\neq \emptyset$
are just relabellings of the $\Theta^{(d)}$ in \BRSTKsevA\ and therefore vanish as well.

Using the results above
the BRST variation of \sevenLie\ is purely anomalous
\eqnn\QKsevAn
$$\eqalignno{
Q\cK_7^{\rm Lie}(\ell)
&= V_1Y^m_{2,3,4,5,6,7}\,\Xi^{(0)\,m}_{1|2,3,4,5,6,7}&\QKsevAn\cr
&\quad{}+ V_{12}Y_{3,4,5,6,7}\,\Xi^{(0)}_{12|3,4,5,6,7} + (2\leftrightarrow3,4,5,6,7)\cr
&\quad{}+V_1Y_{23,4,5,6,7}\,\Xi^{(0)}_{1|23,4,5,6,7} + (2,3|2,3,4,5,6,7)\,,
}$$
and is written entirely using the linear combinations $\Xi^{(0)}$ of \XiSix,
\eqnn\dualtraceseven
$$\eqalignno{
\Xi^{(0)\,m}_{1|2,3,4,5,6,7} &=
-\half \cZ^{mpp}_{1,2,3,4,5,6,7} + \big[\cZ^m_{2|1,3,4,5,6,7} +
\,(2\leftrightarrow3, \ldots,7)\big]\cong - \cZ^m_{1|2,3,4,5,6,7}\qquad{}&\dualtraceseven\cr
\Xi^{(0)}_{12|3,4,5,6,7} &=
-\half \cZ^{pp}_{12,3,4,5,6,7} + \big[\cZ_{3|12,4,5,6,7} +
\,(3\;\leftrightarrow4,5,6,7)\big]\cong - \cZ_{12|3,4,5,6,7}\cr
\Xi^{(0)}_{1|23,4,5,6,7} &=
-\half \cZ^{pp}_{1,23,4,5,6,7}
+ \big[\cZ_{23|1,4,5,6,7}+ (23\leftrightarrow4,5,6,7)\big]\cong
-\cZ_{1|23,4,5,6,7}\,.
}$$
Similar to \QsixLie, the $\cong$ symbol indicates that boundary terms w.r.t.\ $\tau$
have been discarded in the second step of each line.
The rearrangements of the above sums have the same structure as
the trace relations among non-refined and refined building blocks, see
section~\tracesT. As discussed in section \tracesZsec, the worldsheet functions found
via the bootstrap method of section~\sevenwssec\ satisfy the {\it dual} 
trace relations exploited in \dualtraceseven, and \QKsevAn\ becomes,
\eqnn\QKseven
$$\eqalignno{
Q\cK_7^{\rm Lie}(\ell) &= - V_1Y^m_{2,3,4,5,6,7}\,\cZ^m_{1|2,3,4,5,6,7} &\QKseven\cr
&\quad{} - V_{12}Y_{3,4,5,6,7}\,\cZ_{12|3,4,5,6,7} + (2\leftrightarrow3,4,5,6,7)\cr
&\quad{}- V_{1}Y_{23,4,5,6,7}\,\cZ_{1|23,4,5,6,7} + (2,3|2,3,4,5,6,7)\,.
}$$
By the relations \sevNotorious\ between the three topologies of refined ${\cal Z}$-functions,
the BRST variation \QKseven\ can then be written as,
\eqnn\Qsevnot
$$\eqalignno{
Q\cK_7^{\rm Lie}(\ell) &=
\Big(k_2^m V_1 Y_{2,3,4,5,6,7}^m
	  + V_{21}Y_{3,4,5,6,7}
	  + \big[V_1 Y_{23,4,5,6,7} + (3\leftrightarrow4,5,6,7)\big]\Big)
	  \cZ_{12|3,4,5,6,7}\cr
	  &\qquad{} + (2\leftrightarrow3,4,5,6,7) \, .&\Qsevnot
}$$
From \manlocOT\ we recognize the terms inside the parenthesis in \Qsevnot\ as the
BRST variation of $\Delta_{1|2|3,4,5,6,7}$, that is,
the expression for $\cK_7^Y(\ell)$ in \cKDelta\ is tailored to cancel
\eqn\tailorcancel{
Q\cK_7^{\rm Lie}(\ell)=
Q \Delta_{1|2|3,4,5,6,7}\cZ_{12|3,4,5,6,7}+ (2\leftrightarrow3,4,5,6,7)
=-Q\cK_7^Y(\ell)\,.
}
Therefore the full correlator \sevenLie\ is BRST invariant up to
total derivatives,
\eqn\KsevInv{
Q\big(\cK_7^{\rm Lie}(\ell) + \cK_7^Y(\ell)\big) = Q\cK_7(\ell)\cong0\,.
}
Before showing that \sevenLie\ is also monodromy invariant, it
will be convenient to rewrite it using the pseudo-invariants of
section~\BRSTpseudosec, as that will simplify the proof considerably.

%***************************************************************************************
\newsubsubsubsec\sevenCZsec The $C\cdot \cZ$ representation: manifesting BRST invariance

Given that the correlator \sevenLie\ is BRST invariant, it is rewarding to
rewrite it in terms of BRST pseudo-invariants. This can be done following the
same procedure applied in detail for the six-point correlator in
subsection~\sixCZsec, so it will only be sketched here again; we rewrite
$M_1M^{m \ldots}_{A,B, \ldots} = C^{m \ldots}_{1|A,B, \ldots} + \cdots $ and
$M_1\cJ^{m \ldots}_{A,B, \ldots} = P^{m \ldots}_{1|A,B, \ldots} + \cdots $ and
collect the terms containing a factor of $M_{1P}$ with $P\neq\emptyset$. A
long but straightforward analysis using integration-by-parts relations \genTheta\ for the
$\cZ$-functions shows that all terms proportional to $M_{1P}$ vanish and we
arrive at
\eqnn\pseudoSeven
$$\eqalignno{
\cK_7(\ell) &=
{1\over 6} C^{mnp}_{1|A_1, \ldots,A_6} Z^{(s),mnp}_{1,A_1, \ldots,A_6}
+\big[234567|A_1, \ldots,A_6\big] &\pseudoSeven\cr
&+\hskip1pt \half C^{mn}_{1|A_1, \ldots,A_5} Z^{(s),mn}_{1,A_1, \ldots,A_5}
+\big[234567|A_1, \ldots,A_5\big] \cr
&{}+\hskip4pt C^m_{1|A_1, \ldots,A_4}\hskip1.5pt Z^{(s)m}_{1,A_1, \ldots,A_4}
\hskip2.6pt+ \big[234567|A_1, \ldots,A_4\big]\cr
&{}+\hskip4pt C_{1|A_1, \ldots,A_3}\hskip1.5pt Z^{(s)}_{1,A_1, \ldots,A_3}
\hskip2.6pt+ \big[234567|A_1, \ldots,A_3\big]\cr
&-\big[P^m_{1|A_1|A_2, \ldots,A_6} Z^{(s)\,m}_{A_1|1,A_2, \ldots,A_6}+
(A_1\leftrightarrow A_2, \ldots,A_6)\big]
+\big[234567|A_1, \ldots,A_6\big]\cr
&-\big[P_{1|A_1|A_2, \ldots,A_5} Z^{(s)}_{A_1|1,A_2, \ldots,A_5}+
(A_1\leftrightarrow A_2, \ldots,A_5)\big]
+\big[234567|A_1, \ldots,A_5\big]\cr
&-\Delta_{1|2|3,4,5,6,7}\cZ_{12|3,4,5,6,7} + (2\leftrightarrow3,4,5,6,7)\,.
}$$
Note that only six legs participate in the Stirling permutations, and $Z^{(s)}_{\ldots}$
are defined in \cZToZ. To compute the BRST variation of \pseudoSeven\ it will be 
convenient to recall that \partI
\eqnn\QDeltaGamma
$$\eqalignno{
QP^m_{1|2|3,4,5,6,7} &=-\Gamma^m_{1|2,3,4,5,6,7}\,, \hskip20pt QC^{mnp}_{1|2,3,4,5,6,7} =-\d^{(mn}\Gamma^{p)}_{1|2,3,4,5,6,7} \cr
QP_{1|23|4,5,6,7} &=-\Gamma_{1|23,4,5,6,7}\,, \hskip25pt QC^{mn}_{1|23,4,5,6,7} =-\d^{mn}\Gamma_{1|23,4,5,6,7}\cr
QP_{1|2|34,5,6,7} &=-\Gamma_{1|2,34,5,6,7}\,, \hskip26pt QC^m_{1|A,B,C,D} =QC_{1|A,B,C}=0 &\QDeltaGamma\cr
Q\Delta_{1|2|3,4,5,6,7} &= k_2^m\Gamma^m_{1|2,3,4,5,6,7}
+ \big[ s_{23}\Gamma_{1|23,4,5,6,7} + (3\leftrightarrow 4,5,6,7)
\big]\,,
}$$
%\eqn\QDeltaGamma{
%\eqalign{
%QP^m_{1|2|3,4,5,6,7} &=-\Gamma^m_{1|2,3,4,5,6,7}\,,\cr
%QP_{1|23|4,5,6,7} &=-\Gamma_{1|23,4,5,6,7}\,,\cr
%QP_{1|2|34,5,6,7} &=-\Gamma_{1|2,34,5,6,7}\,,\cr
%}\qquad
%\eqalign{
%QC^{mnp}_{1|2,3,4,5,6,7} &=-\d^{(mn}\Gamma^{p)}_{1|2,3,4,5,6,7}\,,\cr
%QC^{mn}_{1|23,4,5,6,7} &=-\d^{mn}\Gamma_{1|23,4,5,6,7}\,,\cr
%QC^m_{1|A,B,C,D} &=QC_{1|A,B,C}=0\,,\cr
%}}
%\vskip-18pt
%$$\eqalignno{
%Q\Delta_{1|2|3,4,5,6,7} &= k_2^m\Gamma^m_{1|2,3,4,5,6,7}
%+ \big[ s_{23}\Gamma_{1|23,4,5,6,7} + (3\leftrightarrow 4,5,6,7)
%\big]\,,\hskip33.9pt{}
%}$$
see \GAMa\ for the anomaly invariants $\Gamma_{1|\ldots}$.
After straightforward algebra and using the trace relations \dualtraceseven\ we obtain,
\eqnn\QCZseven
$$\eqalignno{
Q\cK_7(\ell) &= \Gamma^m_{1|2,3,4,5,6,7}\big(
\cZ^m_{1|2,3,4,5,6,7} + \big[k_2^m\cZ_{12|3,4,5,6,7} +
(2\leftrightarrow3,4,5,6,7)\big]
\big)&\QCZseven\cr
&\ \ \ - s_{23}\Gamma_{1|23,4,5,6,7}(\cZ_{1|23,4,5,6,7} +\cZ_{12|3,4,5,6,7}
-\cZ_{13|2,4,5,6,7}) + (2,3|2,3,4,5,6,7) \cr
&\cong 0\,.
}$$
The linear combinations of worldsheet functions
in \QCZseven\ correspond to the BRST-exact anomalous kinematic
factors displayed in section~\DeltaZsec\ and, as we have seen in \constS,
they vanish up to total derivatives. Therefore, BRST invariance of the 
representation \pseudoSeven\ is indeed confirmed.

%******************************
\subsubsec Single-valuedness

We will take the manifestly BRST-invariant representation \pseudoSeven\ of the seven-point
correlator as a starting point to verify monodromy invariance.
Using the monodromy variations of the seven-point $\cZ$-functions discussed in
section~\sevenwssec\ and in the appendix~\bootstrapapp,
a long but straightforward calculation implies,
\eqn\Dseven{
D\cK_7(\ell) = \Omega_1\d\cK_7^{(1)} + \cdots + \Omega_7\d\cK_7^{(7)}\, ,
}
where ($E^{(s)m \ldots}_{1|A, \ldots}$ is defined in \KLTE)
\eqnn\dKsevenone
$$\eqalignno{
\d\cK_7^{(1)} &= \half k_1^m C^{mnp}_{1|A_2, \ldots,A_7}E^{(s)np}_{1|A_2,
\ldots,A_7} +\big[234567|A_2, \ldots,A_7\big] &\dKsevenone\cr
&\hskip8.35pt{}+k_1^m C^{mn}_{1|A_2, \ldots,A_6}E^{(s)n}_{1|A_2,
\ldots,A_6} +\big[234567|A_2, \ldots,A_6\big]\cr
&\hskip8.35pt{}+k_1^m C^{m}_{1|A_2, \ldots,A_5}E^{(s)}_{1|A_2,
\ldots,A_5} +\big[234567|A_2, \ldots,A_5\big]\cr
&\quad{}-\big[k_1^m
P^m_{1|2|3,4,5,6,7}+\Delta_{1|2|3,4,5,6,7}\big]E^{(s)}_{1|2|3,4,5,6,7}
+ (2\leftrightarrow3,4,5,6,7)\,.
}$$
The other $\d\cK_7^{(j)}$ for $j=2, \ldots,7$ can be obtained
from $\d\cK_7^{(1)}$ by relabeling of $1\leftrightarrow j$
in both the kinematics and GEIs of \dKsevenone. To verify
this last statement one uses the change-of-basis identities
for pseudo-invariants derived in \partI. This is because the relabeling of
$\d\cK_7^{(j)}$ for $j=2, \ldots,7$ involves pseudo-invariants outside
of the {\it canonical} basis $C^{m \ldots}_{1| \ldots}$
(i.e.\ $C^{m \ldots}_{j| \ldots}$ with $j\neq 1$),
whereas the monodromy variation of \Dseven\
obviously contains only elements in the canonical basis.
See the analogous six-point analysis described in section~\Dsixproofsec\
for more details.

The appearance of momentum contractions in \dKsevenone\ signals the need to
use the BRST cohomology identities derived in \partI\ and reviewed in section \CJACOBIsec. In 
addition, one also needs their elliptic dual identities involving momentum contractions 
of $k_1^m E^{m \ldots}$ (cf.\ section \thissubsec) and the trace identity
\eqn\traceEmm{
{1\over 2}E^{mm}_{1|2|3,4,5,6,7} = \big[ E_{1|2|3,4,5,6,7} + (2\leftrightarrow3,4,5,6,7)\big]
+ 2\pi i{\p\over\p\tau}\log{\cal I}_7(\ell)\,.
}
After long but straightforward manipulations one finally concludes that the
monodromy variation \dKsevenone\ is BRST-exact and given by
\eqnn\dKsevenDeltas
$$\eqalignno{
\d\cK_7^{(1)} &= \half 
\Delta^{mn}_{1|A_2, \ldots,A_7}E^{(s)mn}_{1|A_2,
\ldots,A_6} + \big[234567|A_2, \ldots,A_7\big]&\dKsevenDeltas\cr
&\hskip8.35pt{}+\Delta^{m}_{1|A_2, \ldots,A_6}E^{(s)m}_{1|A_2,
\ldots,A_6} + \big[234567|A_2, \ldots,A_6\big]\cr
&\hskip8.35pt{}+\Delta_{1|A_2, \ldots,A_5}E^{(s)}_{1|A_2,
\ldots,A_5} + \big[234567|A_2, \ldots,A_5\big]\cong0\,.
}$$
It is crucial to note that only {\it unrefined} building blocks $\Delta_{1|\ldots}$
arise, whose BRST exactness is discussed in section~\Deltasec. Since
the other $\d\cK_7^{(j)}$ are relabellings of \dKsevenDeltas, it follows
that the complete monodromy variation $D\cK_7(\ell)$ in \Dseven\ is
BRST-exact and therefore vanishes in the cohomology; $D\cK_7(\ell)\cong0$.

%***************************************************************************************
\newsubsubsubsec\sevenCEsec The $C\cdot E$ representation: manifesting BRST invariance \& single-valuedness

Having derived the $C\cdot\cZ$ representation and shown that it is
single valued, we can re-express it to manifest
both BRST and monodromy invariance. We proceed similarly as in the six-point
case by inserting $\cZ^{m \ldots}_{1,A,B, \ldots} = E^{m \ldots}_{1|A,B,
\ldots} + \cdots$ into the $C\cdot\cZ$ representation \pseudoSeven\ and using a
long sequence of BRST cohomology identities described in \partI. Doing this
leads to a manifestly BRST-invariant and single-valued expression
neatly summarized by the following Stirling permutation sums
\eqnn\pseudoEllipticSeven
$$\eqalignno{
\cK_7(\ell) &=
\sum_{r=0}^3 C^{m_1 \ldots m_r}_{1|A_1, \ldots,A_{r+3}} E^{(s)\,m_1 \ldots m_r}_{1|A_1, \ldots,A_{r+3}}
+ \big[234567|A_1, \ldots,A_{r+3}\big]&\pseudoEllipticSeven\cr
%%%
&-\sum_{r=0}^1 \big[P^{m_1 \ldots m_r}_{1|A_1|A_2, \ldots,A_{r+5}} E^{(s)\,m_1 \ldots m_r}_{1|A_1|A_2 \ldots,A_{r+5}}
+ (A_1\leftrightarrow A_2, \ldots,A_{r+5})\big]
+ \big[2 \ldots7|A_1, \ldots,A_{r+5}\big]\,,\cr
}$$
where $E^{(s)m \ldots}_{1|A, \ldots}$ is defined in \KLTE, and the
terms proportional to $\Delta_{1|2|3,4,5,6,7}$ drop out by the trace relations
\dualtraceseven. Expanding the Stirling permutation sums in
\pseudoEllipticSeven\ yields
\eqnn\sevenCE
$$\eqalignno{
\cK_{7}(\ell) &= {1\over6}C^{mnp}_{1|2,3,4,5,6,7}E^{(s)mnp}_{1|2,3,4,5,6,7}&\sevenCE\cr
&+\half C^{mn}_{1|23,4,5,6,7} E^{(s)mn}_{1|23,4,5,6,7} + (2,3|2,3,4,5,6,7)\cr
&+\big[C^{m}_{1|234,5,6,7} E^{(s)m}_{1|234,5,6,7}
+ C^{m}_{1|243,5,6,7} E^{(s)m}_{1|243,5,6,7} \big]+ (2,3,4|2,3,4,5,6,7)\cr
%&+\big[C^{m}_{1|23,45,6,7} E^{(s)m}_{1|23,45,6,7} + (4,5|4,5,6,7)\big] + (2,3|2,3,4,5,6,7)\cr
&+ \big[C^{m}_{1|23,45,6,7} E^{(s)m}_{1|23,45,6,7} + {\rm cyc}(2,3,4)\big]+(6,7|2,3,4,5,6,7)\cr
&+\big[C_{1|2345,6,7} E^{(s)}_{1|2345,6,7} + {\rm perm}(3,4,5)\big]+(2,3,4,5|2,3,4,5,6,7)\cr
&+\big[C_{1|234,56,7} E^{(s)}_{1|234,56,7} + C_{1|243,56,7} E^{(s)}_{1|243,56,7} + {\rm cyc}(5,6,7)\big]
+ (2,3,4|2,3,4,5,6,7)\cr
&+\big[ C_{1|23,45,67} E^{(s)}_{1|23,45,67} + {\rm cyc}(4,5,6) \big]+ (3\leftrightarrow 4,5,6,7)\cr
&- P^m_{1|2|3, 4,5,6,7} E^{(s)m}_{1|2|3,4,5,6,7} + (2\leftrightarrow3,4,5,6,7)\cr
&- P_{1|23|4, 5,6,7} E^{(s)}_{1|23|4, 5,6,7}+(2,3|2,3,4,5,6,7)\cr
&- \big[P_{1|2|34,5,6,7} E^{(s)}_{1|2|34,5,6,7}+{\rm cyc}(2,3,4)\big]+(2,3,4|2,3,4,5,6,7)\,,
}$$
for a total number of $326$ terms with pseudo-invariants $C$ and $81$ terms with $P$.
This is the double-copy expression for the seven-point correlator implicitly proposed in \MafraIOJ.
Similar to \pseudoSeven, only six legs participate in the Stirling permutations, but there is no
analogue of the terms $\Delta_{1|2|3,\ldots,7}{\cal Z}_{12|3,\ldots,7}$ in the last line of the 
$C\cdot {\cal Z}$ representation.

%***************************************************************************************************
\newsubsubsubsec\sevenTEsec The $T\cdot E$ representation: manifesting locality \& single-valuedness

From the $C\cdot E$ representation \sevenCE\ one can derive a manifestly
local and single-valued representation following the same ideas as
explained for the six-point case in section~\sixTEsec. The end result is given by,
\eqnn\augH
$$\eqalignno{
\cK_7(\ell) &= {1\over 6} V_1T^{mnp}_{2,3,\ldots,7}E^{mnp}_{1|2,3,\ldots,7} &\augH \cr
&+ {1\over 2}V_1T^{mn}_{23,4,5,6,7} E^{mn}_{1|23,4,5,6,7} + (2,3|2,3,4,5,6,7)\cr
& +\big[ V_1T^m_{234,5,6,7} E^m_{1|234,5,6,7} +V_1T^m_{243,5,6,7} E^m_{1|243,5,6,7} \big]+
(2,3,4|2,3,4,5,6,7) \cr
%
%& +\big[V_1T^m_{23,45,6,7} E^m_{1|23,45,6,7} + (4,5|4,5,6,7)\big] + (2,3|2,3,4,5,6,7)\cr
&+ \big[V_1 T^{m}_{23,45,6,7} E^{m}_{1|23,45,6,7} + {\rm cyc}(2,3,4)\big]+(6,7|2,3,4,5,6,7)\cr
& +\big[V_1T_{2345,6,7} E_{1|2345,6,7} + {\rm perm}(3,4,5)\big] +(2,3,4,5|2,3,4,5,6,7)\cr
& + \big[V_1T_{234,56,7} E_{1|234,56,7} + V_1T_{243,56,7} E_{1|243,56,7}
+ {\rm cyc}(5,6,7)\big] + (2,3,4|2,3,4,5,6,7)\cr
&+ \big[ V_1T_{23,45,67} E_{1|23,45,67} +{\rm cyc}(4,5,6) \big]+ (3\leftrightarrow 4,5,6,7)\cr
& -  V_1J^m_{2|3,4,5,6,7} E^m_{1|2|3,4,5,6,7} +(2\leftrightarrow3,4,5,6,7) \cr
&- V_1J_{23|4,5,6,7} E_{1|23|4,5,6,7}  + (2,3|2,3,4,5,6,7) \cr
&- \big[ V_1J_{2|34,5,6,7} E_{1|2|34,5,6,7}+{\rm cyc}(2,3,4)\big]+(2,3,4|2,3,4,5,6,7)\,.
}$$
Similar to the six-point case \augF, this $T\cdot E$ representation is related to the $C \cdot {\cal Z}$
representation through the duality between kinematics and worldsheet functions, up to the fact that \augH\
does not exhibit any dual of the terms $\Delta_{1|2|3,\ldots,7}{\cal Z}_{12|3,\ldots,7}$ in \pseudoSeven. 
Moreover, the combinatorial structure of \augH\ is identical to that of the $C \cdot E$ representation \sevenCE. 
In addition, proving BRST invariance of the representation \augH\ requires the same elliptic 
worldsheet identities used to generate \augH\ from \sevenCE.

%************************************
\newsubsec\EightPointsec Eight points

Following the general structure of one-loop correlators presented
in \experK\ and \experQ, the manifestly local Lie-series part 
of the eight-point correlator is proposed to be
\eqn\eightLiecorr{
\cK^{\rm Lie}_8(\ell) \equiv \cK_8^{(0)}(\ell) - \cK_8^{(1)}(\ell) +
\cK_8^{(2)}(\ell)\,,
}
which will later receive a purely anomalous correction $\cK_8^Y(\ell)$.
The unrefined part with  $d=0$
follows the general pattern indicated in \experJ,
\eqnn\eightLie
$$\eqalignno{
\cK_8^{(0)}(\ell) &=
{1\over 4!}V_{A_1} T^{mnpq}_{A_2, \ldots,A_8}
\cZ^{mnpq}_{A_1, \ldots,A_8} + \big[12345678|A_1, \ldots,A_8\big] &\eightLie\cr
%%%
&{}+{1\over 3!}V_{A_1} T^{mnp}_{A_2, \ldots,A_7}
\cZ^{mnp}_{A_1, \ldots,A_7}\hskip.6pt + \big[12345678|A_1, \ldots,A_7\big] \cr
&{}+{1\over 2!}V_{A_1} T^{mn}_{A_2, \ldots,A_6}
\cZ^{mn}_{A_1, \ldots,A_6}\hskip.6pt+ \big[12345678|A_1, \ldots,A_6\big]\cr
&{} +\hskip5pt
V_{A_1}\hskip.8pt T^m_{A_2, \ldots,A_5}\hskip2pt
\cZ^m_{A_1, \ldots,A_5}\hskip3pt+ \big[12345678|A_1, \ldots,A_5\big]\cr
&{}+\hskip5pt
V_{A_1}\hskip.8pt T_{A_2, \ldots,A_4}\hskip2pt \cZ_{A_1, \ldots,A_4}
\hskip3pt+ \big[12345678|A_1,\ldots,A_4\big]\,,
}$$
and contains
${8\stirling8}+{8\stirling7}+{8\stirling6}+{8\stirling5}+{8\stirling4}=1+28+322+1960+6769=9080$
terms, where we recall that ${n\stirling p}$ denotes the Stirling cycle number.
The correlator \eightLiecorr\ also contains
$7{8\stirling8}+6{8\stirling7}+5{8\stirling6}=7+168+1610=1785$ terms with
refinement $d=1$,
\eqnn\doneeight
$$\eqalignno{
\cK_8^{(1)}(\ell) &=
{1\over 2!}\big[V_{A_1} J^{mn}_{A_2|A_3, \ldots,A_8}
\cZ^{mn}_{A_2|A_1,A_3, \ldots,A_8}
+ (A_2\leftrightarrow A_3,\ldots,A_8)\big]
+ [1 \ldots8|A_1, \ldots,A_8]\cr
&\quad{} + \big[V_{A_1} J^m_{A_2|A_3, \ldots,A_7}\cZ^m_{A_2|A_1,A_3, \ldots,A_7}
+ (A_2\leftrightarrow A_3,\ldots,A_7)\big]
+ [1 \ldots8|A_1, \ldots,A_7]\cr
&\quad{}+\big[V_{A_1} J_{A_2|A_3, \ldots,A_6}
\cZ_{A_2|A_1,A_3, \ldots,A_6}
+ (A_2\leftrightarrow A_3,\ldots,A_6)\big]
+ [1 \ldots8|A_1, \ldots,A_6]\,,\cr
&&\doneeight
}$$
and ${7\choose2}=21$ terms with refinement $d=2$,
\eqn\dtwoeight{
\cK_8^{(2)}(\ell) =
V_1 J_{2,3|4,5,6,7,8}\cZ_{2,3|1,4,5,6,7,8} + (2,3|2,3,4,5,6,7,8)\,.
}
The worldsheet functions appearing in the expansions above can be obtained solving
the system of monodromy variations described in section~\bootsec,
and their explicit expressions can be found in the
appendix~\bootstrapapp.

One can also show using the trace relations among local building blocks that the
overall correlator \eightLiecorr\ is unchanged when using
trace-satisfying worldsheet functions $\hat\cZ$ defined in \deformdef\ instead of the naive ones from
the solutions to the monodromy variations,
\eqn\unchanged{
\hat\cK_8^{\rm Lie}(\ell) =\cK_8^{\rm Lie}(\ell)\,.
}

%***********************************
\newsubsubsec\brstsec BRST variation

The computation of $Q\cK_8^{\rm Lie}$ can be performed in a
straightforward fashion using the variations of the local superfields given in
section~\LocalBBsec\ and is given by the general identity \QcKgeneral\ with
$n=8$ (see its $n=7$ instance in \QsevLie). To check whether the correlator is
BRST invariant, it suffices to analyze a few distinct linear combinations of
worldsheet functions encompassed in the definitions of $\Theta^{(d)}$ and
$\Xi^{(d)}$ in \ThetaSix\ and \XiSix.

One can show that the eight-point ${\cal Z}$-functions derived via the bootstrap
approach (cf.\ appendix \appCeight) imply the vanishing of all $\Theta^{(d)}$
topologies of worldsheet functions, see \genTheta. For some of these topologies,
more than ten $\cZ$-functions conspire in a highly non-trivial way to yield
total Koba--Nielsen derivatives that integrate to zero. The full list of
inequivalent topologies can be found in the appendix~\allThetasapp.

However, the combinations $\Xi^{(d)}$ defined in \XiSix\ do not vanish when the
solutions to the monodromy equations are plugged in. For instance, the contributions 
to $Q\cK_8^{\rm Lie}$ proportional to $V_1 Y_{23,45,6,7,8}$ are given by
\eqnn\eitr
$$\eqalignno{
\Xi^{(0)}_{23|1,45,6,7,8}&\equiv
-\half \cZ_{23,1,45,6,7,8}^{pp}
+ \big[\cZ_{23|1,45,6,7,8} + (23\leftrightarrow45,6,7,8)\big]\cr
&=- \cZ_{1|23,45,6,7,8} + R_{1,23,45,6,7,8} &\eitr\cr
&= -\hat\cZ_{1|23,45,6,7,8}\,,
}$$
where the $R$-functions defined in \RemainGfour\
are proportional to ${\rm G}_4$ -- they will be written down below in
\RemainGfourAgain\ for convenience -- and we used the
definition \deformdef\ of $\hat \cZ$ in passing to the last line. The analysis for
the other eight-point building blocks is similar,
\eqn\genThetaXi{
\Xi^{(0)\,m_1 m_2\ldots m_r}_{A_1|B_1,\ldots,B_{r+6}}\cong
- \hat \cZ_{A_1|B_1,\ldots,B_{r+6}}\,, \ \ \ \ 
\Xi^{(1)\,m_1 m_2\ldots m_r}_{A_1|A_2|B_1,\ldots,B_{r+6}}\cong
\hat \cZ_{A_1,A_2|B_1, \ldots,B_{r+6}}\,, \ \ \ \ n=8 \, ,
}
and the BRST variation of \eightLiecorr\ becomes
\eqnn\QKeightFirst
$$\eqalignno{
Q\cK_8^{\rm Lie}(\ell)&=
 -\half V_{A_1}Y^{mn}_{A_2, \ldots,A_8}\,
 \hat\cZ^{mn}_{A_1|A_2,\ldots,A_8}
+ [12 \ldots 8|A_1, \ldots,A_8] &\QKeightFirst\cr
&\hskip8.2pt{}  - V_{A_1}Y^{m}_{A_2, \ldots,A_7}\,\hat\cZ^{m}_{A_1|A_2,\ldots,A_7}
+ [12 \ldots 8|A_1, \ldots,A_7]\cr
&\hskip8.2pt{}  - V_{A_1}Y_{A_2, \ldots,A_6}\,\hat\cZ_{A_1|A_2,\ldots,A_6}
+ [12 \ldots 8|A_1, \ldots,A_6]\cr
&\hskip8.2pt{}+ \big[ V_1Y_{2|3,4,5,6,7,8}\,
\hat\cZ_{1,2|3,4,5,6,7,8}+(2\leftrightarrow3,4,5,6,7,8)\big]\,,
}$$
which can be written more explicitly as
\eqnn\QKeight
$$\eqalignno{
Q\cK_8^{\rm Lie}(\ell)&=
 -\half V_{1}Y^{mn}_{2,3,4,5,6,7,8}\,\hat\cZ^{mn}_{1|2,3,4,5,6,7,8}&\QKeight\cr
& - \big[ V_{1}Y^m_{23,4,5,6,7,8}\,\hat\cZ^m_{1|23,4,5,6,7,8} + (2,3|2,3,4,5,6,7,8) \big]  \cr
& - \big[ V_{12}Y^m_{3,4,5,6,7,8}\,\hat\cZ^m_{12|3,4,5,6,7,8} + (2\leftrightarrow3,4,5,6,7,8) \big] \cr
& - \big[ V_{123}Y_{4,5,6,7,8}\,\hat\cZ_{123|4,5,6,7,8}+V_{132}Y_{4,5,6,7,8}\,\hat\cZ_{132|4,5,6,7,8} + (2,3|2,3,4,5,6,7,8) \big]\cr
& - \big[ V_1Y_{234,5,6,7,8}\,\hat\cZ_{1|234,5,6,7,8} +V_1Y_{243,5,6,7,8}\,\hat\cZ_{1|243,5,6,7,8} + (2,3,4|2,3,4,5,6,7,8) \big] \cr
& - \big[ \big( V_{12}Y_{34,5,6,7,8}\,\hat\cZ_{12|34,5,6,7,8} +{\rm cyc}(2,3,4) \big)+ (2,3,4|2,3,4,5,6,7,8) \big] \cr
&-  \big[ \big( V_1Y_{2,3,4,56,78}\,\hat\cZ_{1|2,3,4,56,78} + {\rm cyc}(5,6,7) \big) + (2,3,4|2,3,4,5,6,7,8)\big]  \cr
& + \big[ V_1Y_{2|3,4,5,6,7,8}\, \hat\cZ_{1,2|3,4,5,6,7,8}+(2\leftrightarrow3,4,5,6,7,8)\big]\,.
}$$
In terms of the undeformed functions, the BRST variation is given by
\eqnn\QKNoGfour
$$\eqalignno{
Q\cK_8^{\rm Lie}(\ell)&=
- \half V_{A_1}Y^{mn}_{A_2, \ldots,A_8}\,
 \big(\cZ^{mn}_{A_1|A_2,\ldots,A_8} - R^{mn}_{A_1, \ldots,A_8}\big)
+ [12 \ldots 8|A_1, \ldots,A_8] &\QKNoGfour\cr
&\hskip8.2pt{}  - V_{A_1}Y^{m}_{A_2, \ldots,A_7}\,
\big(\cZ^{m}_{A_1|A_2,\ldots,A_7} - R^{m}_{A_1, \ldots,A_7}\big)
+ [12 \ldots 8|A_1, \ldots,A_7]\cr
&\hskip8.2pt{}  - V_{A_1}Y_{A_2, \ldots,A_6}\,
\big(\cZ_{A_1|A_2,\ldots,A_6} - R_{A_1, \ldots,A_6}\big)
+ [12 \ldots 8|A_1, \ldots,A_6]\cr
&\hskip8.2pt{}+ \big[ V_1Y_{2|3,4,5,6,7,8}\,
\cZ_{1,2|3,4,5,6,7,8}+(2\leftrightarrow3,4,5,6,7,8)\big]\,,
}$$
where the $R$-functions are all proportional to ${\rm G}_4$,
\eqnn\RemainGfourAgain
$$\eqalignno{
R_{12,34,5,6,7,8} & = 3{\rm G}_4\big(
	 s_{13}
        - s_{14}
        - s_{23}
        + s_{24}\big)\,,&\RemainGfourAgain\cr
R_{123,4,5,6,7,8} & = 3{\rm G}_4\big(s_{12}-2s_{13}+s_{23}\big)\,,\cr
R^m_{12,3,4,5,6,7,8} &= 3{\rm G}_4\big(
s_{12} (k_2^m - k_1^m)
+ \big[ k_3^m (s_{13}-s_{23}) + (3\leftrightarrow4,5,6,7,8)\big]\big)\,,\cr
R^{mn}_{1,2,3,4,5,6,7,8} &=3{\rm G}_4 k_1^{(m} k_2^{n)} s_{12} +(1,2|1,2, \ldots,8)\,.
}$$
Note that
the trace relation $Y^{mm}_{2,3, \ldots,8} = 2Y_{2|3, \ldots,8} +
(2\leftrightarrow3, \ldots,8)$ implies that the contributions of $R^{aa}_{1,2,\ldots,8}$
in \GfourRef\ and \doublyDeform\ cancel. The remaining task is to compensate the leftover variation
\QKNoGfour\ by adding an anomaly sector ${\cal K}_8^Y(\ell)$ to the eight-point correlator.

%****************************************************
\newsubsubsec\Gfourfailuresec Purely anomalous sector

The strategy to cancel the terms \QKeight\ in a bid to achieve BRST invariance
is similar to the seven-point case; we propose to add a purely anomalous
contribution to the eight-point correlator \eightLiecorr,
\eqn\almostF{
\cK_8(\ell) = \cK_8^{\rm Lie}(\ell) + \cK_8^{Y}(\ell)\,.
}
By analogy with the expression \cKDelta\ for ${\cal K}_7^Y(\ell)$, we
start from an ansatz comprising anomalous $\Delta$ superfields of \eightDeltas\
and some unknown worldsheet functions $U$,
\eqnn\deltaJ
$$\eqalignno{
\cK^Y_8(\ell) &= \big[ \Delta^m_{1|2|3,4,\ldots,8}\,U^m_{1|2|3,4,\ldots,8}  + (2\leftrightarrow 3,4,\ldots,8) \big]\cr
&+ \big[ \Delta_{1|23|4,\ldots,8}\,s_{23}U_{1|23|4,\ldots,8}  + (2,3|2, 3,4,\ldots,8) \big] &\deltaJ\cr
&+ \big[ \big( \Delta_{1|2|34,\ldots,8}\,s_{34}U_{1|2|34,\ldots,8} +{\rm cyc}(2,3,4) \big)+
(2,3,4|2,3,4,5,6,7)\big]\,.
}$$
In fact, \deltaJ\ is the most general linear combination of anomalous building blocks
such that their BRST variations are expressible in terms of $V_{1A} Y^{m_1\ldots}_{\ldots}$
rather than $V_{B} Y^{m_1\ldots}_{1\ldots}$ with $1 \notin B$. Any other combination of $Y^{m_1\ldots}_{\ldots}$
in \deltaJ\ would lead to terms $V_{B}, \ 1 \notin B $ in $Q\cK^Y_8(\ell) $ that cannot be cancelled by \QKNoGfour.
In contrast to their seven-point counterpart $\Delta_{1|2|3,4,5,6,7}$, the eight-point instances of the
$\Delta$ superfields exhibit kinematic poles (cf.\ appendix \Deltaapp), so \deltaJ\ amounts to a mild
violation of manifest locality.

In order to determine the $U$-functions in \deltaJ\
we start by noting that $Q^2 \cK^{\rm Lie}_8(\ell) = 0$ implies
that $Q\cK^{\rm Lie}_8(\ell)$ is BRST closed. Therefore all the
ghost-number-four superfields $V_A Y^{m_1 m_2\ldots}_{B_1,B_2,\ldots}$ in
\QKNoGfour\ must combine to ghost-number four BRST invariants given by
$\Gamma$ defined in \partI\ (also see the alternative algorithm
in appendix \Unifiedclikeapp\ for the unrefined cases). This can be
seen by rewriting the local superfields in \QKeight\ in terms of
Berends--Giele currents using \BGmap\ followed by
$M_A {\cal Y}^{m_1
m_2\ldots}_{B_1,B_2,\ldots} \rightarrow \delta_{|A|,1} \Gamma^{m_1
m_2\ldots}_{A|B_1,B_2,\ldots}$ where
\eqnn\Gammasdef
$$\eqalignno{
\Gamma^{mn}_{1|2,3,4,5,6,7,8} &=M_1 \cY^{mn}_{2, \ldots,8}
+  \big[ k_2^{m} M_{12}
\cY^{n}_{3,4,5,6,7,8}
+k_2^{n} M_{12} \cY^{m}_{3,4,5,6,7,8} + (2\leftrightarrow 3, \ldots,8) \big] \cr
& -  \big[ (k_2^{m} k_3^{n} +k_2^{n} k_3^{m}) M_{312} \cY_{4,5,6,7,8} +
(2,3|2,3,4,5,6,7,8) \big]&\Gammasdef\cr
\Gamma^m_{1|23,4,5,6,7,8} &=
    M_{1} \cY^{m}_{23 , 4 , 5 , 6 , 7 , 8}
 + \big[ M_{12} \cY^{m}_{3 , 4 , 5 , 6 , 7 , 8}
 +  M_{123} \cY_{4 , 5 , 6 , 7 , 8} k_{3}^{m}
 - (2\leftrightarrow3)\big]\cr
& + \big[k_{4}^{m} \big(M_{14} \cY_{23 , 5 , 6 , 7 , 8}
 -  M_{214} \cY_{3 , 5 , 6 , 7 , 8}
 +  M_{314} \cY_{2 , 5 , 6 , 7 , 8}\big) + (4\leftrightarrow5,6,7,8)\big]\cr
\Gamma_{1|234,5,6,7,8} &=
 M_{1} \cY_{234 , 5 , 6 , 7 , 8}
+  M_{12} \cY_{34 , 5 , 6 , 7 , 8}
 +  M_{123} \cY_{4 , 5 , 6 , 7 , 8}\cr
&{} +  M_{214} \cY_{3 , 5 , 6 , 7 , 8}
 -  M_{14} \cY_{23 , 5 , 6 , 7 , 8}
 +  M_{143} \cY_{2 , 5 , 6 , 7 , 8}\cr
\Gamma_{1|23,45,6,7,8} &=
   M_{1} \cY_{23 , 45 , 6 , 7 , 8}
 + \big[ M_{12} \cY_{45 , 3 , 6 , 7 , 8} -(2\leftrightarrow3)\big]\cr
& + \big[ M_{14} \cY_{23 , 5 , 6 , 7 , 8}
 +  M_{215} \cY_{3 , 4 , 6 , 7 , 8}
 -  M_{315} \cY_{2 , 4 , 6 , 7 , 8} -(4\leftrightarrow5)\big] \cr
 \Gamma_{1|2|3,4,5,6,7,8} &= M_1 \cY_{2|3,4,\ldots,8} + M_{12} k_2^m \cY^m_{3,4,\ldots,8}
 +\big[ s_{23} M_{123} \cY_{4,\ldots,8}  +(3\leftrightarrow 4,\ldots,8)\big]
 \,.
}$$
Under these transformations, it is possible to verify that \QKeight\ is identical to
\eqnn\deltaG
$$\eqalignno{
Q\cK^{\rm Lie}_8(\ell) &=
- \half \Gamma^{mn}_{A_1|A_2, \ldots,A_8}\,Z^{(s),mn}_{A_1|A_2,\ldots,A_8} + [12
\ldots 8|A_1, \ldots,A_8] &\deltaG\cr
&\hskip8.2pt{}  - \Gamma^{m}_{A_1|A_2, \ldots,A_7}\,
Z^{(s),m}_{A_1|A_2,\ldots,A_7} + [12 \ldots 8|A_1, \ldots,A_7]\cr
&\hskip8.2pt{}  - \Gamma_{A_1|A_2, \ldots,A_6}\,
Z^{(s)}_{A_1|A_2,\ldots,A_6} + [12 \ldots 8|A_1, \ldots,A_6]\cr
&\hskip8.2pt{} + \big[ \Gamma_{1|2|3,4,5,6,7,8}\,
Z^{(s)}_{1,2|3,4,5,6,7,8}+(2\leftrightarrow3,4,5,6,7,8)\big]\cr
&\hskip8.2pt+ \half V_{A_1}Y^{mn}_{A_2, \ldots,A_8}\,R^{mn}_{A_1, \ldots,A_8}
+ [12 \ldots 8|A_1, \ldots,A_8]\cr
&\hskip8.2pt{}  + V_{A_1}Y^{m}_{A_2, \ldots,A_7}\,R^{m}_{A_1, \ldots,A_7}
+ [12 \ldots 8|A_1, \ldots,A_7]\cr
&\hskip8.2pt{}  + V_{A_1}Y_{A_2, \ldots,A_6}\,R_{A_1, \ldots,A_6}
+ [12 \ldots 8|A_1, \ldots,A_6]\,,
}$$
where $Z^{(s)}$ is defined in \cZToZ. It is beneficial to rewrite \QKeight\ in this way
because the $Q$-variation of the anomalous correlator \deltaJ\ takes the same
form once we insert the expressions for $Q\Delta$ in \eightDQBRST:
\eqnn\deltaL
$$\eqalignno{
Q\cK^Y_8(\ell) &= \Gamma^{mn}_{1|2,3,4,5,6,7,8}\,
\big[ k_2^m U^n_{1|2|3,4,5,6,7,8} + (2\leftrightarrow 3,4,5,6,7,8) \big] &\deltaL\cr
&+\big[s_{23} \Gamma^m_{1|23,4,\ldots,8}\,
\big(U^m_{1|2|3,\ldots,8} - U^m_{1|3|2,\ldots,8}\cr
&\qquad{}+ \big[k_{23}^m U_{1|23|4,\ldots,8}  +
(23\leftrightarrow4,5,6,7,8)\big]\big)
+ (2,3|2,3,4,5,6,7,8) \big]\cr
&+ \big[s_{23}s_{45} \Gamma_{1|23,45,6,7,8}\,\big(U_{1|2|3,45,6,7,8}-U_{1|3|2,45,6,7,8}  \cr
&\qquad{} + U_{1|4|5,23,6,7,8} - U_{1|5|4,23,6,7,8}\big)+ (2,3|4,5|2,3,4,5,6,7,8) \big] \cr
& + \Big[ \Gamma_{1|234,5,6,7,8}\,\big[
s_{23}s_{24}\big(U_{1|23|4,\ldots,8} - U_{1|4|23,\ldots,8} -
U_{1|3|24,\ldots,8}  + U_{1|24|3,\ldots,8}\big)\cr
&\qquad\qquad\qquad\quad{}+ s_{23}s_{34}\big(U_{1|23|4,\ldots,8}
- U_{1|4|23,\ldots,8} +U_{1|2|34,\ldots,8}-U_{1|34|2,\ldots,8}\big)\big]\cr
&\quad{}+ \Gamma_{1|243,5,6,7,8}\,\big[
s_{23}s_{24}\big(U_{1|24|3,\ldots,8} - U_{1|3|24,\ldots,8}
- U_{1|4|23,\ldots,8}  + U_{1|23|4,\ldots,8}\big)\cr
%%%
&\qquad\qquad\qquad\quad{}+s_{24}s_{34}\big(U_{1|24|3,\ldots,8} - U_{1|3|24,\ldots,8} +
U_{1|2|43,\ldots,8}-U_{1|43|2,\ldots,8}\big)\big]\cr
& \ \ \ \ \ \ + (2,3,4|2,3,4,5,6,7,8) \Big]\cr
& - \big[\Gamma_{1|2|3,4,5,6,7,8}\,
(k_2^m U^m_{1|2|3,4,5,6,7,8} + s_{23}U_{1|23|4,5,6,7,8} +s_{24}U_{1|24|3,5,6,7,8}  \cr
& \ \ \ \ \ \ +\cdots +s_{28}U_{1|28|3,4,5,6,7}   )+ (2\leftrightarrow3,4,5,6,7,8)
\big]\,.
}$$
As we will see, the functions $U$ in the anomalous
correlator \deltaJ\ can be chosen such as to cancel all $\Gamma$ terms from \deltaG.
This can be achieved provided the following equations hold:
\eqnn\oneTops
\eqnn\twoTops
\eqnn\threeTops
\eqnn\fourTops
\eqnn\fiveTops
$$\eqalignno{
% Note the different s_ij scaling of the scalar U functions!
 % 	    + Gamma(1,[p],2,[c],3,[c],4,[c],5,[c],6,[c],7,[c],8,[v],N1_?,N2_?) * (
 %          + kk(2,N2_?)*ZU(1,[p],2,[p],3,[c],4,[c],5,[c],6,[c],7,[c],8,[v],N1_?)
 %          + kk(3,N2_?)*ZU(1,[p],3,[p],2,[c],4,[c],5,[c],6,[c],7,[c],8,[v],N1_?)
 %          + kk(4,N2_?)*ZU(1,[p],4,[p],3,[c],2,[c],5,[c],6,[c],7,[c],8,[v],N1_?)
 %          + kk(5,N2_?)*ZU(1,[p],5,[p],3,[c],4,[c],2,[c],6,[c],7,[c],8,[v],N1_?)
 %          + kk(6,N2_?)*ZU(1,[p],6,[p],3,[c],4,[c],5,[c],2,[c],7,[c],8,[v],N1_?)
 %          + kk(7,N2_?)*ZU(1,[p],7,[p],3,[c],4,[c],5,[c],6,[c],2,[c],8,[v],N1_?)
 %          + kk(8,N2_?)*ZU(1,[p],8,[p],3,[c],4,[c],5,[c],6,[c],7,[c],2,[v],N1_?)
 %          - 1/2*ZZO(1,[p],2,[c],3,[c],4,[c],5,[c],6,[c],7,[c],8,[v],N1_?,N2_?)
 %          )
0&\cong\Big(\cZ^{mn}_{1|2,3,4,5,6,7,8} - \big[k_2^{(m} U^{n)}_{1|2|3,4,5,6,7,8} + (2\leftrightarrow
3, \ldots,8)\big]\Big)\Gamma^{mn}_{1|2,3,4,5,6,7,8} &\oneTops\cr
 %       + Gamma(1,[p],2,3,[c],4,[c],5,[c],6,[c],7,[c],8,[v],N1_?) * (
 %          + kk(2,N1_?)*s(2,3)*ZU(1,[p],2,3,[p],4,[c],5,[c],6,[c],7,[c],8)
 %          + kk(3,N1_?)*s(2,3)*ZU(1,[p],2,3,[p],4,[c],5,[c],6,[c],7,[c],8)
 %          + kk(4,N1_?)*s(2,3)*ZU(1,[p],4,[p],2,3,[c],5,[c],6,[c],7,[c],8)
 %          + kk(5,N1_?)*s(2,3)*ZU(1,[p],5,[p],2,3,[c],4,[c],6,[c],7,[c],8)
 %          + kk(6,N1_?)*s(2,3)*ZU(1,[p],6,[p],2,3,[c],4,[c],5,[c],7,[c],8)
 %          + kk(7,N1_?)*s(2,3)*ZU(1,[p],7,[p],2,3,[c],4,[c],5,[c],6,[c],8)
 %          + kk(8,N1_?)*s(2,3)*ZU(1,[p],8,[p],2,3,[c],4,[c],5,[c],6,[c],7)
 %          + s(2,3)*ZU(1,[p],2,[p],3,[c],4,[c],5,[c],6,[c],7,[c],8,[v],N1_?)
 %          - s(2,3)*ZU(1,[p],3,[p],2,[c],4,[c],5,[c],6,[c],7,[c],8,[v],N1_?)
 %          - s(2,3)*ZZO(1,[p],2,3,[c],4,[c],5,[c],6,[c],7,[c],8,[v],N1_?)
 %          )
0&\cong \Big(\cZ^m_{1|23,4, \ldots,8}- U^m_{1|2|3,\ldots,8} + U^m_{1|3|2,\ldots,8}
- \big[k_{23}^m U_{1|23|4,\ldots,8} + (23\leftrightarrow4,
\ldots,8)\big]\Big)\Gamma^{m}_{1|23,4, \ldots,8}\cr
&&\twoTops\cr
 %       + Gamma(1,[p],2,3,[c],4,5,[c],6,[c],7,[c],8) * (
 %          + s(2,3)*s(4,5)*ZU(1,[p],2,[p],4,5,[c],3,[c],6,[c],7,[c],8)
 %          - s(2,3)*s(4,5)*ZU(1,[p],3,[p],4,5,[c],2,[c],6,[c],7,[c],8)
 %          + s(2,3)*s(4,5)*ZU(1,[p],4,[p],2,3,[c],5,[c],6,[c],7,[c],8)
 %          - s(2,3)*s(4,5)*ZU(1,[p],5,[p],2,3,[c],4,[c],6,[c],7,[c],8)
 %          - s(2,3)*s(4,5)*ZZO(1,[p],2,3,[c],4,5,[c],6,[c],7,[c],8)
 %          )
0&\cong\Big(\cZ_{1|23,45,6,7,8} - \big[U_{1|4|23,5,6,7,8} + U_{1|5|23,4,6,7,8} -
(23\leftrightarrow45)\big]\Big)\Gamma_{1|23,45,6,7,8}
&\threeTops\cr
 	% + Gamma(1,[p],2,3,4,[c],5,[c],6,[c],7,[c],8) * (
        %   + s(2,3)*s(2,4)*ZU(1,[p],2,3,[p],4,[c],5,[c],6,[c],7,[c],8)
        %   + s(2,3)*s(2,4)*ZU(1,[p],2,4,[p],3,[c],5,[c],6,[c],7,[c],8)
        %   - s(2,3)*s(2,4)*ZU(1,[p],3,[p],2,4,[c],5,[c],6,[c],7,[c],8)
        %   - s(2,3)*s(2,4)*ZU(1,[p],4,[p],2,3,[c],5,[c],6,[c],7,[c],8)
%
        %   + s(2,3)*s(3,4)*ZU(1,[p],2,[p],3,4,[c],5,[c],6,[c],7,[c],8)
        %   + s(2,3)*s(3,4)*ZU(1,[p],2,3,[p],4,[c],5,[c],6,[c],7,[c],8)
        %   - s(2,3)*s(3,4)*ZU(1,[p],3,4,[p],2,[c],5,[c],6,[c],7,[c],8)
        %   - s(2,3)*s(3,4)*ZU(1,[p],4,[p],2,3,[c],5,[c],6,[c],7,[c],8)
%
        %   - s(2,3)*s(3,4)*ZZO(1,[p],2,3,4,[c],5,[c],6,[c],7,[c],8)
	%   - s(2,3)*s(2,4)*ZZO(1,[p],2,3,4,[c],5,[c],6,[c],7,[c],8)
        %   - s(2,3)*s(2,4)*ZZO(1,[p],2,4,3,[c],5,[c],6,[c],7,[c],8)
        %   )
%
0&\cong\Big(s_{23}(s_{24}+s_{34})\cZ_{1|234,5,6,7,8}
+ s_{23} s_{24} \cZ_{1|243,5,6,7,8} &\fourTops\cr
&\qquad{}- s_{23}s_{24}\big(
  U_{1|23|4,\ldots,8}
+ U_{1|24|3,\ldots,8}
- U_{1|3|24,\ldots,8}
- U_{1|4|23,\ldots,8}
\big)\cr
&\qquad{}- s_{23}s_{34}\big(
  U_{1|2|34,\ldots,8}
+ U_{1|23|4,\ldots,8}
- U_{1|34|2,\ldots,8}
- U_{1|4|23,\ldots,8}
\big)\Big)\Gamma_{1|234,5,6,7,8}\cr
%
 % 	+ Gamma(1,[p],2,[p],3,[c],4,[c],5,[c],6,[c],7,[c],8) * (
 %          - kk(2,N1_?)*ZU(1,[p],2,[p],3,[c],4,[c],5,[c],6,[c],7,[c],8,[v],N1_?)
 %          + ZZO(1,[c],2,[p],3,[c],4,[c],5,[c],6,[c],7,[c],8)
 %          - s(2,3)*ZU(1,[p],[p],4,[c],5,[c],6,[c],7,[c],8)
 %          - s(2,4)*ZU(1,[p],[p],3,[c],5,[c],6,[c],7,[c],8)
 %          - s(2,5)*ZU(1,[p],[p],3,[c],4,[c],6,[c],7,[c],8)
 %          - s(2,6)*ZU(1,[p],[p],3,[c],4,[c],5,[c],7,[c],8)
 %          - s(2,7)*ZU(1,[p],[p],3,[c],4,[c],5,[c],6,[c],8)
 %          - s(2,8)*ZU(1,[p],[p],3,[c],4,[c],5,[c],6,[c],7)
 %          )
%%
0&\cong\Big( \cZ_{1,2|3, \ldots,8}
- k_2^m U^m_{1|2|3, \ldots,8} - \big[s_{23}U_{1|23|4, \ldots,8} +
(3\leftrightarrow4, \ldots,8)\big]\Big)\Gamma_{1|2|3,4,5,6,7,8} \, .
&\fiveTops
}$$
To solve these equations it will be convenient to exploit the vanishing of
$\cZ^\Delta$ according to \noGfour. For instance,
$0\cong\cZ^{\Delta, mn}_{1|2,3,4,5,6,7,8}$ can be used to rewrite
\eqn\rewZ{
\cZ^{mn}_{1|2,3,4,5,6,7,8} = - \big[ k_2^{(m} \cZ^{n)}_{12|3,4,5,6,7,8} +
(2\leftrightarrow3, \ldots,8)\big] + \big[k_2^{(m}k_3^{n)}\cZ_{213|4,5,6,7,8}
+ (2,3|2, \ldots,8)\big]
}
allowing \oneTops\ to be solved for $U^m$. In turn, plugging $U^m$ into \twoTops\ and using
the vanishing of $\cZ^\Delta_{1|23,4,5,6,7,8}$ from \DelZs\ allows the
determination of the other two topologies of $U$ in \deltaJ. The resulting expressions
\eqnn\vecUsol
$$\eqalignno{
U^m_{1|2|3,4,5,6,7,8} &= -\cZ^m_{12|3, \ldots,8} + \half\big[k_3^m\cZ_{213|4,
\ldots,8} + (3\leftrightarrow4, \ldots,8)\big]\,,&\vecUsol\cr
U_{1|23|4,5,6,7,8} &= \half \big(\cZ_{132|4,5,6,7,8} -
\cZ_{123|4,5,6,7,8}\big)\,, \cr
U_{1|2|34,5,6,7,8} &= \half \big(\cZ_{312|4,5,6,7,8} -
\cZ_{412|3,5,6,7,8}\big)\,, 
}$$
are consistent with the remaining equations \threeTops\ to \fiveTops. One could wonder if 
the trace relation $\Gamma^{mn}_{1|2,3,\ldots,8}= 2\Gamma_{1|2|3,\ldots,8}+(2\leftrightarrow 3,\ldots,8)$ 
among the anomaly invariants might generate corrections to the
last equation \fiveTops\ from tensor traces in \oneTops. This is not the case because the
chosen representation of $\cZ^{mn}$ in the tensorial equation \oneTops\ does not feature
any $\d^{mn}{\rm G}_4$ deformations.

Given that the expressions \vecUsol\ for the $U$-functions in $\cK^Y_8(\ell)$ solve 
all of \oneTops\ to \fiveTops, the BRST variation of the overall correlator \almostF\ reduces to
\eqnn\aimQK
$$\eqalignno{
% QK8+QKDelta_VY_local.h
Q(\cK^{\rm Lie}_8(\ell) + \cK^Y_8(\ell)) &=
\half V_{A_1}Y^{mn}_{A_2, \ldots,A_8}\,R^{mn}_{A_1, \ldots,A_8}
+ [12 \ldots 8|A_1, \ldots,A_8] &\aimQK\cr
&\hskip8.2pt{}  + V_{A_1}Y^{m}_{A_2, \ldots,A_7}\,R^{m}_{A_1, \ldots,A_7}
+ [12 \ldots 8|A_1, \ldots,A_7]\cr
&\hskip8.2pt{}  + V_{A_1}Y_{A_2, \ldots,A_6}\,R_{A_1, \ldots,A_6}
+ [12 \ldots 8|A_1, \ldots,A_6]\,.
}$$
The $R$-functions from \RemainGfourAgain\ are all proportional to the
holomorphic Eisenstein series ${\rm G}_4$, i.e.\ any dependence of the BRST variation 
\aimQK\ on $\ell$ or the $z_j$ has cancelled.

Unfortunately, we explicitly checked \FORM\ that there is no manifestly local deformation
of the correlator that can be used to cancel the remaining terms in \aimQK.
Therefore, even though the BRST variation of $\cK_8^{\rm Lie}(\ell)+\cK_8^Y(\ell)$ turns out to be a
local expression, its component expansion is non-local, see appendix \Deltaapp\ for the
kinematic poles of the $\Delta_{1|\ldots}$ superfields in \deltaJ. This
suggests that there may be another non-local sector whose BRST variation cancels
\aimQK, although we have not been able to pinpoint it yet. We leave the quest for
finding such a completion to
future investigations.

%*********************************************************************
\newnewsec\loopintsec Modular forms: Integrating out the loop momentum

This section is dedicated to the integration over the loop momentum which will lead to
manifestly single-valued one-loop correlators. In this way, the correlators
acquire well-defined weights under modular transformations, namely holomorphic
weight $n{-}4$ for the loop integral of ${\cal K}_n(\ell)$.

At the same time, closed-string correlators are no longer chirally split after
integration over the loop momenta \refs{\verlindes,\DHokerPDL,\xerox}. We will
describe the systematics of the interactions between left- and right movers that
arises from loop integration of the holomorphic squares $|{\cal K}_n(\ell)|^2$.
The setup of $\cZ$-functions and GEIs turns out to provide an efficient handle
on the vector contractions between left- and right movers and the loss of
meromorphicity of the open-string contributions after integration over $\ell$.

Let us briefly summarize the notation of part I \& II. As detailed in sections \chiralsplitsec\ 
and \intellmom, the net effect of loop integration on the Koba--Nielsen factor \IIIKNfactor\ 
is captured by
\eqnn\KNKNBag
$$\eqalignno{
\hat {\cal I}^{\rm open}_n &= {(2\pi i)^D \over (\Im \tau)^{{D\over 2}}}
\exp \Big( \sum^n_{i<j}s_{ij} \Big[ \log \big|  \theta_1(z_{ij},\tau) \big|
- { \pi (\Im z_{ij})^2 \over \Im \tau}  \Big]\Big) \, , &\KNKNBag  \cr
\hat {\cal I}_n &= {(2\pi i)^D\over (2 \Im \tau)^{{D\over 2}}}
\exp\Big(\sum^n_{i<j}s_{ij} \Big[\log\big| \theta_1(z_{ij},\tau)  \big|^2
- {2\pi \over \Im \tau} (\Im z_{ij})^2 \Big]\Big)\, . 
}$$
After factorizing these universal quantities in the worldsheet integrand of
open- and closed-string amplitudes \againopen\ and \againclosed,
\eqnn\againnewamprep
$$\eqalignno{
{\cal A}_n  &=
\sum_{\rm top} C_{\rm top} \int_{D_{\rm top}}\!\!\!\!
d\tau \, dz_2 \, d z_3 \, \ldots \, d z_{n} \,
\hat {\cal I}^{\rm open}_n \, [[ \, \langle {\cal K}_n(\ell)\rangle\, ]]\,,
&\againnewamprep\cr
{\cal M}_n  &=
 \int_{{\cal F}}
d^2\tau \, d^2z_2 \, d^2 z_3 \, \ldots \, d^2 z_{n} \, \hat {\cal I}_n \, [[ \,
\langle {\cal K}_n(\ell)\rangle \, \langle\tilde{\cal K}_n(-\ell)\rangle \, ]]
\,,
}$$
the leftover integrand w.r.t. the punctures $z_j$ and modular parameters $\tau$
is furnished by ``loop-integrated'' correlators $[[{\cal K}_n(\ell)]]$ and $ [[{\cal K}_n(\ell)
\tilde{\cal K}_n(-\ell)  ]]$ in combination with the zero-mode prescription
$\langle \ldots \rangle$ of the pure-spinor formalism \psf. Hence, the notation
$[[\ldots]]$ in \againnewamprep\ addresses the net effect of shifting
the loop momentum as a Gaussian integration variable, cf. section \intellmom.
The normalization is chosen as $[[1]]=1$, and the simplest non-trivial examples
$ [[\ell^m]] = L_{0}^m$ and $ [[\ell^m \ell^n ]] =L_{0}^m L_{0}^n -{\pi\over \Im \tau} \delta^{mn}$
are most compactly written in terms of the non-meromorphic quantity $L_{0}^m  
= \sum_{j=2}^n k_j^m\nu_{1j}$ with $\nu_{ij} \equiv 2\pi i {\Im z_{ij} \over \Im \tau}$, see \KNKNC\ for
integration over higher powers of $\ell$. The contribution $-{\pi\over \Im \tau} \delta^{mn}$
to $[[\ell^m \ell^n ]]$ is the first instance of the aforementioned interactions between left- and right movers.

%********************************************************
\newsubsec\FiveLoopsec Five-point open-string correlators

Starting from this section, we apply the techniques of integrating the loop
momentum to the correlators ${\cal K}_n(\ell)$ of section \explsec. We will
complement the direct integration of GEIs with a study of the $T\cdot {\cal Z}$
and $C\cdot {\cal Z}$ representations where the origin of the kinematic factors
from the OPEs is more transparent.

%************************************************************************************************
\newsubsubsec\FiveTFsec The $T\cdot\cF$ representation: manifesting locality \& single-valuedness

As discussed in \xerox, integration over the loop momentum leads to manifestly
single-valued representations of chirally-split correlators. We therefore
integrate out the loop momentum from the representation \fivecorr\ using \KNKNC\
to obtain
\eqnn\intfive
$$\eqalignno{
[[\cK_5(\ell)]] &= \big[k_2^m V_1T^m_{2,3,4,5}\nu_{12} +
V_{12}T_{3,4,5}g_{12}^{(1)} + (2\leftrightarrow3,4,5)\big]
+ \big[V_1 T_{23,4,5}g_{23}^{(1)} + (2,3|2,3,4,5)\big]\cr
&=\big[V_{12}T_{3,4,5}(g_{12}^{(1)}{+}\nu_{12}) {+} (2{\leftrightarrow}3,4,5)\big]
{+} \big[V_1 T_{23,4,5}(g_{23}^{(1)}{+}\nu_{23}) {+} (2,3|2,3,4,5)\big]\,,&\intfive
}$$
where to arrive in the second line we used the cohomology identity \Vtwoid\ as
$k_2^m V_1T^m_{2,3,4,5} \cong V_{12}T_{3,4,5} - \big[ V_{1}T_{23,4,5} +
(3\leftrightarrow 4,5)\big]$ and $\nu_{13}{-}\nu_{12}=\nu_{23}$. So we see that
the single-valued functions $f^{(1)}_{ij}=g^{(1)}_{ij}+\nu_{ij}$ are
constructively obtained and we get the following correlator
\eqnn\singfivef
$$\eqalignno{
[[\cK_5(\ell)]] &= V_{12}T_{3,4,5}\,\cF_{12,3,4,5} +
(2\leftrightarrow3,4,5) &\singfivef\cr
&\quad{}+V_1T_{23,4,5}\cF_{1,23,4,5} + (2,3|2,3,4,5)\,,
}$$
in terms of manifestly single-valued functions $\cF_{12,3,4,5}\equiv f^{(1)}_{12}$.
Given that the functions $f^{(w)}_{ij}$ defined by \NHKron\ carry $w$ units of holomorphic
modular weight, see \modKron, the correlator \singfivef\ is a modular form of weight one.

%*******************************************************************************************************
\newsubsubsec\FiveCFsec The $C\cdot\cF$ representation: manifesting BRST invariance \& single-valuedness

It is also possible to obtain a representation without the loop momentum which
manifests both BRST invariance and single-valuedness. This can be achieved in at
least two ways: integrating out the loop momentum from the $C\cdot\cZ$
representation \brstfivecorr\ or using integration-by-parts identities to
eliminate all $f^{(1)}_{1j}$ integrands with $j=2,3,4,5$ from \singfivef.

First, integrating out the loop
momentum from \brstfivecorr\ using the identity,
\eqn\loopint{
L_0^m C^m_{1|2,3,4,5} = -\sum_{j=1}^{5}
\nu_j k_j^m C^m_{1|2,3,4,5} = \big[ s_{23} \nu_{23} C_{1|23,4,5} +
(2,3|2,3,4,5) \big]\,,
}
leads to the manifestly single-valued and BRST-invariant form of the five-point
correlator
\eqn\loopK{
[[{\cal K}_5(\ell) ]]=  C_{1|23,4,5} s_{23}f^{(1)}_{23} + (2,3|2,3,4,5)\,.
}
This form reproduces the five-point one-loop correlator proposed in \fiveptNMPS\
and rederived in \refs{\oneloopbb,\MafraIOJ}. Alternatively, one can arrive at
the representation \loopK\ using integration-by-parts identities \newzderivzero\
in the local and single-valued representation \singfivef. In fact, this is how
\loopK\ was originally derived in \oneloopbb.  The derivations of this paper are
based on single-valuedness and BRST-invariance constraints, and one obtains a
much richer perspective on the correlators. In summary, the integration over the
loop momentum yields two additional representations of the five-point correlator
from \allfives,
\eqnn\unnumbB
$$\eqalignno{
[[{\cal K}_5(\ell)]]&=
\big[ V_{12}T_{3,4,5} \cF_{12,3,4,5} + (2\leftrightarrow3,4,5) \big]
+ \big[ V_1 T_{23,4,5} \cF_{1,23,4,5} + (2,3|2,3,4,5) \big]\,,\cr
[[{\cal K}_5(\ell)]]&= C_{1|23,4,5}F^{(s)}_{1,23,4,5} + (2,3|2,3,4,5)\,,&\unnumbB
}$$
where we used the shorthand $F^{(s)}_{1,23,4,5}=s_{23}
\cF_{1,23,4,5}=s_{23}f^{(1)}_{23}$ in the second line.
More generally, by analogy with the $Z^{(s)}$ in \cZToZ, we define the following Lie-symmetry
satisfying analogues of the shuffle-symmetric ${\cal F}^{m_1m_2\ldots}_{A,B,\ldots}$-functions,
\eqnn\KLTF
$$\eqalignno{
F^{(s)m_1 \ldots}_{aA,bB, \ldots} &\equiv \sum_{A',B', \ldots} S(A|A')_a
S(B|B')_b \cdots {\cal F}^{m_1 \ldots}_{aA',bB', \ldots}\,,&\KLTF
}$$
which will be tensorial at higher multiplicity. We see that integrating out the
loop momentum from the functions $\cZ$ in \fivecorr\ has the same effect as
sending $\ell\to0$ and $g^{(1)}_{ij}\to f^{(1)}_{ij}$. However, these
replacement rules are tied to the present open-string context and no longer
apply to the closed-string five-point correlators of section \fiveclloc.

%*********************************************************************************
\newsubsubsec\FiveEEsec Lessons from the $T\cdot E$ and $C\cdot E$ representations

As an alternative to the earlier computations, one can start from the
representations \fiveTE\ or \covK\ of ${\cal K}_5(\ell)$ in terms of GEIs and
insert the results \scalarGEIf\ and \intellA\ for their loop integral. The
manifestly local $T\cdot E$ representation \fiveTE\ yields
\eqnn\fromTEfive
$$\eqalignno{
[[{\cal K}_5(\ell)]] &= V_1 T_{23,4,5} f^{(1)}_{23} + (2,3|2,3,4,5) &\fromTEfive \cr
&+ V_1 (k_2^m T^m_{2,3,4,5}+T_{23,4,5}+T_{24,3,5}+T_{25,3,4}) f_{12}^{(1)} + (2\leftrightarrow 3,4,5)
}$$
after reorganizing terms which agrees with the $T\cdot {\cal F}$ representation \singfivef\
up to the cohomology identity \Vtwoid. Similarly, the manifestly BRST-invariant $C\cdot E$
representation \covK\ yields
\eqnn\fromCEfive
$$\eqalignno{
[[{\cal K}_5(\ell)]] &= s_{23} C_{1|23,4,5} f^{(1)}_{23} + (2,3|2,3,4,5) &\fromTEfive \cr
&+ (k_2^m C^m_{1|2,3,4,5}+s_{23}C_{1|23,4,5}+s_{24}C_{1|24,3,5}
+s_{25}C_{1|25,3,4}) f_{12}^{(1)} + (2\leftrightarrow 3,4,5)\, ,
}$$
after reorganizing terms. This in turn matches the $C\cdot {\cal F}$ representation \loopK,
because the second line is BRST exact by \deltasecE.

%******************************************************
\newsubsec\SixLoopsec Six-point open-string correlators

%***********************************************************************************************
\newsubsubsec\sixTFsec The $T\cdot\cF$ representation: manifesting locality \& single-valuedness

We already know that the six-point correlator \sixLie\ is single-valued, and in
this section this will be manifested by integrating out the loop momentum
and checking that the generated variables $\nu_{ij}$
combine into single-valued functions $f^{(n)}_{ij}$ according to \lopF
\eqn\gtofsix{
g^{(1)}_{ij} + \nu_{ij} = f^{(1)}_{ij},\qquad
g^{(2)}_{ij} + \nu_{ij}g^{(1)}_{ij} + \half \nu_{ij}^2  = f^{(2)}_{ij}\,.
}
Indeed, integrating out the loop momentum in the representation \sixLie\ using \lopF\ 
yields
\eqnn\sixLieF
$$\eqalignno{
[[\cK_6(\ell)]] &=
\half V_{A_1} T^{mn}_{A_2, \ldots,A_6} \cF^{mn}_{A_1, \ldots,A_6}
+[123456|A_1, \ldots,A_6] &\sixLieF\cr
&\quad{}+ V_{A_1} T^m_{A_2, \ldots,A_5} \cF^m_{A_1, \ldots,A_5}
+ [123456|A_1, \ldots,A_5]\cr
&\quad{}+ V_{A_1} T_{A_2, \ldots,A_4} \cF_{A_1, \ldots,A_4}
+ [123456|A_1, \ldots,A_4]\,,
}$$
with manifestly single-valued worldsheet functions given by
\eqnn\newfs
$$\eqalignno{
\cF_{123,4,5,6}&\equiv f^{(1)}_{12}f^{(1)}_{23} + f^{(2)}_{12} +
f^{(2)}_{23} - f^{(2)}_{13}\,, &\newfs\cr
\cF_{12,34,5,6}&\equiv f^{(1)}_{12}f^{(1)}_{34}
+ f^{(2)}_{13} + f^{(2)}_{24}
- f^{(2)}_{14} - f^{(2)}_{23}\,,\cr
\cF^m_{12,3,4,5,6}&\equiv
(k_2^m - k_1^m)f^{(2)}_{12}
+ \big[ k_3^m (f^{(2)}_{13} - f^{(2)}_{23}) + (3\leftrightarrow
4,5,6)\big]\,,\cr
\cF^{mn}_{1,2,3,4,5,6}&\equiv
\bigl[( k_1^{m}k_2^{n}+k_1^{n}k_2^{m}) f^{(2)}_{12} + (1,2|1,2,3,4,5,6)
\bigr]\,.
}$$
To see this, we use the integration-by-parts identity \newzderivzero\ obtained from
$\p_2(\nu_2  \hat {\cal I}_6)=0$,
\eqn\ibpnu{
\nu_2 f_{12}^{(1)} \cong {1\over s_{12}}\Big({\pi\over \Im \tau} +
 \big[s_{23} \nu_2f^{(1)}_{23} +(3\leftrightarrow 4,5,6)\big]\Big)\,,
}
and drop the BRST-exact linear combinations given by \localClosedSix\ and
\manlocM. Given that an additional summand of ${\pi\over \Im \tau}$ arises from
the loop integral over ${1\over 2} \ell_m \ell_n V_1T_{2,3,4,5,6}^{mn}$, the
coefficient of the modular anomaly cancels by the building-block trace relation
\HOb. Similarly as in the five-point open-string calculations, the functions
$\cF$ in \newfs\ are related to $\cZ$ from \newgsAgain\ by $\ell\to0$ and
$g^{(n)}_{ij}\to f^{(n)}_{ij}$. Furthermore, we see from \modKron\ that the
non-holomorphic six-point correlator \sixLieF\ is a modular form of weight two.

%******************************************************************************************************
\newsubsubsec\sixCFsec The $C\cdot\cF$ representation: manifesting BRST invariance \& single-valuedness

There are several alternatives to deriving a manifestly BRST-invariant form of
the correlator without the loop momentum. The most straightforward way is to use
integration-by-parts identities \newzderivzero\ in the representation \sixLieF.
A long calculation very similar to the derivation of \tediousSix\ in
section~\SixCompsec\ leads to
\eqnn\PSanom
$$\eqalignno{
[[\cK_6(\ell)]]&=
\big[ \big( s_{23}s_{34} f^{(1)}_{23}  f^{(1)}_{34}C_{1|234,5,6} + {\rm cyc}(2,3,4)\big)
+ (2,3,4 | 2,3,4,5,6)  \big]&\PSanom\cr
& + \big[ \big(s_{23}s_{45} f^{(1)}_{23} f^{(1)}_{45} C_{1|23,45,6}
+ {\rm cyc}(3,4,5)\big)
+ (6 \leftrightarrow 5,4,3,2)  \big]\cr
&+  \big[ f^{(2)}_{12} C_{1|2|3,4,5,6} + (2\leftrightarrow 3,4,5,6) \big]
+ \big[f^{(2)}_{23} C_{1|(23)|4,5,6} + (2,3|2,3,4,5,6) \big]\,,
}$$
see \Ponetwo\ or \equivOld\ for the pseudo-invariant kinematic factors $C_{1|2|3,4,5,6}$ and $C_{1|(23)|4,5,6}$.
This representation reproduces the correlator proposed in \MafraNWR\ based
on BRST cohomology properties together with an anomaly-cancellation analysis.
At the same time, \PSanom\ can be easily checked to be equivalent to
\eqnn\PSanomequiv
$$\eqalignno{
[[\cK_6(\ell)]]&={1\over 2}C^{mn}_{1|2,3,4,5,6} F^{(s)mn}_{1,2,3,4,5,6}+
\big[  C^{m}_{1|23,4,5,6} F^{(s)m}_{1,23,4,5,6}
+ (2,3 | 2,3,4,5,6)  \big]&\PSanomequiv \cr
& + \big[ \big( C_{1|234,5,6} F^{(s)}_{1,234,5,6}
+ C_{1|243,5,6} F^{(s)}_{1,243,5,6}\big)+ (2,3,4|2,3,4,5,6)  \big]\cr
& + \big[ \big( C_{1|23,45,6} F^{(s)}_{1,23,45,6}
+ {\rm cyc}(3,4,5)\big)
+ (6 \leftrightarrow 5,4,3,2)  \big]\,,
}$$
using the Lie-symmetric version \KLTF\ of the functions ${\cal F}^{m\ldots}_{A,B,\ldots}$ in \newfs.

%*********************************************************************************
\newsubsubsec\sixxEEsec Lessons from the $T\cdot E$ and $C\cdot E$ representations

Again, one can combine the above results for the loop integrals over six-point
GEIs with the $T\cdot E$ and $C\cdot E$ representations of ${\cal K}_6(\ell)$ in
\augF\ and \lowenE, respectively. Based on the loop integration
$[[E^{mn}_{1|2,\ldots}]]= -{\pi \over \Im \tau} \delta^{mn}+\ldots$ and
$[[E_{1|2|3,\ldots}]]= -{\pi \over \Im \tau}+\ldots$ in \intellA\ and \intellC,
the cancellation of the modular anomalies is transparent in both
representations: Either by the trace relation \HOb\ among local building blocks
or by the trace relation \pseudoTRA\ among pseudo-invariants.

In particular, the terms $\cK_{6}(\ell) ={1\over 2} 
C^{mn}_{1|2,\ldots}E^{mn}_{1|2,\ldots} - \big[ P_{1|2|3,\ldots}E_{1|2|3,\ldots}
+ (2\leftrightarrow 3,\ldots) \big]+\ldots$ in the $C\cdot E$ representations
\lowenE\ illustrate the duality between BRST anomalies and modular anomalies: In
the same way as the modular anomaly of $[[\cK_{6}(\ell) ]]$ cancels by the trace
relation \pseudoTRA\ between $C^{mn}_{1|2,\ldots}$ and $P_{1|2|3,\ldots}$, the
BRST anomaly localizes to a boundary term in moduli space since the GEIs
$E^{mn}_{1|2,\ldots}$ and $E_{1|2|3,\ldots}$ satisfy the dual trace relation
\traceA\ (or \newtrex\ after integration over $\ell$).

Also, note that the $C\cdot {\cal F}$ representation \PSanomequiv\ results from
straightforward regroupings of terms in the integrated $C\cdot E$
representations \lowenE: There is no need to perform integration by parts on the
$f^{(n)}_{ij}$, and the coefficients of $f^{(1)}_{1j}, \ j=2,3,4,5,6$ are easily
seen to vanish after using Fay relations and cohomology identities of section
\CJACOBIsec.

%*********************************************
\newsubsec\closedsec Closed-string correlators

One of the major motivations for chiral splitting is that closed-string
correlators are literally the square of open-string correlators before
integration over $\ell$, cf.\ \againclosed. Performing the loop integral reveals
modular invariance of the closed-string amplitude representation
\againnewamprep, at the expense of introducing interactions between left- and
right-movers. We will now illustrate these interactions based on examples up to
six points.

Most obviously, the expressions of section \intellmom\ for integrated GEIs are
augmented by additional terms involving ${\pi \over \Im \tau}$ when the
opposite-chirality sector contributes additional loop momenta, e.g.
\eqnn\EellA
$$\eqalignno{
[[\ell^n E^m_{1|2,3,4,5}]] &= -{ \pi \over \Im \tau} \delta^{mn}
+ L_0^n \big[ k^m_2 f^{(1)}_{12} + (2\leftrightarrow 3,4,5) \big]\,,&\EellA \cr
[[\ell^n E^m_{1|23,4,5,6}]] &= -{ \pi \over \Im \tau} \delta^{mn} V_1(1,2,3)
+ L_0^n \Big( k^m_3 f^{(1)}_{12} f^{(1)}_{23} +k^m_2 f^{(1)}_{13} f^{(1)}_{23}
+ k_{23}^m (f^{(2)}_{12} - f^{(2)}_{13})
\cr
&\quad{} + (k_3^m - k_2^m)f^{(2)}_{23}
+ \big[k_4^m f^{(1)}_{14}V_1(1,2,3)+(4\leftrightarrow 5,6) \big] \Big)\,, \cr
[[\ell^p \ell^q E^{mn}_{1|2,3,4,5,6}]] &=
\big({\pi \over \Im \tau} \Big)^2 \delta^{m(n} \delta^{pq)}
-  {\pi\over\Im\tau} L_0^{(p}\delta^{q)(m} \big[ k_2^{n)} f^{(1)}_{12}
+ (2\leftrightarrow 3,4,5,6) \big]\cr
&\quad{}  - {\pi\over\Im\tau}\delta^{mn} L_0^p L_0^q
+ 2 \Big( L_0^p L_0^q - {\pi\over\Im\tau}\delta^{pq} \Big)
\big[ k_2^m k_2^n f^{(2)}_{12} + (2\leftrightarrow 3,4,5,6) \big]\cr
&\quad{} + \Big( L_0^p L_0^q - {\pi\over\Im\tau}\delta^{pq} \Big)
\big[ (k_2^{m} k_3^{n}{+}k_2^{m} k_3^{n}) f^{(1)}_{12} f^{(1)}_{13} + (2,3|2,3,4,5,6)\big]\,, \cr
[[\ell^m E_{1|2|3,4,5,6}]] &= - { \pi \over \Im \tau} f^{(1)}_{12} k_2^m + L_0^m \Big( {-} {\pi\over\Im\tau}
 - 2 s_{12} f^{(2)}_{12} + f^{(1)}_{12} \big[s_{23}f^{(1)}_{23}+(3\leftrightarrow 4,5,6)\big] \Big) \cr
 &= L_0^m \Big( {-} {\pi\over\Im\tau} + \partial f_{12}^{(1)}
 + s_{12}(f^{(1)}_{12})^2  - 2 s_{12} f^{(2)}_{12} \Big)\,.
}$$
Once these additional loop momenta are regrouped into complex conjugate GEIs,
the net effect of the additional $L_0^m$ is to recombine the $\bar g^{(n)}$ functions to
\eqn\cclopF{
\bar f^{(n)}(z,\tau) \equiv \sum_{k=0}^n {(-\nu)^k \over k!}
\bar g^{(n-k)}(z,\tau) \, .
}
The minus signs relative to \lopF\ are due to $\nu_{ij} \rightarrow - \nu_{ij}$
under complex conjugation. Likewise, our normalization conventions for the loop
momentum transforms $\ell \rightarrow -\ell$ in passing from GEIs to their
complex conjugates, as reflected in the notation $\tilde {\cal K}_n(-\ell)$ for
right-moving correlators in \againnewamprep. For instance, the vectorial GEI in
$\tilde {\cal K}_5(-\ell)= \bar E^m_{1|2,3,4,5} \tilde C^m_{1|2,3,4,5}+\ldots$
reads $\bar E^m_{1|2,3,4,5}= -\ell + \big[ k_2^m \bar
g^{(1)}_{12}+(2\leftrightarrow 3,4,5) \big]$, and the loop integral of its
holomorphic square can be performed via \EellA,
\eqn\exnonamon{
[[E^m_{1|2,3,4,5} \bar E^n_{1|2,3,4,5}]] =  { \pi \over \Im \tau} \delta^{mn}+
\big[ k^m_2 f^{(1)}_{12} + (2\leftrightarrow 3,4,5) \big]
\big[ k^n_2\bar f^{(1)}_{12} + (2\leftrightarrow 3,4,5) \big] \ .
}
The first term exemplifies that factors of ${\pi \over \Im \tau}$ are not
necessarily associated with modular anomalies in a closed-string setup: Both
${\pi \over \Im \tau}$ and the remaining terms $f^{(1)}_{ij}\bar f^{(1)}_{kl}$
in \exnonamon\ have modular weights $(1,1)$, in lines with modular invariance of
the five-point amplitude \againnewamprep. In fact, the cancellation of modular
anomalies in integrated open-string six-point correlators applies separately to
both chiral halves of the closed-string calculation.

%**************************************
\newsubsubsec\closedfivesec Five points

Starting from the $T\cdot {\cal Z}$ representation \fivecorr\ of the
open-string five-point correlator, loop integration over its holomorphic square yields
\eqnn\fiveclloc
$$\eqalignno{
[[{\cal K}_5(\ell) \tilde {\cal K}_5(-\ell) ]] &= \Big|
\big[ V_{12}T_{3,4,5}  {\cal F}_{12,3,4,5}{+} (2\leftrightarrow 3,4,5) \big]
+ \big[ V_1T_{23,4,5}  {\cal F}_{1,23,4,5}{+} (2,3|2,3,4,5) \big] \Big|^2 \cr
& +{\pi  \over \Im \tau}  V_1 T^m_{2,3,4,5} \tilde V_1 \tilde T^m_{2,3,4,5} \, . &\fiveclloc
}$$
The second line augments the square of the integrated open-string correlator in its $T\cdot {\cal F}$
representation \singfivef\ by a left-right contraction. The recombination of $g^{(1)}_{ij}+\nu_{ij}=f^{(1)}_{ij}$
and $\bar g^{(1)}_{ij}-\nu_{ij}= \bar f^{(1)}_{ij}$ follows the mechanism of the open-string context, see \intfive.

The local form \fiveclloc\ of the five-point closed-string correlator has been spelled
out in \oneloopMichael. As already emphasized in the reference, integrations by
parts \newzderivzero\ are more subtle in presence of both $f^{(n)}_{ij}$
and $\bar f^{(n)}_{ij}$: Additional terms ${\pi \over \Im \tau}$ may arise in trading
$s_{12} f^{(1)}_{12}$ for $s_{23} f^{(1)}_{23}+(3\leftrightarrow 4,5)$ on the left-moving
side, depending on the labels of the accompanying right-moving $\bar f^{(1)}_{ij}$,
see e.g.\ \exLR. Hence, one cannot just replace the left-moving terms in the first line of
\fiveclloc\ by their manifestly BRST-invariant counterparts \loopK\ without inspecting
the respective right-movers and altering the coefficient of ${\pi  \over \Im \tau}$.

Instead, a manifestly BRST-invariant rewriting of \fiveclloc\ can be conveniently
found by integrating the $C\cdot {\cal Z}$ representation of ${\cal K}_5(\ell) \tilde {\cal K}_5(-\ell)$,
\eqnn\fivecl
$$\eqalignno{
[[{\cal K}_5(\ell) \tilde {\cal K}_5(-\ell) ]] &= \big| s_{23}f^{(1)}_{23} C_{1|23,4,5} {+} (2,3|2,3,4,5) \big|^2 +
{\pi  \over \Im \tau}  C^m_{1|2,3,4,5} \tilde C^m_{1|2,3,4,5} \, . &\fivecl
}$$
This representation has been firstly given in \MafraNWR, based on a long sequence of
integration-by-parts identities in \fiveclloc\ and carefully tracking all $\partial_i \bar f^{(1)}_{ij}
= \bar \partial_i f^{(1)}_{ij} = -{\pi \over \Im \tau}$ \oneloopMichael.

%************************************
\newsubsubsec\closedsixsec Six points

A manifestly BRST invariant closed-string six-point correlator has been proposed in \MafraNWR
\eqnn\closedZ
$$\eqalignno{
[[{\cal K}_6(\ell) \tilde {\cal K}_6(-\ell) ]] &= {\cal K}^{\rm open}_6 \tilde {\cal K}^{\rm open}_6
+ { \pi \over \Im\tau}   \big| s_{23}f^{(1)}_{23} C^m_{1|23,4,5,6} + (2,3|2,3,4,5,6) \big|^2 &\closedZ \cr
&\! \! \! +\Big( {\pi \over \Im \tau} \Big)^2 \Big( {1\over 2}  C^{mn}_{1|2,3,4,5,6} \tilde C^{mn}_{1|2,3,4,5,6}
 -   \big[  | P_{1|2|3,4,5,6}  |^2 + (2\leftrightarrow 3,4,5,6) \big]  \Big)\, ,
}$$
where ${\cal K}^{\rm open}_6$ is essentially the representation of $[[{\cal K}_6(\ell)  ]]$ given in \PSanom,
\eqnn\PSanomagain
$$\eqalignno{
{\cal K}^{\rm open}_6&=
\big[ \big( s_{23}s_{34} f^{(1)}_{23}  f^{(1)}_{34}C_{1|234,5,6} + {\rm cyc}(2,3,4)\big)
+ (2,3,4 | 2,3,4,5,6)  \big]&\PSanomagain\cr
& + \big[ \big(s_{23}s_{45} f^{(1)}_{23} f^{(1)}_{45} C_{1|23,45,6}
+ {\rm cyc}(3,4,5)\big)
+ (6 \leftrightarrow 5,4,3,2)  \big]\cr
&+  \big[ f^{(2)}_{12} C_{1|2|3,4,5,6} + (2\leftrightarrow 3,4,5,6) \big]
+ \big[f^{(2)}_{23} C_{1|(23)|4,5,6} + (2,3|2,3,4,5,6) \big]\,,
}$$
up to Koba--Nielsen derivatives to be detailed below. The pseudo-invariants $
C_{1|2|3,4,5,6}$ and $C_{1|(23)|4,5,6}$ have been defined in \Ponetwo. The
second line of \closedZ\ has not yet been derived from first principles but was
inferred by indirect arguments including properties of the low-energy limit
\MafraNWR. In appendix \prfsixclosed, we will demonstrate the terms $|
P_{1|2|3,4,5,6} |^2$ in \closedZ\ to follow from a careful analysis of
integration-by-parts identities.

Our derivation of \closedZ\ starts from the $C\cdot E$ representation \lowenE\ of ${\cal K}_6(\ell) $
and a convenient organization of the loop integrals in the closed-string case according to the number of contractions
$\ell^m \ell^n \rightarrow -{\pi \over \Im \tau}$ between left- and right-movers
\eqnn\prfeqA
$$\eqalignno{
[[{\cal K}_6(\ell) \tilde {\cal K}_6(-\ell) ]] &= [[{\cal K}_6(\ell)]] \cdot [[ \tilde {\cal K}_6(-\ell) ]]  &\prfeqA \cr
&+ {\pi   \over \Im \tau} \Big[\Big[{\partial {\cal K}_6(\ell) \over \partial \ell_m} \Big] \Big]   \delta_{mn}   \Big[\Big[ {\partial \tilde {\cal K}_6(-\ell) \over \partial(-\ell_n)} \Big] \Big]  \cr
&+{1\over 2} \Big( {\pi \over \Im \tau} \Big)^2 C^{mn}_{1|2,3,4,5,6} \tilde C^{mn}_{1|2,3,4,5,6} \, .
}$$
The double contractions between left- and right movers are sensitive to no contribution
to other than $ {\cal K}_6(\ell) = {1\over 2} \ell_m \ell_n C^{mn}_{1|2,3,4,5,6}+\ldots$ and lead to
the last line. For the vectorial open-string constituents of \prfeqA,
the representation \lowenE\ gives rise to
\eqnn\prfeqCshort
$$\eqalignno{
 \Big[\Big[{\partial {\cal K}_6(\ell) \over \partial \ell_m} \Big] \Big]  &=
 \big[ C^m_{1|23,4,5,6} s_{23} f_{23}^{(1)} + (2,3|2,3,4,5,6) \big]\cr
& + \big[ P_{1|2|3,4,5,6} k_2^m \nu_{12} + (2 \leftrightarrow 3,4,5,6)\big]\,,&\prfeqCshort
}$$
see appendix \prfsixclosed\ for intermediate steps. Finally, the scalar contributions to \prfeqA
 \eqnn\prfeqE
$$\eqalignno{
[[{\cal K}_6(\ell) ]] &=
{\cal K}^{\rm open}_6 - \big[ N_{1|2|3,4,5,6} P_{1|2|3,4,5,6} + (2\leftrightarrow 3,4,5,6) \big] &\prfeqE
}$$
augment \PSanomagain\ by total derivatives
 \eqnn\prfeqF
$$\eqalignno{
N_{1|2|3,4,5,6} \hat {\cal I}_6&= \Big( {\pi \over \Im \tau}
+ \nu_{12} (s_{12}f^{(1)}_{12}
- \big[ s_{23}f^{(1)}_{23}{+}(3{\leftrightarrow} 4,5,6) \big]) \Big)
\hat {\cal I}_6 =- {\partial \over \partial z_2}( \nu_{12}\hat {\cal I}_6)\cr
\tilde N_{1|2|3,4,5,6} \hat {\cal I}_6&=  \Big( {\pi \over \Im \tau}
- \nu_{12} (s_{12}\bar f^{(1)}_{12}
-\big[ s_{23} \bar f^{(1)}_{23}{+}(3{\leftrightarrow} 4,5,6) \big])
 \Big) \hat {\cal I}_6 ={\partial \over \partial \bar z_2}( \nu_{12}  \hat {\cal I}_6 ) &\prfeqF
}$$
that have been dropped in the open-string context of \PSanom, see \intellC. In
the present closed-string context, however, the quantity ${\cal K}^{\rm open}_6$
in \PSanomagain\ cannot be replaced by integration-by-parts equivalent
representations of $[[{\cal K}_6(\ell)]]$, say the local expression in \sixLieF.
The factors of $\bar f^{(w)}_{ij}$ in the accompanying $\tilde {\cal K}^{\rm
open}_6$ in \closedZ\ are affected by holomorphic total derivatives via
\ccnonhol.

Upon insertion into \prfeqA, the first term of \prfeqCshort\ and the holomorphic
square of ${\cal K}^{\rm open}_6$ in \prfeqE\ explain the first line of the
final result \closedZ. The second line of \closedZ, however, arises from the
factors of $\nu_{1j}$ in \prfeqCshort\ and \prfeqF\ through a sequence of
integrations by parts, see appendix \prfsixclosed\ for details. Note that the
modular anomalies of \prfeqA\ cancel separately in both $[[{\cal K}_6(\ell)]]$
and $[[ \tilde {\cal K}_6(-\ell) ]]$, following the mechanisms of section
\sixxEEsec. The BRST anomaly of \closedZ\ was shown in \MafraNWR\ to yield a
boundary term in $\tau$, based on a special case of \newtauderivzero.

%**********************************************
\newsubsubsec\closedhighsec Higher multiplicity

The organization of the closed-string loop integration in the six-point example \prfeqA\
readily generalizes to higher multiplicity. The left-right contractions in the seven-point
correlator can be captured via
\eqnn\prfeqAseven
$$\eqalignno{
&[[{\cal K}_7(\ell) \tilde {\cal K}_7(-\ell) ]] =
[[{\cal K}_7(\ell)]] \cdot [[ \tilde {\cal K}_7(-\ell) ]]
+ {\pi \over\Im\tau}  \Big[\Big[{\p{\cal K}_7(\ell)\over\p\ell_m} \Big] \Big]
\delta^{mn}\Big[\Big[{\p\tilde {\cal K}_7(-\ell)\over\p(-\ell_n)} \Big] \Big] &\prfeqAseven \cr
&  + {1\over 2!}\Big( {\pi \over \Im \tau} \Big)^2
\Big[\Big[{\p^2 {\cal K}_7(\ell) \over\p\ell_m \p\ell_n}\Big]\Big]
\delta^{mp} \delta^{nq}\Big[\Big[ {\p^2 \tilde {\cal K}_7(-\ell) \over\p(-\ell_p)\p(-\ell_q)}\Big]\Big]\cr
& +{1\over 3!} \Big( {\pi \over \Im \tau} \Big)^3
\Big[\Big[{\p^3 {\cal K}_7(\ell) \over \p \ell_m \p \ell_n \p \ell_p} \Big] \Big]
\delta^{mq} \delta^{nr}\delta^{ps}\Big[\Big[ {\p^3 \tilde {\cal K}_7(-\ell) \over \p(-\ell_q)\p(-\ell_r)\p(-\ell_s)} \Big] \Big]
 \, ,
}$$
where the last line evaluates to ${1\over 3!} ( {\pi \over \Im \tau} )^3
C^{mnp}_{1|2,\ldots} \tilde C^{mnp}_{1|2,\ldots}$ when the $C\cdot E$
representation \sevenCE\ of ${\cal K}_7(\ell)$ is used.  By adapting the
techniques of appendix \prfsixclosed, it should be possible to bring
\prfeqAseven\ into a form similar to \closedZ\ where all the $g^{(n)}_{ij}$ and
$\bar g^{(n)}_{ij}$ are completed to $f^{(n)}_{ij}$ and $\bar f^{(n)}_{ij}$
through the loop integration. The most laborious part of this calculation might
be to identify the seven-point generalization of the terms like $-( {\pi \over
\Im \tau} )^2 |P_{1|2|3,\ldots}|^2$ in \closedZ.

Note that the all-multiplicity generalization of \prfeqAseven\ reads
\eqnn\prfeqAn
$$\eqalignno{
&[[{\cal K}_n(\ell) \tilde {\cal K}_n(-\ell) ]] =
\sum_{r=0}^{n-4} {1\over r!} \Big( {\pi \over \Im \tau} \Big)^{r}
\Big[\Big[{\p^r {\cal K}_r(\ell) \over\p\ell_{m_1}\p\ell_{m_2} \ldots \p\ell_{m_{r}}}\Big]\Big]
 &\prfeqAn \cr
& \ \times \delta^{m_1 p_1} \delta^{m_2 p_2} \ldots \delta^{m_r p_r}
\Big[\Big[ {\p^r \tilde {\cal K}_n(-\ell) \over \p(-\ell_{p_1})\p(-\ell_{p_2})\ldots
\p(-\ell_{p_r})} \Big] \Big] \, .
}$$

%*********************************************************************************
\newsubsec\closedLE Closed-string low-energy limits versus open-string correlators

The one-loop low-energy effective action\foot{See \GreenTV\ for the exact
coefficient of the $R^4$ operator in the type-IIB effective action, including
all perturbative and non-perturbative contributions.} of type-IIB and type-IIA
superstrings features a supersymmetrized higher-curvature operator\foot{While
the $R^4$ operators in the tree-level effective action of the type-IIA and
type-IIB theories are identical, at one loop they differ
by a contribution proportional to $\epsilon_{10}\epsilon_{10} R^4$
\refs{\GrisaruPX, \GrisaruVI}.  As detailed in
\oneloopMichael, the type-IIB matrix elements of this section are proportional
to the $\alpha'^3 \zeta_3$-order of the respective tree amplitudes, where the
proportionality constant depends on the R-symmetry charge of the components
(say gravitons or dilatons).} $R^4$ at its leading order in $\alpha'$
\GreenSW.  Hence, the low-energy limit of one-loop closed-string amplitudes
yields matrix elements with a single insertion of a supersymmetrized $R^4$
operator. By inspection of their $(n\leq 7)$-point examples, these matrix
elements will be shown to relate to open-string correlators by the duality
between pseudo-invariants and GEIs \MafraIOJ.

%****************************************
\newsubsubsec\closedLEaa Up to six points

Once the $z_j$-dependence of closed-string correlators $[[{\cal K}_n(\ell)
\tilde {\cal K}_n(-\ell) ]]$ is expressed in terms of $f^{(n)}_{ij}$ and $\bar
f^{(n)}_{ij}$, their low-energy limit can be conveniently extracted through
the techniques of \refs{\Richards, \oneloopMichael}. The idea is to perform
the $\alpha'$-expansion of the integrals in \againnewamprep\ over the punctures
while keeping $\tau$ finite\foot{This approach yields a power series in
$\alpha'$ that is tailored to infer the one-loop low-energy effective action.
The branch cuts of the overall amplitude due to the $\tau \rightarrow i
\infty$ limit of the moduli-space integral are disentangled when integrating
over the $z_j$ at fixed values $\tau$. Still, in analyzing effective
interactions beyond the low-energy limit, a subtle interplay between the
branch cuts and the power-series part has to be taken into account
\refs{\GreenUJ, \DHokerGMR}.}
in this process. Then, the leftover $\tau$-integration at the leading order
in $\alpha'$ straightforwardly yields the volume ${\pi \over 3}$ of moduli space.

In this setup, the representations \fivecl\ and \closedZ\ of the closed-string
correlators are tailored to extract the following low-energy limits \MafraNWR\
\eqnn\lowenA
$$\eqalignno{
{\cal M}_4^{R^4} &= C_{1|2,3,4}\tilde C_{1|2,3,4}  &\lowenA \cr
{\cal M}_5^{R^4} &= C^m_{1|2,3,4,5}\tilde C^m_{1|2,3,4,5}
+ \big[C_{1|23,4,5}\,s_{23}\tilde C_{1|23,4,5}
+ (2,3|2,3,4,5) \big] \cr
{\cal M}_6^{R^4} &= {1\over 2}  C^{mn}_{1|2,3,4,5,6} \tilde C^{mn}_{1|2,3,4,5,6}
 -   \big[   P_{1|2|3,4,5,6} \tilde P_{1|2|3,4,5,6}   + (2\leftrightarrow 3,4,5,6) \big] \cr
&{} + \big[  C^m_{1|23,4,5,6}\, s_{23}\tilde C^m_{1|23,4,5,6}
+ (2,3|2,3,4,5,6) \big] \cr
&{} + \big[ \big(C_{1|23,45,6}\,s_{23} s_{45} \tilde C_{1|23,45,6} + {\rm
cyc}(3,4,5)\big)+(6\leftrightarrow5,4,3,2)] \cr
&{} + \big[\big(C_{1|234,5,6}\,s_{23} s_{34} \tilde C_{1|234,5,6} + {\rm cyc}(2,3,4)\big)
 + (2,3,4|2,3,4,5,6)\big] \, ,
}$$
also see \GreenSW\ and \refs{\Richards, \oneloopMichael} for earlier discussions
of the four- and five-point examples.

As highlighted in \MafraIOJ, the expressions \lowenA\ for the matrix elements
of $R^4$ are related to open-string correlators ${\cal K}_n(\ell)$ by the
duality between kinematics and worldsheet functions. More precisely, the above
${\cal M}_n^{R^4}$ can be mapped to the $C\cdot E$ representations of the
open-string correlators by trading the right-moving pseudo-invariants for the
GEIs with the same slot structure \MafraIOJ,
\eqn\thecorresp{
{\cal K}_n(\ell) = {\cal M}^{R^4}_n \,\big|_{\tilde C,\tilde P \rightarrow E} \, .
}
This can be checked from the formal rewriting
${\cal K}_4(\ell) = C_{1|2,3,4} E_{1|2,3,4}$ of \fourptcorr\ and as well as the expressions
\covK\ and \lowenE\ for ${\cal K}_5(\ell)$ and ${\cal K}_6(\ell)$, respectively.

In the same way as open-string correlators admit a variety of representations,
one can rewrite the matrix elements \lowenA\ such as to manifest their
locality properties. The idea is to aim for a kinematic analogue of the
$T\cdot E$ representations ${\cal K}_4(\ell) = V_1T_{2,3,4} E_{1|2,3,4}$ as
well as \fiveTE\ and \augF\ of the open-string correlators. The duality
between pseudo-invariants and GEIs translates these manifestly local
representations of ${\cal K}_n(\ell)$ into
\eqnn\lowenloc
$$\eqalignno{
{\cal M}_4^{R^4} &= V_{1} T_{2,3,4}\tilde C_{1|2,3,4}  &\lowenloc \cr
{\cal M}_5^{R^4} &= V_1 T^m_{2,3,4,5}\tilde C^m_{1|2,3,4,5}
+ V_1\big[ T_{23,4,5}\,\tilde C_{1|23,4,5}
+ (2,3|2,3,4,5) \big] \cr
{\cal M}_6^{R^4} &= {1\over 2}  V_1 T^{mn}_{2,3,4,5,6} \tilde C^{mn}_{1|2,3,4,5,6}
 - V_1  \big[  J_{2|3,4,5,6} \tilde P_{1|2|3,4,5,6}  + (2\leftrightarrow 3,4,5,6) \big] \cr
&{} +V_1 \big[  T^m_{23,4,5,6}\, \tilde C^m_{1|23,4,5,6}
+ (2,3|2,3,4,5,6) \big] \cr
&{} +V_1 \big[ \big(T_{23,45,6}\, \tilde C_{1|23,45,6} + {\rm
cyc}(3,4,5)\big)+(6\leftrightarrow5,4,3,2)] \cr
&{} + V_1 \big[ T_{234,5,6}\, \tilde C_{1|234,5,6} +T_{243,5,6}\, \tilde C_{1|243,5,6}
 + (2,3,4|2,3,4,5,6)\big] \, .
}$$
These representations of the matrix elements are tailored to connect with the
Feynman diagrams of the effective action proportional to $R+R^4$: All the propagators
stem from the right-moving pseudo-invariants whose expansion in terms of
Berends--Giele currents is reviewed in section \BRSTpseudosec. These
Berends--Giele constituents manifest that each term in \lowenloc\ has at most
$n{-}4$ propagators, reflecting at least one vertex of valence $\geq 4$ in
each diagram.

The equivalence of \lowenA\ and \lowenloc\ can be checked without any further
calculation by exploiting the duality between kinematics and worldsheet
functions: Given that ${\cal K}_n(\ell)$ and ${\cal M}_4^{R^4}$ are related by
exchange of pseudo-invariants and GEIs, the manipulations that connect the
$T\cdot E$ and $C\cdot E$ representations of the correlators apply in
identical form to the matrix elements of $R^4$. This follows from the
observations of section \thissubsec\ that all the integration-by-parts
identities among GEIs at $n\leq 6$ points have a counterpart in the BRST
cohomology, relating pseudo-invariants of different tensor rank.

In summary, the low-energy limit of closed-string one-loop amplitudes results
in supersymmetrized matrix elements of $R^4$ that share the structure of
open-string correlators, cf.\ \thecorresp. Like this, the duality between
kinematics and worldsheet functions connects the representations \lowenA\ and
\lowenloc\ and implies that the matrix elements are both local and BRST
invariant.

%************************************
\newsubsubsec\closedLEbb Seven points

As explained in section \closedhighsec, the low-energy limit of the
closed-string seven-point amplitude may not be readily available from the
expression \prfeqAseven\ for the loop-integrated correlator. Still, it is
tempting to invoke the connection between matrix elements of $R^4$ and
open-string correlators to propose a candidate expression on the basis
of the $C\cdot E$ representation \sevenCE\ of ${\cal K}_7(\ell)$:
\eqnn\sevenCLE
$$\eqalignno{
{\cal M}_{7}^{R^4} &= {1\over6}C^{mnp}_{1|2,3,4,5,6,7}\tilde C^{(s)mnp}_{1|2,3,4,5,6,7}&\sevenCLE\cr
&+\half C^{mn}_{1|23,4,5,6,7} \tilde C^{(s)mn}_{1|23,4,5,6,7} + (2,3|2,3,4,5,6,7)\cr
&+\big[C^{m}_{1|234,5,6,7} \tilde C^{(s)m}_{1|234,5,6,7}
+ C^{m}_{1|243,5,6,7} \tilde C^{(s)m}_{1|243,5,6,7} \big]+ (2,3,4|2,3,4,5,6,7)\cr
%&+\big[C^{m}_{1|23,45,6,7} E^{(s)m}_{1|23,45,6,7} + (4,5|4,5,6,7)\big] + (2,3|2,3,4,5,6,7)\cr
&+ \big[C^{m}_{1|23,45,6,7} \tilde C^{(s)m}_{1|23,45,6,7} + {\rm cyc}(2,3,4)\big]+(6,7|2,3,4,5,6,7)\cr
&+\big[C_{1|2345,6,7} \tilde C^{(s)}_{1|2345,6,7} + {\rm perm}(3,4,5)\big]+(2,3,4,5|2,3,4,5,6,7)\cr
&+\big[C_{1|234,56,7} \tilde C^{(s)}_{1|234,56,7} + C_{1|243,56,7}  \tilde C^{(s)}_{1|243,56,7} + {\rm cyc}(5,6,7)\big]
+ (2,3,4|2,3,4,5,6,7)\cr
&+\big[ C_{1|23,45,67} \tilde C^{(s)}_{1|23,45,67} + {\rm cyc}(4,5,6) \big]+ (3\leftrightarrow 4,5,6,7)\cr
&- P^m_{1|2|3,4,5,6,7} \tilde P^{(s)m}_{1|2|3,4,5,6,7} + (2\leftrightarrow3,4,5,6,7)\cr
&- P_{1|23|4, 5,6,7} \tilde P^{(s)}_{1|23|4,5,6,7}+(2,3|2,3,4,5,6,7)\cr
&- \big[P_{1|2|34,5,6,7} \tilde P^{(s)}_{1|2|34,5,6,7}+{\rm cyc}(2,3,4)\big]+(2,3,4|2,3,4,5,6,7)\,.
}$$
The superscripts $^{(s)}$ of the right-moving pseudo-invariants instruct to
perform the matrix multiplications with $S(A|A')_a$ as in the definitions
\cZToZ\ and \KLTE\ of $Z^{(s)}$ and $E^{(s)}$, e.g.\ $\tilde
C^{(s)}_{1|23,45,67}=s_{23}s_{45} s_{67} \tilde C_{1|23,45,67}$. 

In order to validate the proposal \sevenCLE, we shall verify that the
BRST invariant expression is at the same time compatible with the
locality properties of an $R^4$ matrix element. Since the seven-point
pseudo-invariants obey the relations of the dual GEIs up to the anomalous
$\Delta^{\ldots}_{1|\ldots}$ superfields, cf.\ section \EjacC, we can
apply the manipulations of the correlator ${\cal K}_7(\ell)$ to the above
expression for ${\cal M}_{7}^{R^4}$. In the same way as integration-by-parts
relations among GEIs yield the $T\cdot E$ representation \augH\ for ${\cal
K}_7(\ell)$, \sevenCLE\ must be equivalent to
\eqnn\augHLE
$$\eqalignno{
{\cal M}_7^{R^4} &= {1\over 6} V_1T^{mnp}_{2,3,4,5,6,7}  \tilde C^{mnp}_{1|2,3,4,5,6,7} &\augHLE \cr
&+ {1\over 2}V_1T^{mn}_{23,4,5,6,7} \tilde C^{mn}_{1|23,4,5,6,7} + (2,3|2,3,4,5,6,7)\cr
& +\big[ V_1T^m_{234,5,6,7} \tilde C^m_{1|234,5,6,7} +V_1T^m_{243,5,6,7} \tilde C^m_{1|243,5,6,7} \big]+
(2,3,4|2,3,4,5,6,7) \cr
%
%& +\big[V_1T^m_{23,45,6,7} E^m_{1|23,45,6,7} + (4,5|4,5,6,7)\big] + (2,3|2,3,4,5,6,7)\cr
&+ \big[V_1 T^{m}_{23,45,6,7} \tilde C^{m}_{1|23,45,6,7} + {\rm cyc}(2,3,4)\big]+(6,7|2,3,4,5,6,7)\cr
& +\big[V_1T_{2345,6,7} \tilde C_{1|2345,6,7} + {\rm perm}(3,4,5)\big] +(2,3,4,5|2,3,4,5)\cr
& + \big[V_1T_{234,56,7} \tilde C_{1|234,56,7} + V_1T_{243,56,7} \tilde C_{1|243,56,7}
+ {\rm cyc}(5,6,7)\big] + (2,3,4|2,3,4,5,6,7)\cr
&+ \big[ V_1T_{23,45,67} \tilde C_{1|23,45,67} +{\rm cyc}(4,5,6) \big]+ (3\leftrightarrow 4,5,6,7)\cr
& -  V_1J^m_{2|3,4,5,6,7} \tilde P^m_{1|2|3,4,5,6,7} +(2\leftrightarrow3,4,5,6,7) \cr
&- V_1J_{23|4,5,6,7} \tilde P_{1|23|4,5,6,7}  + (2,3|2,3,4,5,6,7) \cr
&- \big[ V_1J_{2|34,5,6,7} \tilde P_{1|2|34,5,6,7}+{\rm cyc}(2,3,4)\big]+(2,3,4|2,3,4,5,6,7)\,,
}$$
and we have made a separate check that the BRST non-exact
$\Delta_{1|2|3,\ldots}$ are absent. Given that the candidate expression
\sevenCLE\ is both BRST invariant and local, we expect it to match with the
seven-point matrix element of $R^4$. This corroborates the correspondence
\thecorresp\ up to multiplicity seven \MafraIOJ.

%************************************
\newsubsubsec\closedLEcc Eight points

At eight points, the analysis of section \EightPointsec\ led to obstacles in
constructing a BRST-invariant and local open-string correlator from the
methods of this work. A closely related problem is the availability of the
holomorphic Eisenstein series ${\rm G}_4$ as a deformation of eight-point
GEIs. Any addition of ${\rm G}_4$ is compatible with the defining property
\MonFinv\ of GEIs and the desired modular weight four upon loop integration.
While the construction of eight-point GEIs subject to trace relation is left
for the future, we shall propose an eight-point candidate for ${\cal
M}_8^{R^4}$,
\eqnn\pseudoEllipticEight
$$\eqalignno{
 {\cal M}_8^{R^4}&=
\sum_{r=0}^4 {1\over r!} C^{m_1 \ldots m_r}_{1|A_1, \ldots,A_{r+3}}
\tilde C^{(s)\,m_1 \ldots m_r}_{1|A_1, \ldots,A_{r+3}}
+ \big[2345678|A_1, \ldots,A_{r+3}\big]&\pseudoEllipticEight\cr
%%%
& \! \! \! \! \! \! \! \! \! \! \! \! \! \! -\sum_{r=0}^2 {1\over r!}  \big[P^{m_1 \ldots m_r}_{1|A_1|A_2, \ldots,A_{r+5}}
\tilde P^{(s)\,m_1 \ldots m_r}_{1|A_1|A_2 \ldots,A_{r+5}}
+ (A_1\leftrightarrow A_2, \ldots,A_{r+5})\big]
+ \big[2 3\ldots 8|A_1, \ldots,A_{r+5}\big]\cr
%%%
&\! \! \! \! \! \! \!  \! \! \! \! \! \! \! + \big[  P_{1|2,3|4,5,6,7,8} \tilde P_{1|2,3|4,5,6,7,8} + (2,3|2,3,4,5,6,7,8)\big] \,,
}$$
see section \secexperC\ and appendix \stirlingapp\ for the notation $[2345678|A_1,
\ldots,A_{j}]$. The BRST variation of \pseudoEllipticEight\ vanishes by the
trace relations \pseudoTRA\ and \pseudoTRB\ among the pseudo-invariants, and
we expect it to be equivalent to the following local representation,
\eqnn\pseudolocalEight
$$\eqalignno{
 {\cal M}_8^{R^4}&=
\sum_{r=0}^4 {1\over r!}  V_1 T^{m_1 \ldots m_r}_{A_1, \ldots,A_{r+3}} \tilde C^{m_1 \ldots m_r}_{1|A_1, \ldots,A_{r+3}}
+ \big[2345678|A_1, \ldots,A_{r+3}\big]&\pseudolocalEight\cr
%%%
&\! \! \! \! \! \! \! \! \! \! \! \! \! \! -\sum_{r=0}^2 {1\over r!}  V_1\big[J^{m_1 \ldots m_r}_{A_1|A_2, \ldots,A_{r+5}} \tilde P^{m_1 \ldots m_r}_{1|A_1|A_2 \ldots,A_{r+5}}
+ (A_1\leftrightarrow A_2, \ldots,A_{r+5})\big]
+ \big[2 3\ldots 8|A_1, \ldots,A_{r+5}\big]\cr
%%%
&\! \! \! \! \! \! \! \! \! \! \! \! \! \! +V_1 \big[ J_{2,3|4,5,6,7,8} \tilde P_{1|2,3|4,5,6,7,8}+ (2,3|2,3,4,5,6,7,8)
\big] \,.
}$$
By the cohomology identities and trace relations of the right-moving
pseudo-invariants, also the local representation \pseudolocalEight\ is BRST
invariant, see the detailed argument below.  And since all the left-moving
superspace building blocks of \pseudolocalEight\ appear in
\pseudoEllipticEight\ with the same right-moving coefficient, the two
expressions should be equivalent.

Since the trace relations \traceeightA\ among eight-point GEIs exhibit the
inhomogeneities proportional to ${\rm G}_4$, the duality between kinematics and
worldsheet functions does not generate any BRST-invariant candidate
correlators from \pseudolocalEight. But it is encouraging to see that it is
only the constant $ {\rm G}_4$ and none of the vast set of $z_j$- and
$\ell$-dependent eight-point $\cZ$-functions or GEIs that obstructs the
construction of local and BRST-invariant correlators.

%*******************************************
\newsubsubsec\closedLEdd Higher multiplicity

In order to obtain a more general perspective on its BRST invariance, we note 
that \pseudolocalEight\ is closely related to the Lie-series contributions ${\cal K}_8^{\rm Lie}(\ell)$ 
of \experK: By replacing the $\cZ$-functions in \ddGeneral\ according to
\eqnn\replZs
$$\eqalignno{
{\cal Z}^{m_1\ldots m_r}_{1A,B_{1},B_2,\ldots,B_{r+3}} &\rightarrow \delta_{A,\emptyset} 
\tilde C^{m_1\ldots m_r}_{1|B_1,B_2,\ldots ,B_{r+3}} &\replZs \cr
{\cal Z}^{m_1\ldots m_r}_{B_1,\ldots ,B_d|1A,B_{d+1},\ldots,B_{r+d+3}} &\rightarrow \delta_{A,\emptyset} 
\tilde P^{m_1\ldots m_r}_{1|B_1,\ldots ,B_d|B_{d+1},\ldots,B_{r+d+3}}
}$$
such that all of the accompanying $V_{1A}$ with $A\neq \emptyset$ are set to zero, we recover 
\pseudolocalEight\ from ${\cal K}_8^{\rm Lie}(\ell)$. Then, the coefficients
$\Theta^{(d)}$ and $\Xi^{(d)}$ in  \ThetaSix\ and \XiSix\ of the ghost-number
four superfields in $Q{\cal K}_8^{\rm Lie}(\ell)$ are mapped to 
\eqnn\ThetaSixmp
$$\eqalignno{
\Theta^{(0)\,m_1 m_2\ldots m_r}_{A|1,B_1,\ldots,B_{r+3}}&\rightarrow
k_A^p \tilde C^{pm_1 m_2\ldots m_r}_{1|A,B_1,B_2,\ldots,B_{r+3}} + \big[
\tilde C^{m_1 m_2\ldots m_r}_{1|S[A,B_1],B_2,\ldots,B_{r+3}}
+(B_1\leftrightarrow B_2,\ldots,B_{r+3})\big]\cr
&- k_A^{(m_1}\tilde P^{m_2 \ldots m_r)}_{1|A|B_1, \ldots,B_{r+3}}
- \sum_{A=XY}\big(\tilde P^{m_1 m_2 \ldots m_r}_{1|X|Y,B_1, \ldots,B_{r+3}}
- (X\leftrightarrow Y)\big)\,, \cr
\Theta^{(1)\,m_1 m_2\ldots m_r}_{A|B|1,B_1,\ldots,B_{r+4}}&\rightarrow
- k_A^p \tilde P^{pm_1 \ldots m_r}_{1|B|A,B_1, \ldots,B_{r+4}}
- \tilde P^{m_1 \ldots m_r}_{1|S[A,B]|B_1, \ldots,B_{r+4}}  &\ThetaSixmp \cr
&- \big[\tilde P^{m_1 \ldots m_r}_{1|B|S[A,B_1],B_2, \ldots,B_{r+4}}
+ (B_1\leftrightarrow B_2,\ldots,B_{r+4})\big]\cr
& + k_A^{(m_1}\tilde P^{m_2 \ldots m_r)}_{1|A,B|B_1, \ldots,B_{r+4}}
+ \sum_{A=XY}\big(\tilde P^{m_1 m_2 \ldots m_r}_{1|X,B|Y,B_1, \ldots,B_{r+4}}
- (X\leftrightarrow Y)\big)\,,
}$$
as well as
\eqnn\XiSixmp
$$\eqalignno{
\Xi^{(0)\,m_1 m_2\ldots m_r}_{1|B_1,\ldots,B_{r+5}}&\rightarrow
-\half \tilde C^{ppm_1 \ldots m_r}_{1|B_1, \ldots,B_{r+5}}
+ \big[\tilde P^{m_1\ldots m_r}_{1|B_1|B_2, \ldots,B_{r+5}} + (B_1\leftrightarrow B_2, \ldots
B_{r+5})\big]\,,\quad{}&\XiSixmp\cr
\Xi^{(1)\,m_1 m_2\ldots m_r}_{1|A|B_1,\ldots,B_{r+6}}&\rightarrow
\half \tilde P^{ppm_1 \ldots m_r}_{1|A|B_1, \ldots,B_{r+6}}%\cr
%&\qquad{}
- \big[\tilde P^{m_1\ldots m_r}_{1|A,B_1|B_2, \ldots,B_{r+6}}
+ (B_1\leftrightarrow B_2, \ldots B_{r+6})\big]\,.\quad{}%
}$$
At multiplicity eight, \XiSixmp\ vanishes, and all instances of \ThetaSixmp\ boil down to
the anomalous $\Delta_{1|\ldots}$ superfields by the results of section \CJACOBIsec.
The BRST non-exact $\Delta_{1|\ldots}$ in turn drop out from $Q {\cal M}_8^{R^4}$ 
by the trace relations of the local superfields at ghost-number four, confirming BRST 
invariance of \pseudolocalEight. Moreover, all of $\Xi^{(0)},\Xi^{(1)}$ and $\Theta^{(0)}$
at arbitrary higher multiplicity are mapped to zero or $\Delta_{1|\ldots}$ 
under \replZs\ -- see e.g. \JRal.

%At multiplicity eight, all instances of \ThetaSixmp\ and \XiSixmp\ vanish by the results of
%section \CJACOBIsec, confirming the BRST invariance of \pseudolocalEight.
%Moreover, all of $\Theta^{(0)}$, $\Xi^{(0)}$ and $\Xi^{(1)}$
%at arbitrary higher multiplicity are mapped to zero under \replZs\ -- see e.g. \JRal.

For the images of $\Theta^{(1)}$, by contrast, only their $(n\leq 8)$-point
instances are known to reduce to $\Delta_{1|\ldots}$  in the BRST cohomology, see \deltasecF. It is
an open question whether the same is true at $n\geq 9$ points and for
generalizations $\Theta^{(d)}$ with higher refinement $d\geq 2$. 

Note, however, that the vanishing kinematic factors on the right-hand side of \XiSixmp\
are the result of translating $\cZ$-functions to kinematic factors via \replZs. At the level of
open-string correlators, i.e.\ before applying the (non-invertible) map \replZs, the $\Xi^{(d)}$ 
are generically non-zero, cf. \genThetaXi.

Up to these open questions on the pseudo-invariants, it appears likely to arrive at BRST-invariant 
and local expressions for $n$-point matrix elements of $R^4$ by applying the map \replZs\ to 
${\cal K}_n^{\rm Lie}(\ell)$. Then, the leftover task to
generate BRST-invariant and local correlators in a $T\cdot E$ representation would be
to identify a suitable system of GEIs: Such $T\cdot E$ representations of ${\cal K}_n(\ell)$
would follow from ${\cal M}_n^{R^4}$ through the duality between pseudo-invariants and 
GEIs if the latter can be made to
\medskip
\item{$\bullet$} satisfy all the trace relations dual to those of the pseudo-invariants
\item{$\bullet$} obey the analogue of the condition $\Theta^{(d)}=0$ (possibly up to 
analogues of the BRST non-exact anomaly superfields $\Delta_{1|\ldots}$, cf.\ the 
objects $G_{1|\ldots}$ in \morenotRext\ and \allJacvE),
\eqnn\dream
$$\eqalignno{
0&\cong
- k_A^p E^{pm_1 \ldots m_r}_{1|B_1,\ldots,B_d|A,C_1, \ldots,C_{r+d+3}}
- \big[  E^{m_1 \ldots m_r}_{1|S[A,B_1],B_2,\ldots,B_d|C_1, \ldots,C_{r+d+3}}
+(B_1\leftrightarrow B_2,\ldots,B_d) \big]
\cr
&\ \ \ - \big[ E^{m_1 \ldots m_r}_{1|B_1,\ldots,B_d|S[A,C_1],C_2, \ldots,C_{r+d+3}}
+ (C_1\leftrightarrow C_2,\ldots,C_{r+d+3})\big] &\dream\cr
&\ \ \ + k_A^{(m_1} E^{m_2 \ldots m_r)}_{1|A,B_1,\ldots,B_d|C_1, \ldots,C_{r+d+3}}
+ \sum_{A=XY}\big( E^{m_1 m_2 \ldots m_r}_{1|X,B_1,\ldots,B_d|Y,C_1, \ldots,C_{r+d+3}} - (X\leftrightarrow Y)\big)\,.
}$$
\medskip
\noindent
In case one succeeds in generating local and BRST invariant $T\cdot E$
representations for ${\cal K}_n(\ell)$ in this way, one would still have to find
and inspect the corresponding $T\cdot \cZ$ form: For consistency with the OPEs
among vertex operators, the correlators must admit a representation, where the
slots of the multiparticle superfields in $V_A,T^{m_1,\ldots}_{B,C,\ldots}$ and
$J^{m_1,\ldots}_{B_1,\ldots,B_d|C,\ldots}$ line up with the singularity
structure of the accompanying $g^{(1)}_{ij}$. As a drawback of the GEIs in
$T\cdot E$ representations, their slot structure does not expose the
singularities of the $g^{(1)}_{ij}$.

In spite of the large list of open questions, we are optimistic that the above ideas will
on the long run guide a path towards an explicit $n$-point open-string correlator.

%%%%%%%%%%%%%%%%%%%
\newsec Conclusions

It is appropriate to summarize the achievements and future directions
arising from this series of three papers \wipI.
We have presented a method to determine manifestly local
one-loop correlators of the pure-spinor superstring. Their dependence on the
external polarizations is organized in terms of BRST-covariant building blocks
discussed in part I.

A bootstrap procedure is introduced to assemble the
accompanying worldsheet functions from loop momenta and coefficients of the
Kronecker--Eisenstein series. As a key input of the bootstrap, the monodromies
of the worldsheet function around the $B$-cycle are taken to mirror the BRST
variations of the associated kinematic factors. This is a first example for a
multifaceted {\it duality} between kinematics and worldsheet functions 
described in part II.

The bootstrap approach results in shuffle-symmetric worldsheet functions that
conspire with the Lie symmetries of the kinematic factors: The two kinds of
ingredients combine into a Lie-polynomial structure which leads to a natural
ansatz for manifestly local $n$-point correlators. Up to six points, the Lie
polynomials are BRST-invariant by themselves and reproduce the non-local
correlators known from earlier work \MafraNWR. At seven points, the
Lie-polynomial ansatz exhibits a simple BRST variation which can be cancelled
by adding a local collection of certain {\it anomalous} superfields to the
full correlator. Starting from eight points, however, an anomalous BRST 
variation along with the holomorphic Eisenstein series ${\rm G}_4$ remains
uncancelled. Like this, we can only give an incomplete proposal for the
eight-point correlator, leaving a single kinematic factor along with ${\rm
G}_4$ undetermined. We leave it as an open problem for the future to
understand the systematics of Eisenstein series ${\rm G}_k$ in
$(n{\geq}8)$-point correlators.

Further aspects of the duality between kinematics and worldsheet functions
concern the BRST-(pseudo-)invariants obtained from certain non-local
combinations of kinematic building blocks \partI. By exporting their underlying
combinatorial pattern to the shuffle-symmetric worldsheet functions, one is
led to the notion of {\it generalized elliptic integrands} (GEIs) whose
$B$-cycle monodromies cancel upon integration over the loop momentum. GEIs are
observed to share the relations of the dual kinematic factors up to seven
points, but the preliminary definition of eight-point GEIs are found to
violate certain trace relations by a factor of ${\rm G}_4$. Hence, it remains
to incorporate ${\rm G}_k$ into $(n{\geq}8)$-point GEIs in order to realize
the duality between kinematics and worldsheet functions at all multiplicities.

We rewrite the $(n{\leq} 7)$-point correlators in terms of (pseudo-)invariants
and/or GEIs such as to manifest the respective kinds of invariances. When both
of BRST-invariance and monodromy invariance are manifested, the
(pseudo-)invariants and GEIs are found to enter on completely symmetric
footing.  This kind of exchange symmetry between kinematics and worldsheet
functions is reminiscent of the disk amplitudes of \refs{\nptString,
\BroedelTTA}, where gauge-theory trees and Parke--Taylor integrands are freely
interchangeable. Hence, the observed duality between kinematics and
worldsheet functions up to and including seven points induces a double-copy structure in
one-loop open-superstring amplitudes \MafraIOJ. In the same way as disk
amplitudes are dual to supergravity trees when replacing worldsheet integrals
by kinematics, the duality maps one-loop open-superstring amplitudes to matrix
elements of the supersymmetrized higher-curvature operator $R^4$.

The results of this work result suggest a variety of follow-up directions.

{\bf Higher genus:} Most obviously, the systems of BRST-covariant kinematic building blocks and
shuffle-symmetric worldsheet functions call for an extension to higher genus.
First instances of BRST-covariant vectorial superfields have been studied in
the low-energy regime of two-loop five-point \refs{\GomezUHA, \MafraMJA} and
three-loop four-point amplitudes \refs{\threeloop, \MafraGIA}. The principle
of BRST-covariance should guide their systematic generalizations to higher
tensor ranks as well as analogues of the refined and anomalous building blocks
of this work.

As to the worldsheet functions, one would need to identify a higher-genus
generalization of the Kronecker--Eisenstein series and its expansion
coefficients, where the elliptic functions of \TsuchiyaJOO\ may play a role.
It would be interesting to extend the duality between worldsheet functions and
kinematics -- in particular between monodromy and BRST variations -- to the
multiloop level.

{\bf Gravitational operators versus open-string correlators:}
There is an intuitive reason to find the matrix elements of $R^4$ and no other
gravitational operator as the kinematic duals to one-loop open-string
correlators: The supersymmetrized higher-curvature operator $R^4$ governs the
low-energy limit of the corresponding closed-string amplitudes. Accordingly,
the supersymmetrized matrix elements of\foot{The shorthands $D^4R^4$ and $D^6
R^4$ are understood to comprise the companion terms $D^2R^5+R^6$ and
$D^4R^5+D^2R^6+R^7$ of the same mass dimension determined by non-linear
supersymmetry.} $D^4R^4$ and $D^6 R^4$ are likely to imprint their double-copy
structure on two-loop and three-loop open-string correlators.

 In one-loop string amplitudes with reduced supersymmetry in turn, the
closed-string low-energy limit results in matrix elements of $R^2$ \GregoriHI.
Hence, the open-string one-loop correlators with half- and quarter-maximal
supersymmetry of \BergWUX\ should share the double-copy structure of $R^2$
involving GEIs similar to the ones in this work.

{\bf Field-theory limits and ambitwistors:}
The framework of chiral splitting is a natural starting point to determine
loop integrands of super-Yang--Mills and supergravity in momentum space from the
field-theory limit. We leave it to follow-up work to investigate the $\tau
\rightarrow i \infty$ degeneration of GEIs relevant to field-theory amplitudes
and the emergence of new color-kinematics dual representations.

Moreover, the superstring correlators of this work can be exported to the
one-loop amplitudes of the ambitwistor string \refs{\GeyerBJA, \GeyerJCH}. It
will then be interesting to explore the interplay of GEIs with the
color-kinematics dual field-theory amplitudes obtained from the methods of
\refs{\HeSPX, \GeyerELA}. The same questions will arise at higher genus
\refs{\GeyerWJX, \GeyerXWU}.

{\bf GEIs and scalar amplitudes:}
The double-copy structure of open-string tree amplitudes \refs{\nptString,
\BroedelTTA} motivated the interpretation of Parke--Taylor-type disk integrals
as scattering amplitudes in effective field theories of scalars. Indeed, the
low-energy limit of disk integrals reproduces the tree amplitudes of
bi-adjoint scalars with a $\phi^3$ interaction \DPellis\ and Goldstone bosons
\refs{\NLSM, \CarrascoYGV}. Similarly, higher orders in their
$\alpha'$-expansion suggest higher-mass-dimension deformations of the
respective Lagrangians collectively referred to as {\it Z-theory} \refs{\NLSM,
\MafraMCC, \CarrascoYGV}.

In one-loop string amplitudes, GEIs are found to play a role similar to the
Parke--Taylor factors at tree level. Hence, it is tempting to compare the
moduli-space integrals of GEIs with loop integrands in scalar field theories
-- for worldsheets of both toroidal and cylinder topology. Also, it will be
interesting to compare such integrated GEIs with the forward limits of
Z-theory amplitudes.

{\bf Connections with combinatorics:} After observing that several patterns and identities obeyed by
the BRST pseudo-invariants are also satisfied by the GEIs, one is left
wondering if these {\it kinematic} and {\it worldsheet-function} invariants
could be a manifestation of a more fundamental mathematical property
of objects constructed from building blocks subject to the shuffle symmetries.
After all, the combinatorics of these ``invariants'' can be generated by linear
maps acting on words that also feature prominently in the free-Lie-algebra
literature. We suspect that many combinatorial algorithms on words
have direct relevance to the study of scattering amplitudes and in particular
string-theory correlators, and that many relations among amplitudes
can be understood in terms of free-Lie-algebra structures.

\bigskip \noindent{\bf Acknowledgements:}
We are indebted to the IAS Princeton and to Nima Arkani-Hamed for kind
hospitality during an inspiring visit which initiated this project. This
research was supported by the Munich Institute for Astro- and Particle Physics
(MIAPP) of the DFG cluster of excellence ``Origin and Structure of the
Universe'', and we are grateful to the organizers for creating a stimulating
atmosphere. We are also grateful to the Hausdorff Institute Bonn for
hospitality through the Hausdorff Trimester Program ``Periods in Number Theory,
Algebraic Geometry and Physics'' when parts of this work have been done.
CRM is supported by a University Research Fellowship from the Royal
Society. The research of OS was supported in part by Perimeter Institute for
Theoretical Physics. Research at Perimeter Institute is supported by the
Government of Canada through the Department of Innovation, Science and Economic
Development Canada and by the Province of Ontario through the Ministry of
Research, Innovation and Science.

%********************************************
\appendix{A}{Stirling cycle permutation sums}
\applab\stirlingapp

\noindent In order to explain the Stirling cycle permutation sums used
throughout this work it is convenient to start by briefly recalling the
definition, using the notation and terminology proposed in \tnn, of the {\it
Stirling cycle numbers} ${n\stirling p}$ and {\it Stirling set numbers}
${n\stirlingtwo p}$.

The Stirling set number ${n\stirlingtwo p}$
represents the number of ways to partition a set of $n$ elements into
$p$ non-empty sets \knuthconcrete. For example, ${4\stirlingtwo 2}=7$ because there
are seven ways to split the set $\{1,2,3,4\}$ into two non-empty subsets:
\eqnn\stirsetEx
$$\eqalignno{
&\{1,2,3\}\cup\{4\},\quad
\{1,2,4\}\cup\{3\},\quad
\{1,3,4\}\cup\{2\},\quad
\{2,3,4\}\cup\{1\}, &\stirsetEx\cr
&\{1,2\}\cup\{3,4\},\quad
\{1,3\}\cup\{2,4\},\quad
\{1,4\}\cup\{2,3\}.
}$$
The Stirling cycle number ${n\stirling p}$ is closely related and represents
the number of ways to split $n$ objects into $p$ cycles\foot{A cycle
is defined up to cyclic rearrangements;
$(12\ldots k)=(23\ldots k1)$.}.
It is easy to write
down the different arrangements of cycles once the Stirling
set partitions have been worked out: simply convert a given $k$-element
subset into its $(k{-}1)!$ distinct cycles as $\{1,2, \ldots,k\}\rightarrow
(123 \ldots k)+{\rm perm}(2,3,\ldots,k)$. For example, using the above
subset decomposition
of ${4\stirlingtwo2}$ we obtain ${4\stirling2}=11$:
\eqnn\stircycEx
$$\eqalignno{
&(123)(4),\quad(132)(4),\quad
(124)(3),\quad(142)(3),\quad
(134)(2),\quad(143)(2),&\stircycEx\cr
&(1)(234),\quad(1)(243),\quad
(12)(34),\quad(13)(24),\quad
(14)(23)\,.
}$$
Since there is no unique way of representing a product of disjoint
cycles we fix this ambiguity by ordering the cycles as follows:
\eqnn\cycord
\medskip
\item {\it i.} each cycle is written with its smallest element first,
\hfill\cycord\hfilneg
\item {\it ii.} the cycles are written in increasing order of its smallest element.
\medskip
\noindent For example, $(65)(471)(23)$ becomes $(147)(23)(56)$.
Given the above conventions we can now define:
\medskip
\proclaim Definition 2. The Stirling cycle permutation sum of a
generic object $S_{A_1,A_2, \ldots,A_p}$ with $p$ slots
is denoted by
\eqn\Sperm{
S_{A_1,A_2, \ldots,A_p} + [1,2, \ldots,n|A_1,A_2, \ldots,A_p]\, ,
}
and it represents the sum over all $n\stirling p$ ways to
partition the set $\{1,2, \ldots,n\}$ into $p$ cycles, ordered according
to \cycord, and that
are distributed
to $S_{A_1, \ldots,A_p}$ as follows,
$$
(a_1\ldots a_{n_a})(b_1\ldots b_{n_b})\cdots
(p_1\ldots p_{n_p})\rightarrow S_{a_1\ldots a_{n_a}, b_1\ldots b_{n_b}, \ldots,
p_1\ldots p_{n_p}}\,.
$$
\par
\medskip
\noindent To illustrate the above definition, let us consider
$C_{1|A,B,C} + \big[2,3,4,5,6|A,B,C\big]$.
In this case, all the $35$ partitions of
$\{2,3,4,5,6\}$ into $3$ cycles ordered according to the above convention
are given by
$$\eqalignno{
&(2)(3)(456),\;
 (2)(3)(465),\;
 (2)(356)(4),\;
 (2)(365)(4),\;
 (2)(346)(5),\cr
& (2)(364)(5),\;
 (2)(345)(6),\;
 (2)(354)(6),\;
  (256)(3)(4),\;
  (265)(3)(4),\cr
&  (246)(3)(5),\;
 (264)(3)(5),\;
 (245)(3)(6),\;
 (254)(3)(6),\;
 (236)(4)(5),\cr
& (263)(4)(5),\;
 (235)(4)(6),\;
 (253)(4)(6),\;
 (234)(5)(6),\;
 (243)(5)(6),\cr
& (2)(34)(56),\;
 (2)(35)(46),\;
 (2)(36)(45),\;
 (24)(3)(56),\;
 (25)(3)(46),\cr
& (26)(3)(45),\;
 (23)(4)(56),\;
 (25)(36)(4),\;
 (26)(35)(4),\;
 (23)(46)(5),\cr
& (24)(36)(5),\;
 (26)(34)(5),\;
 (23)(45)(6),\;
 (24)(35)(6),\;
 (25)(34)(6)\,.
}$$
Therefore
\eqnn\CABCsix
$$\displaylines{
C_{1|A,B,C} + \big[2,3,4,5,6|A,B,C\big] = \hfil\CABCsix\hfilneg\cr
C_{1|2,3,456} +
 C_{1|2,3,465} +
 C_{1|2,356,4} +
 C_{1|2,365,4} +
 C_{1|2,346,5}
+ C_{1|2,364,5} +
 C_{1|2,345,6}\,\phantom{.}\cr +
 C_{1|2,354,6} +
  C_{1|256,3,4} +
  C_{1|265,3,4}
+  C_{1|246,3,5} +
 C_{1|264,3,5} +
 C_{1|245,3,6} +
 C_{1|254,3,6}\,\phantom{.}\cr +
 C_{1|236,4,5}
+ C_{1|263,4,5} +
 C_{1|235,4,6} +
 C_{1|253,4,6} +
 C_{1|234,5,6} +
 C_{1|243,5,6}
 +C_{1|2,34,56}\,\phantom{.}\cr +
 C_{1|2,35,46} +
 C_{1|2,36,45} +
 C_{1|24,3,56} +
 C_{1|25,3,46}
 +C_{1|26,3,45} +
 C_{1|23,4,56} +
 C_{1|25,36,4} \,\phantom{.}\cr +
 C_{1|26,35,4} +
 C_{1|23,46,5}
 + C_{1|24,36,5} +
 C_{1|26,34,5} +
 C_{1|23,45,6} +
 C_{1|24,35,6} +
 C_{1|25,34,6}\,.
}$$
For some typical numbers appearing in this work, we note that
the total number of terms in the local representation \dzeroGeneral\ of $\cK_n^{(0)}(\ell)$
is given by $T_n\equiv {n\stirling 4} + {n\stirling 5} + \cdots + {n\stirling n}$,
while in the corresponding manifestly BRST-invariant representation they become
$C_n\equiv {n-1\stirling 3} + {n-1\stirling 4} + \cdots + {n-1\stirling n-1}$.
For example, $T_n = 1,11,101,932,9080,94852,1066644$ and
$C_n = 1,7,46,326,2556,22212,212976$ for $n=4,5,6,7,8,9,10$.

%*************************************************************************
\subsubsec Stirling cycle permutations of the seven-point $d=0$ correlator

For convenience and to provide yet another explicit example of a Stirling cycle
permutation sum, we write down the complete expansion
of the unrefined Lie polynomials in \sevenLie,
\eqnn\sevenloc
$$\eqalignno{
{\cal K}^{(0)}_7(\ell) &= {1\over6}V_1
T^{mnp}_{2,3,4,5,6,7}\cZ^{mnp}_{1,2,3,4,5,6,7}   &\sevenloc\cr
&+ {1\over 2}\big[V_{12}
T^{mn}_{3,4,5,6,7}\cZ^{mn}_{12,3,4,5,6,7}  + (2\leftrightarrow 3,\ldots,7) \big]\cr
&+ {1\over 2}\big[ V_1T^{mn}_{23,4,5,6,7}\cZ^{mn}_{1,23,4,5,6,7}  + (2,3|2,3,\ldots,7) \big] \cr
&+\big[
(V_{123}T^m_{4,5,6,7}\cZ^m_{123,4,5,6}+V_{132}T^m_{4,5,6,7}\cZ^m_{132,4,5,6}) + (2,3|2,3,\ldots,7)\big]\cr
&+ \big[ \big(V_1T^m_{234,5,6,7}\cZ^m_{1,234,5,6,7}+V_1T^m_{243,5,6,7}\cZ^m_{1,243,5,6,7}\big) + (2,3,4|2,3,\ldots,7)\big]\cr
&+\big[ \big(V_{12}T^m_{34,5,6,7}\cZ_{12,34,5,6,7}^m  + (3,4|3,4,\ldots,7)\big)
+ (2\leftrightarrow 3,\ldots,7) \big] \cr
&+\big[\big( V_{1}T^m_{23,45,6,7}\cZ^m_{1,23,45,6,7} + {\rm cyc}(2,3,4)\big) +(6,7|2,3,\ldots,7) \big] \cr
&+  \big[ \big(V_{1234}T_{5,6,7}\cZ_{1234,5,6,7} + {\rm perm}(2,3,4) \big)  + (2,3,4|2,3,\ldots,7) \big]\cr
&+ \big[ \big(V_{123}T_{45,6,7}\cZ_{123,45,6,7}+V_{132}T_{45,6,7}\cZ_{132,45,6,7}
+(4,5|4,5,6,7) \big) +(2,3|2,\ldots,7) \big] \cr
&+ \big[ \big(V_{12}T_{345,6,7}\cZ_{12,345,6,7} +V_{12}T_{354,6,7}\cZ_{12,354,6,7}+{\rm cyc}(2,3,4,5)\big) +(6,7|2,\ldots,7)  \big]\cr
&+  \big[\big( V_{12}T_{3,45,67}\cZ_{12,3,45,67}+V_{13}T_{2,45,67}\cZ_{13,2,45,67} + {\rm cyc}(4,5,6)\big) +(2,3|2,\ldots,7) \big]\cr
&+ \big[ \big(V_1T_{2345,6,7}\cZ_{1,2345,6,7}+ {\rm perm}(3,4,5) \big) + (2,3,4,5|2,3,\ldots,7) \big] \cr
&+ \big[\big(V_{1}T_{23,4,567}\cZ_{1,23,4,567} +V_{1}T_{23,4,576}\cZ_{1,23,4,576} + {\rm cyc}(2,3,4)\big) + (2,3,4|2,\ldots,7)\big]\cr
&+  \big[ \big( V_{1}T_{23,45,67}\cZ_{1,23,45,67}  + {\rm cyc}(4,5,6) \big) + (3\leftrightarrow 4,5,6,7) \big] \, .
}$$
It is straightforward but tedious to see that there are $932$ terms above,
reproducing the number $T_7=932$ discussed above.

%***********************************
\newsubsec\liepolsec Lie polynomials

There are several characterizations of a Lie polynomial in the mathematics
literature, see for example \reutenauer. For our purposes, Lie polynomials are composed by linear
combinations of nested commutators in a given set of non-commutative
indeterminates. For example, if $t^{a_1}, t^{a_2}, t^{a_3}$ are non-commutative,
$P=[t^{a_1},[t^{a_2},t^{a_3}]] + 3\mkern1mu[[t^{a_1},t^{a_2}],t^{a_3}]$ is a Lie polynomial
while $N=t^{a_1}t^{a_2}t^{a_3}$ is
not.

The identification of a Lie-polynomial structure within the correlators of this
work stems from a theorem proved by Ree \Ree. Using the notation of
section~\wordssec,
the theorem states that if $M_A$ satisfies shuffle
symmetries (i.e., $M_{R\shuffle S}=0$, for any $R,S\neq\emptyset$) and
$t^{a_i}$ are non-commutative indeterminates with $t^A \equiv t^{a_1}t^{a_2}
\ldots t^{a_p}$,
a sum over all words $A$ of length $p$ of the form
\eqn\ReeTheo{
P=\sum_A M_A t^A\,
}
gives rise to a Lie polynomial of degree $p$. For example, at degree two the
shuffle symmetry on $M_{a_1a_2}$ implies that $M_{a_2a_1} = - M_{a_1a_2}$ and
the sum \ReeTheo\ becomes $P=M_{a_1a_2}t^{a_1}t^{a_2} +  M_{a_2a_1}t^{a_2}t^{a_1} =
M_{a_1a_2}[t^{a_1},t^{a_2}]$. Hence $P$ is a Lie polynomial.

In general, the sum in \ReeTheo\ can be rewritten as a sum proportional to $\sum M_A
t^{\ell(A)}$ where $\ell(A)$ is the Dynkin map defined in \ellmap. Thus, the Lie
polynomial arising from \ReeTheo\ has the form of a sum over products of objects satisfying
shuffle symmetries and objects satisfying generalized Jacobi symmetries.
Schematically we have $P=\sum(\hbox{shuffle})(\hbox{Lie})$. This is
precisely the structure within each word (slot) in
the local form of the one-loop correlators found in this work, see for 
example \dzeroGeneral.

%%%%%%%%%%%%%%%%%%%%%%%%%%%%%%%%%%%%%%%%%%%%%%%%%%%%%%%%%%%%%%
\appendix{B}{Monodromy invariance of the six-point correlator}
\applab\LocalMonapp

\noindent
In this appendix we demonstrate the monodromy invariance of the
six-point correlator in its local representation \sixLie\ and thereby
provide an alternative to the proof in section \Dsixproofsec\ with 
manifest BRST invariance. It will be convenient to define the following 
shorthands:
\eqnn\shor
$$\eqalignno{
X^{(a)}_{1|23,4,5,6} &\equiv V_1k_1^p T^{p}_{23,4,5,6}- V_{231}T_{4,5,6}
- V_{41}T_{23,5,6} - V_{51}T_{23,4,6} - V_{61}T_{23,4,5} &\shor\cr
X^{(a)m}_{1|2,3,4,5,6} &\equiv
V_1k_1^n T^{mn}_{2,3,4,5,6} - V_{21}T^m_{3,4,5,6}
- V_{31}T^m_{2,4,5,6}
- V_{41}T^m_{2,3,5,6}
- V_{51}T^m_{2,3,4,6}
- V_{61}T^m_{2,3,4,5}
\cr
X^{(b)}_{13|2|4,5,6} &\equiv V_{13}k_2^p T^{p}_{2,4,5,6}- V_{132}T_{4,5,6}
- V_{13}T_{42,5,6} - V_{13}T_{52,4,6} - V_{13}T_{62,4,5}\cr
X^{(b)}_{1|2|34,5,6} &\equiv V_{1}k_2^p T^{p}_{2,34,5,6}- V_{12}T_{34,5,6}
- V_1 T_{342,5,6} - V_1T_{34,52,6} - V_1 T_{34,62,5}\cr
X^{(b)m}_{1|2|3,4,5,6} &\equiv V_1k_2^n T^{mn}_{2,3,4,5,6}
- V_{12}T^m_{3,4,5,6} - V_1T^m_{32,4,5,6}
- V_1T^m_{42,3,5,6}
- V_1T^m_{52,3,4,6}
- V_1T^m_{62,3,4,5}\,.
}$$
Using the monodromy variations of the six-point functions \sixMonApp\
we get the following variation of the correlator \sixLie,
\eqn\DKsix{
D\cK_6(\ell)  = \Omega_1 \d \cK_6^{(1)} + \Omega_2 \d \cK_6^{(2)} + \cdots +
\Omega_6 \d \cK_6^{(6)}\, ,
}
where
\eqnn\omeOne
\eqnn\omeTwo
$$\eqalignno{
\d \cK_6^{(1)} & = E^m_{1|2,3,4,5,6} X^{(a)m}_{1|2,3,4,5,6} 
 + \big[E_{1|23,4,5,6}X^{(a)}_{1|23,4,5,6} + (2,3|2,3,4,5,6)\big] &\omeOne\cr
\d \cK_6^{(2)} & = E^m_{2|1,3,4,5,6}X^{(b)m}_{1|2|3,4,5,6}
 + \big[E_{2|13,4,5,6}X^{(b)}_{13|2|4,5,6} + (3 \leftrightarrow 4,5,6)\big]&\omeTwo\cr
&\ \ \  + \big[E_{2|34,1,5,6}\,X^{(b)}_{1|2|34,5,6}
+ (3,4 |3,4,5,6)\big]\, ,\cr
}$$
and the other $\d \cK_6^{(i)}$ for $i=3,4,5,6$ are obtained by relabeling of
$\d \cK_6^{(2)}$ in \omeTwo. Since the bookkeeping variables $\Omega_i$ are 
independent, all the $\d \cK_6^{(i)}$ must vanish separately.

For the $\Omega_1$ terms in \omeOne, after using the BRST identities
$$\eqalignno{
QJ^m_{1|2,3,4,5,6} &= X^{(a)m}_{1|2,3,4,5,6}
+ \Delta^m_{1,2,3,4,5,6}
+ \big[k_2^m (V_2 J_{1|3,4,5,6} - \cY_{12,3,4,5,6})
+ (2\leftrightarrow 3,4,5,6)\big]\cr
QJ_{1|23,4,5,6} &= X^{(a)}_{1|23,4,5,6} + s_{23}(V_2J_{1|3,4,5,6}
-V_3J_{1|2,4,5,6})  + Y_{1,23,4,5,6}\cr
}$$
together with their elliptic counterpart
$k_2^m E^m_{1|2,3,4,5,6} = - \big[s_{23}E_{1|23,4,5,6} + (3\leftrightarrow4,5,6)\big]$,
one arrives at a BRST-exact variation since the unrefined $\Delta^{m \ldots}_{A|B,C,
\ldots}$ are BRST exact \partI,
\eqnn\varoneS
$$\eqalignno{
\d K_6^{(1)} & = E^m_{1|2,3,4,5,6}\Big(QJ^m_{1|2,3,4,5,6} -
\Delta^m_{1|2,3,4,5,6}\big) &\varoneS\cr
&\ \ \  + \Big[E_{1|23,4,5,6}\big(
QJ_{1|23,4,5,6} - s_{23}\Delta_{1|23,4,5,6}\big)
+ (2,3|2,3,4,5,6)\Big]\,.\cr
}$$
For the $\Omega_2$ terms in \omeTwo, upon using BRST identities (see section 8 of \partI\ for the $D_{\ldots}$)
\eqnn\QXbs
$$\eqalignno{
QD^m_{1|2|3,4,5,6} & = X^{(b)m}_{1|2|3,4,5,6}
 - k_2^m V_1J_{2|3,4,5,6} + \Delta^m_{1|2,3,4,5,6}
 + \Big[ {k_3^m\over s_{13}}X^{(b)}_{13|2|4,5,6} +
(3\leftrightarrow4,5,6) \Big]\cr
%%%
%%%
s_{34}QD_{1|2|34,5,6} &= X^{(b)}_{1|2|34,5,6}
+ {s_{34}\over s_{13}}X^{(b)}_{13|2|4,5,6}
- {s_{34}\over s_{14}}X^{(b)}_{14|2|3,5,6}
+ s_{34}\Delta_{1|34,2,5,6} &\QXbs
}$$
and relabellings of $k_2^m E^m_{1|2,3,4,5,6} = - \big[s_{23}E_{1|23,4,5,6} + (3\leftrightarrow4,5,6)\big]$,
one gets
\eqnn\vartwoS
$$\eqalignno{
\d K_6^{(2)} &= E^m_{2|1,3,4,5,6}(QD^m_{1|2|3,4,5,6} -
\Delta^m_{1|2,3,4,5,6}) &\vartwoS\cr
&\ \ \  + \big[ E_{2|34,1,5,6}(s_{34}QD_{1|2|34,5,6} - s_{34}\Delta_{1|34,2,5,6}) + (3,4|3,4,5,6) \big]
}$$
which vanishes in the cohomology for the same reason as above. Therefore, the
six-point correlator \sixLie\ is confirmed to be single valued.

The above proof can be extended to higher-point correlators, but since it is
simpler to prove monodromy invariance using a non-local representation
with manifest BRST invariance (see section \Dsixproofsec), we will omit 
further discussions.

%******************************************************************
\appendix{C}{Vanishing linear combinations of worldsheet functions}
\applab\allThetasapp

\noindent In this appendix we write down a few explicit expansions of the
vanishing linear combinations of worldsheet functions given by $\Theta^{(d)}$
from \ThetaSix.

At six points, the three topologies of worldsheet functions were expanded in
\DeltaZs\ and are easily checked to be zero.

%*******************
\subsec Seven points

At seven points, the inequivalent topologies of $\Theta^{(0)}$ are given by
\eqnn\gaugeW
$$\eqalignno{
\Theta^{(0)}_{2|1,34,56,7} &= k_2^m \cZ^m_{1,2,34,56,7}
-s_{12} \cZ_{12,34,56,7} + s_{27} \cZ_{1,27,34,56}&\gaugeW\cr
&+ s_{23} \cZ_{1,234,56,7} - s_{24} \cZ_{1,243,56,7}
+ s_{25} \cZ_{1,256,34,7} - s_{26} \cZ_{1,265,34,7}\cr
%%%
%%%
\Theta^{(0)}_{23|1,45,6,7} &= k_{23}^m \cZ^m_{1,23,45,6,7}
- s_{12} \cZ_{123,45,6,7}+s_{13} \cZ_{132,45,6,7} \cr
&-s_{24}\cZ_{1,3245,6,7} + s_{34} \cZ_{1,2345,6,7}
+ s_{25} \cZ_{1,3254,6,7} - s_{35} \cZ_{1,2354,6,7} \cr
&+\big[ s_{36} \cZ_{1,236,45,7} -s_{26} \cZ_{1,326,45,7}  + (6\leftrightarrow 7) \big]\cr
&-\cZ_{2|3,1,45,6,7} +\cZ_{3|2,1,45,6,7}\cr
%%%
%%%
\Theta^{(0)}_{234|1,5,6,7} &= k_{234}^m \cZ^m_{1,234,5,6,7} + \big[
s_{13}(\cZ_{1324,5,6,7} + \cZ_{1342,5,6,7})  \cr
& \ \ \ \ - s_{12} \cZ_{1234,5,6,7} -s_{14} \cZ_{1432,5,6,7}
+ (1\leftrightarrow 5,6,7) \big] \cr
& -\cZ_{2|34,1,5,6,7}-\cZ_{23|4,1,5,6,7}
+\cZ_{34|2,1,5,6,7}+\cZ_{4|23,1,5,6,7}\cr
%%%
%%%
\Theta^{(0)}_{4|123,5,6,7} &= k_4^m \cZ^m_{123,4,5,6,7} + s_{34} \cZ_{4321,5,6,7} + s_{14} \cZ_{4123,5,6,7}\cr
& - s_{24}(\cZ_{4213,5,6,7}+\cZ_{4231,5,6,7})  + \big[ s_{45}\cZ_{123,45,6,7} + (5\leftrightarrow 6,7) \big]\cr
%%%
%%%
\Theta^{(0)\,m}_{2|1,34,5,6,7} &= k_2^n \cZ^{mn}_{1,2,34,5,6,7} - s_{12} \cZ^m_{12,34,5,6,7} + s_{23} \cZ^m_{1,234,5,6,7} - s_{24} \cZ^m_{1,243,5,6,7} \cr
& + \big[  s_{25} \cZ^m_{1,25,34,6,7} + (5\leftrightarrow 6,7) \big]
- k_2^m\cZ_{2|1,34,5,6,7} \cr
\Theta^{(0)\,m}_{23|1,4,5,6,7} &= k_{23}^n  \cZ^{mn}_{1,23,4,5,6,7}
- s_{12} \cZ^{m}_{123,4,5,6,7} + s_{13} \cZ^{m}_{132,4,5,6,7} \cr
&  + \big[s_{34} \cZ^m_{1,234,5,6,7}- s_{24} \cZ^m_{1,324,5,6,7}
+ (4\leftrightarrow 5,6,7) \big]\cr
&- k_{23}^m\cZ_{23|1,4,5,6,7} - \cZ^m_{2|3,1,4,5,6,7} + \cZ^m_{3|2,1,4,5,6,7}\cr
%%%
%%%
\Theta^{(0)\,mn}_{2|1,3,4,5,6,7} &= k_2^p \cZ^{mnp}_{1,2,3,4,5,6,7}
- s_{12} \cZ^{mn}_{12,3,4,5,6,7}
+ \big[s_{23} \cZ^{mn}_{1,23,4,5,6,7}+ (3\leftrightarrow 4,5,6,7)\big]\cr
& - k_2^{(m}\cZ^{n)}_{2|1,3,4,5,6,7}\cr
%%%
%%%
\Theta^{(1)}_{2|3|1,4,5,6,7} &=
- k_2^p \cZ_{3|2,1,4,5,6,7}
- s_{23}\cZ_{23|1,4,5,6,7}
- \big[s_{21}\cZ_{3|21,4,5,6,7}+(1\leftrightarrow4,5,6,7)\big]
}$$
and can also be verified to be zero up to total derivatives.

%*******************
\subsec Eight points

At eight points, the following topologies can be shown to vanish up to total derivatives:
\def\Thetathree#1#2#3{V_{#1}V_{#2}T^{mnp}_{#3}\Theta^{(0)\,mnp}_{#2|#1,#3}}
\def\Thetatwo#1#2#3{V_{#1}V_{#2}T^{mn}_{#3}\Theta^{(0)\,mn}_{#2|#1,#3}}
\def\Thetaone#1#2#3{V_{#1}V_{#2}T^{m}_{#3}\Theta^{(0)\,m}_{#2|#1,#3}}
\def\Thetazero#1#2#3{V_{#1}V_{#2}T_{#3}\Theta^{(0)}_{#2|#1,#3}}
\eqnn\Ztopsol
$$\eqalignno{
\Thetathree{1}{2}{3,4,5,6,7,8}&\cong0,\qquad\Thetatwo{1}{2}{34,5,6,7,8}\cong0 &\Ztopsol\cr
\Thetatwo{1}{23}{4,5,6,7,8}&\cong0,\qquad\Thetaone{1}{2}{34,56,7,8}\cong0\cr
\Thetaone{1}{2}{345,6,7,8}&\cong0,\qquad\Thetaone{1}{23}{45,6,7,8}\cong0\cr
\Thetaone{1}{234}{5,6,7,8}&\cong0,\qquad\Thetazero{1}{23}{45,67,8}\cong0\cr
\Thetazero{1}{234}{56,7,8}&\cong0,\qquad\Thetazero{1}{2345}{6,7,8}\cong0\cr
\Thetazero{1}{2}{34,56,78}&\cong0,\qquad\Thetazero{1}{2}{345,67,8}\cong0\cr
\Thetazero{1}{2}{3456,7,8}&\cong0,\qquad\Thetazero{1}{23}{456,7,8}\cong0\cr
}$$
as well as
\eqnn\eightJs
$$\eqalignno{
V_{1}V_2J_{34|5,6,7,8}\Theta^{(1)}_{2|34|1,5,6,7,8}&\cong0,\qquad
V_{1}V_2J_{3|45,6,7,8}\Theta^{(1)}_{2|3|1,45,6,7,8}\cong0 &\eightJs\cr
V_1V_2J^m_{3|4,5,6,7,8}\Theta^{(1)\,m}_{2|3|1,4,5,6,7,8}&\cong0,\qquad
V_{1}V_{23}J_{4|5,6,7,8}\Theta^{(1)}_{23|4|1,5,6,7,8}\cong0 \, .\cr
}$$
The coefficients of $V_{1A}$ with $A\neq \emptyset$ are just relabellings of the 
$\Theta^{(d)}$ in \Ztopsol\ and \eightJs\ and therefore vanish as well.

The explicit expansion of all eight-point topologies from \Ztopsol\ and
\eightJs\ is somewhat lengthy, so let us display just a couple of examples. It is
not hard to be convinced that their vanishing, up to total derivatives, is a
non-trivial statement:
\eqnn\nontTh
$$\eqalignno{
%+ V(1)*V(2,3)*Ti(4,5,[c],6,7,[c],8) * (
\Theta^{(0)}_{23|1,45,67,8} &=
         k_{23}^m \cZ^m_{1,23,45,67,8}
	+ \big[ s_{38} \cZ_{1,238,45,67}
          + s_{13} \cZ_{132,45,67,8} - (2\leftrightarrow 3) \big]\cr
&          + \big[(s_{24} + s_{34})\cZ_{1,2345,67,8} - (4\leftrightarrow5)\big]
	  + \big[(s_{26} + s_{36})\cZ_{1,2367,45,8} - (6\leftrightarrow7)\big]\cr
&          + \big[s_{24}(\cZ_{1,2435,67,8} + \cZ_{1,2453,67,8}) -(4\leftrightarrow5)\big]\cr
&	  + \big[s_{26}(\cZ_{1,2637,45,8} + \cZ_{1,2673,45,8}) - (6\leftrightarrow7)\big]\cr
&  - \cZ_{2|1,3,45,67,8}
          + \cZ_{3|1,2,45,67,8} \cong 0\,,\cr
\Theta^{(0)}_{2|1,345,67,8} &=
           k_2^m \cZ_{1,2,345,67,8}
          + s_{23} \cZ_{1,2345,67,8}
          - s_{24} (\cZ_{1,2435,67,8} + \cZ_{1,2453,67,8})\cr
&          + s_{25} \cZ_{1,2543,67,8}
          + \big[s_{26}\cZ_{1,267,345,8} - (6\leftrightarrow7)\big]  &\nontTh\cr 
&          + s_{28} \cZ_{1,28,345,67}
          - s_{12} \cZ_{12,345,67,8} \cong 0\,, \cr
\Theta^{(0)\,m}_{234|1,5,6,7,8} &=
k_{234}^p\cZ^{pm}_{234,1,5,6,7,8} - k_{234}^m \cZ_{234|1,5,6,7,8}  \cr
&+ \big[s_{14}\cZ^m_{2341,5,6,7,8}
{-} s_{13}(\cZ^m_{2431,5,6,7,8} {+} \cZ^m_{4231,5,6,7,8}){+} s_{12}\cZ^m_{4321,5,6,7,8}
{+} (1{\leftrightarrow}5,6,7,8)\big] \cr
&-\cZ^m_{2|34,1,5,6,7,8}
-\cZ^m_{23|4,1,5,6,7,8}+\cZ^m_{34|2,1,5,6,7,8}
+\cZ^m_{4|23,1,5,6,7,8}\cong0\,,\cr
\Theta^{(1)}_{3|4|12,5,6,7,8} &=-k_3^p\cZ^p_{4|12,3,5,6,7,8}
- s_{34}\cZ_{34|12,5,6,7,8}
- s_{31}\cZ_{4|312,5,6,7,8}\cr
&+ s_{32}\cZ_{4|321,5,6,7,8}
- \big[s_{35}\cZ_{4|35,12,6,7,8} + (5\leftrightarrow6,7,8)\big]\cong0\ .
}$$

%******************************
\appendix{D}{Proof of \closedZ}
\applab\prfsixclosed

\noindent
The purpose of this appendix is to deliver intermediate steps in deriving
the manifestly BRST-invariant representation \closedZ\ of the
six-point closed-string correlator that has been proposed in \MafraNWR.

%*******************************************************************
\newsubsec\prfpartA Single contraction between left and right movers

The open-string contribution \prfeqCshort\ involving a single vector contraction between
left- and right-movers stems from the derivative
\eqnn\prfeqB
$$\eqalignno{
{\p {\cal K}_6(\ell) \over \p \ell_m}    &=   C^{mn}_{1|2,3,4,5,6} \big( \ell^n + \big[
 g^{(1)}_{12} k_2^n + (2\leftrightarrow 3,4,5,6) \big] \big) &\prfeqB\cr
 &\ \ \ + \big[ C^m_{1|23,4,5,6} s_{23} V_1(1,2,3) + (2,3|2,3,4,5,6) \big] \cr
 &\ \ \  -\big[ P_{1|2|3,4,5,6} k_2^m g_{12}^{(1)}+ (2 \leftrightarrow 3,4,5,6) \big]
}$$
of the $C\cdot E$ representation \lowenE\ (using the loop-momentum-dependent 
form of $E_{1|2|3,4,5,6}$ in the first line of \SixErefined). Upon integration over $\ell$, we obtain
\eqnn\prfeqC
$$\eqalignno{
 \Big[\Big[{\p {\cal K}_6(\ell) \over \p \ell_m} \Big] \Big]  &=C^{mn}_{1|2,3,4,5,6} \big[
 f^{(1)}_{12} k_2^n + (2\leftrightarrow 3,4,5,6) \big] &\prfeqC\cr
 &\ \ \ + \big[ C^m_{1|23,4,5,6} s_{23} V_1(1,2,3) + (2,3|2,3,4,5,6) \big] \cr
 & \ \ \ +\big[ P_{1|2|3,4,5,6} k_2^m (\nu_{12}-f_{12}^{(1)})+ (2 \leftrightarrow 3,4,5,6) \big] \cr
 %%%
 %%%
 &\cong \big[ C^m_{1|23,4,5,6} s_{23} f_{23}^{(1)} + (2,3|2,3,4,5,6) \big] \cr
 &\ \ \ + \big[ P_{1|2|3,4,5,6} k_2^m \nu_{12} + (2 \leftrightarrow 3,4,5,6) \big] \cr
 &\ \ \ +\big[ f^{(1)}_{12} \Big(k_2^n C^{mn}_{1|2,3,\ldots,6}  -k_2^m P_{1|2|3,4,5,6} +s_{23}C^m_{1|23,4,5,6} \cr
 & \ \ \ \ \ \ \ \ + s_{24}C^m_{1|24,3,5,6} +s_{25}C^m_{1|25,3,4,6} +s_{26}C^m_{1|26,3,4,5} \Big) + (2\leftrightarrow 3,4,5,6)\big] \, ,
 }$$
 and BRST-exactness of the coefficient of $f_{12}^{(1)}$ in the last two lines leads to \prfeqCshort. As we will see, 
 the $\nu_{1j}$-dependent terms in the second line of \prfeqA ,
 \eqnn\prfeqD
$$\eqalignno{
&\big[ C^m_{1|23,4,5,6} s_{23} f_{23}^{(1)} + (2,3|2,3,\ldots,6) \big] 
\big[ \tilde C^m_{1|23,4,5,6} s_{23} \bar f_{23}^{(1)} + (2,3|2,3,\ldots,6) \big]  \cr
& + \big[ P_{1|2|3,4,5,6} k_2^m \nu_{12} + (2 \leftrightarrow 3,4,5,6) \big]
\big[ \tilde C^m_{1|23,4,5,6} s_{23} \bar f_{23}^{(1)} + (2,3|2,3,\ldots,6) \big] \cr
&-\big[ C^m_{1|23,4,5,6} s_{23} f_{23}^{(1)} + (2,3|2,3,\ldots,6) \big] 
 \big[ \tilde P_{1|2|3,4,5,6} k_2^m \nu_{12} + (2 \leftrightarrow 3,4,5,6) \big] \cr
&- \big[ P_{1|2|3,4,5,6} k_2^m \nu_{12} + (2 \leftrightarrow 3,4,5,6) \big]
 \big[ \tilde P_{1|2|3,4,5,6} k_2^m \nu_{12} + (2 \leftrightarrow 3,4,5,6) \big] \, ,&\prfeqD
}$$
will cancel in the end (where all of \prfeqD\ is accompanied by a factor of ${ \pi \over \Im \tau}$).

%*************************************************************************************************
\newsubsec\prfpartB Contributions from two (anti-)holomorphic derivatives

We shall now elaborate on the contributions of the (anti-)holomorphic derivatives
$N_{1|2|3,\ldots}$ and $\tilde N_{1|2|3,\ldots}$ \prfeqF\ that arise in the expression \prfeqE\ for 
$[[{\cal K}_6(\ell)]]$. Combinations of both $N_{1|2|3,\ldots}$ and $\tilde N_{1|2|3,\ldots}$
can lead to the following two inequivalent situations,
 \eqnn\prfeqG
 \eqnn\prfeqH
$$\eqalignno{
N_{1|2|3,4,5,6} \tilde N_{1|3|2,4,5,6}&= 
\nu_{12} {\p \over \p z_2} \Big( {\pi \over \Im \tau} - \nu_{13} (s_{13}\bar f^{(1)}_{13}
+s_{23} \bar f^{(1)}_{23}- \big[ s_{34}\bar f^{(1)}_{34} + (4\leftrightarrow 5,6) \big]  ) \Big)
\cr
 &= 
 {\pi \over \Im \tau} \nu_{12} \nu_{13} s_{23} &\prfeqG \cr
 N_{1|2|3,4,5,6} \tilde N_{1|2|3,4,5,6}&= \nu_{12} {\p \over \p z_2} \Big( {\pi \over \Im \tau} - \nu_{12} (s_{12}\bar f^{(1)}_{12}
- \big[ s_{23}\bar f^{(1)}_{23} + (3\leftrightarrow 4,5,6) \big] )\Big)
 \cr
 &= - {\pi \nu_{12} \over \Im \tau}  \Big( s_{12}\bar f^{(1)}_{21}+ \big[ s_{23}\bar f^{(1)}_{23} + (3\leftrightarrow 4,5,6) \big] 
 +\nu_{12}(s_{12}{+}s_{23}{+}\ldots{+}s_{26}) \Big) \cr
 &= - {\pi \nu_{12} \over \Im \tau}   {\p \over \p \bar z_2} \log \hat {\cal I}_6
 \cong {\pi  \over \Im \tau} {\p \over \p \bar  z_2}  \nu_{12} = \Big( {\pi  \over \Im \tau}  \Big)^2\,,
 &\prfeqH
}$$
using $ {\p \over \p \bar z_2} \bar f^{(1)}_{2j} = -{\pi \over \Im \tau}$ and momentum conservation.
Hence, the part of $[[{\cal K}_6(\ell)]] \cdot [[\tilde {\cal K}_6(-\ell)]]$ with two factors of
$N_{1|2|3,\ldots}$ and $\tilde N_{1|2|3,\ldots}$ adds up to 
 \eqnn\prfeqI
$$\eqalignno{
&\big[ N_{1|2|3,4,5,6} P_{1|2|3,4,5,6} + (2\leftrightarrow 3,4,5,6) \big] \big[ \tilde N_{1|2|3,4,5,6} \tilde P_{1|2|3,4,5,6} + (2\leftrightarrow 3,4,5,6) \big]  \cr
&=  {\pi  \over \Im \tau}  \big[ P_{1|2|3,4,5,6} k_2^m \nu_{12} + (2 \leftrightarrow 3,4,5,6) \big]
 \big[ \tilde P_{1|2|3,4,5,6} k_2^m \nu_{12} + (2 \leftrightarrow 3,4,5,6) \big]  \cr
 &\ \ \ + \Big( {\pi  \over \Im \tau}  \Big)^2 \big[ |P_{1|2|3,4,5,6}|^2+ (2\leftrightarrow 3,4,5,6) \big]  \, .&\prfeqI
}$$
Note that the first line of the right-hand side cancels the last line in \prfeqD,
and the last term of \prfeqI\ will interfere with the crossterms to be discussed next.

%***********************************************************************
\newsubsec\prfpartC Contributions from one (anti-)holomorphic derivative

Finally, there are crossterm contributions to $[[{\cal K}_6(\ell)]] \cdot [[\tilde {\cal K}_6(-\ell)]]$
 \eqnn\prfeqJ
$$\eqalignno{
&N_{1|2|3,4,5,6} \tilde {\cal K}_6^{\rm open} =  \nu_{12}  {\p \over \p z_2}  \Big\{ {1\over 2} \tilde C^{mn}_{1|2,\ldots,6}
[[\bar E^{mn}_{1|2,\ldots,6}]] + \big( s_{23} \tilde C^{m}_{1|23,4,5,6} [[\bar E^{mn}_{1|23,4,5,6}]] + (2,3|2,\ldots,6) \big)\cr
 &
 \ \ + \Big[ \tilde P_{1|2|3,4,5,6} \Big( {\pi \over \Im \tau} + 2 s_{12} \bar f^{(2)}_{12}
  - \bar f^{(1)}_{12}\big[ s_{23}\bar f^{(1)}_{23}
  +(3\leftrightarrow 4,5,6) \big]
  \Big) + (2\leftrightarrow 3,4,5,6) \Big]   \Big\} \,,&\prfeqJ
}$$
where $ \tilde {\cal K}_6^{\rm open} $ given by \PSanomagain\ is obtained from
the $C\cdot E$ representation \lowenE\ of $\tilde {\cal K}_6(-\ell)$. The coefficient of
$\tilde P_{1|2|3,4,5,6} $ is most subtle to evaluate since the $z_2$-derivative of
the $\bar f^{(n)}_{ij}$ functions generates a Koba--Nielsen derivative w.r.t. $\bar z_2$:
 \eqnn\prfeqK
$$\eqalignno{
&\nu_{12}  {\p \over \p z_2}  \Big( {\pi \over \Im \tau} + 2 s_{12} \bar f^{(2)}_{12}
- \bar f^{(1)}_{12}\big[ s_{23}\bar f^{(1)}_{23}
  +(3\leftrightarrow 4,5,6) \big]
  \Big) \cr
  & \ \ \ =  {\pi \nu_{12} \over \Im \tau} \Big( s_{12}\bar f^{(1)}_{12}- \big[ s_{23}\bar f^{(1)}_{23}
  +(3\leftrightarrow 4,5,6)\big] \Big)  \cr
  & \ \ \ 
  = -  {\pi \nu_{12} \over \Im \tau}  {\p\over\p\bar z_2}
  \log \hat {\cal I}_6
  = {\pi\over\Im\tau}{\p\over\p\bar z_2}\nu_{12} = \big( {\pi\over\Im\tau}\Big)^2  &\prfeqK
}$$
Given that the only contributions to ${\p \over \p z_2} [[\bar
E^{\ldots}_{1|\ldots}]]$ arise from the quantity $L_0^m$ in \earlynu, the
remaining terms in \prfeqJ\ reduce to vector contractions of ${\p L_0^m
\over \p z_2} = -{\pi \over \Im \tau} k_2^m$,
\eqnn\prfeqL
$$\eqalignno{
N_{1|2|3,4,5,6} \tilde {\cal K}_6^{\rm open} &=
\big( {\pi  \over \Im \tau} \Big)^2 \tilde P_{1|2|3,4,5,6}
+ {\pi \nu_{12} \over \Im \tau} \Big\{ k_2^m \tilde C^{mn}_{1|2,3,4,5,6}
\big[ k_2^n \bar f^{(1)}_{12}+ (2\leftrightarrow 3,4,5,6) \big]\cr
&\ \ \ + k_2^m \big[ s_{23} \tilde C^m_{1|23,4,5,6}
(\bar f^{(1)}_{12}+\bar f^{(1)}_{23}+\bar f^{(1)}_{31}) + (2,3|2,3,4,5,6) \big]\cr
&\ \ \ - \big[ s_{23} \tilde P_{1|3|2,4,5,6} \bar f^{(1)}_{13}
+ (3\leftrightarrow 4,5,6) \big] \Big\}  &\prfeqL \cr
&=  \big( {\pi  \over \Im \tau} \Big)^2 \tilde P_{1|2|3,4,5,6}
+ {\pi \nu_{12} \over \Im \tau} k_2^m
\big[ s_{23} \tilde C^m_{1|23,4,5,6}  \bar f^{(1)}_{23} + (2,3|2,3,4,5,6) \big]\,,
}$$
where we have repeated the simplifications of \prfeqC\ in the last step. By adjoining permutations
and the complex conjugate of \prfeqL, one arrives at
\eqnn\prfeqM
$$\eqalignno{
&- \big[ N_{1|2|3,4,5,6} P_{1|2|3,4,5,6} + (2\leftrightarrow 3,4,5,6) \big]
\tilde {\cal K}_6^{\rm open}
-\big[ \tilde N_{1|2|3,4,5,6} \tilde P_{1|2|3,4,5,6}
+ (2\leftrightarrow 3,4,5,6) \big]  {\cal K}_6^{\rm open} \cr
&= -  {\pi  \over \Im \tau}   \big[ P_{1|2|3,4,5,6} k_2^m \nu_{12}
+ (2 \leftrightarrow 3,4,5,6) \big] \big[ s_{23} \tilde C^m_{1|23,4,5,6}
\bar f^{(1)}_{23} + (2,3|2,3,4,5,6) \big]
\cr
&\ \ \ + {\pi  \over \Im \tau} \big[ s_{23}C^m_{1|23,4,5,6} f^{(1)}_{23}
+ (2,3|2,3,4,5,6) \big]  \big[ \tilde P_{1|2|3,4,5,6} k_2^m \nu_{12}
+ (2 \leftrightarrow 3,4,5,6) \big] \cr
&\ \ \ - 2\big({\pi\over\Im\tau} \Big)^2 \big[ |P_{1|2|3,4,5,6}|^2
+ (2\leftrightarrow 3,4,5,6) \big]\,,&\prfeqM
}$$
where the last term interferes with \prfeqI, and the remaining terms on the
right-hand side cancel the crossterms in the second and third line of \prfeqD.
In summary, by combining \prfeqD, \prfeqI\ and \prfeqM, one arrives at the most
subtle contributions $\sim C^m_{1|23,\ldots} \tilde C^m_{1|ij,\ldots}$ and
$|P_{1|i|j,\ldots}|^2$ in \closedZ, concluding the derivation of the proposal in
\MafraNWR.

\listrefs

\bye